\documentclass[sigconf, anonymous=false, review=false]{acmart} % sigconf class ensures the correct formatting for SIGMOD including the letter-size pages (8,5'' x 11'')

\usepackage{amsmath}
\usepackage{bm}
\usepackage[ruled,vlined,linesnumbered]{algorithm2e}
\usepackage{caption}
\usepackage{subcaption}
\usepackage[toc,page]{appendix}

\newcommand{\cut}[1]{}
\newcommand{\rev}[1]{#1}
\newcommand{\veat}[1]{\vspace{#1}}

\newboolean{show-tr}
\setboolean{show-tr}{true}
\newcommand{\techreport}[2]{\ifthenelse{\boolean{show-tr}}{{#1}}{{#2}}}

\newcommand{\graph}{\ensuremath{\boldsymbol{\mathcal{G}}}}
\newcommand{\sys}{Spade}
\newcommand{\ouralgo}{MVDCube}
\newcommand{\ouralgoshort}{MVD}
\newcommand{\pgcube}{PGCube}
\newcommand{\pgstar}{PGCube$^*$}
\newcommand{\pgdist}{PGCube$^d$}

\DeclareMathOperator{\Var}{Var}
\DeclareMathOperator{\Cov}{Cov}

\newtheorem{problemst}{Problem}
\newtheorem{lemma}{Lemma}
\newtheorem{theorem}{Theorem}

\newcommand{\ra}[1]{\renewcommand{\arraystretch}{#1}}

\AtBeginDocument{%
	\providecommand\BibTeX{{%
			\normalfont B\kern-0.5em{\scshape i\kern-0.25em b}\kern-0.8em\TeX}}}

\setcopyright{acmlicensed}
\copyrightyear{2021}
\acmYear{2021}
\acmDOI{10.1145/3448016.3457307}

\acmConference[SIGMOD '21]{Proceedings of the 2021 International Conference on Management of Data}{June 20--25, 2021}{Virtual Event, China}
\acmBooktitle{Proceedings of the 2021 International Conference on Management of Data (SIGMOD '21), June 20--25, 2021, Virtual Event, China}
\acmPrice{15.00}
\acmISBN{978-1-4503-8343-1/21/06}

\begin{document}
	\fancyhead{} % command requested by the editor
	
	\title{Efficient Exploration of Interesting Aggregates in RDF Graphs}

	\author{Yanlei Diao$^{1,2}$\qquad Pawe\l{} Guzewicz$^{1,2}$\qquad Ioana Manolescu$^{2,1}$\qquad Mirjana Mazuran$^{2,1}$}
	\affiliation{
		\institution{$^1$ Ecole Polytechnique, Institut Polytechnique de Paris, France\qquad $^2$ Inria, France}
		\country{}
	}
	\email{yanlei.diao@polytechnique.edu, {pawel.guzewicz, ioana.manolescu, mirjana.mazuran}@inria.fr}
	
%	\renewcommand{\shortauthors}{Trovato and Tobin, et al.}
	\renewcommand{\shortauthors}{Diao and Guzewicz, et al.}
	
	\begin{abstract}
As large Open Data are increasingly shared as RDF graphs today, there is a growing demand to help users discover the most interesting facets of a graph, which are often hard to grasp without automatic tools.
We consider the problem of \emph{automatically identifying the $k$ most interesting aggregate queries} that can be evaluated on an RDF graph, given an integer $k$ and a user-specified \emph{interestingness function}.
Our problem departs from analytics in relational data warehouses in that
($i$)~in an RDF graph we are not \emph{given} but we must \emph{identify} the facts, dimensions, and measures of candidate aggregates; % and aggregate functions; 
($ii$)~the classical approach to efficiently evaluating multiple aggregates breaks in the face of multi-valued dimensions in RDF data\cut{:  facts may have zero, one or more values for dimensions}.
In this work, we propose an \emph{extensible end-to-end framework} that enables the identification and evaluation of interesting aggregates based on a new \emph{RDF-compatible one-pass algorithm for efficiently evaluating a lattice of aggregates} and a novel \emph{early-stop technique} (with probabilistic guarantees) that can prune uninteresting %nonviable 
aggregates. %reduce the aggregate evaluation cost. 
Experiments using both real and synthetic graphs demonstrate the ability of our framework to find interesting aggregates in a large 
search space, the efficiency of our algorithms (with up to $2.9\times$ speedup over a similar pipeline based on existing algorithms), and scalability as the data size and complexity grow.
%\YD{More concrete results?}\IM{I found it hard to put more here}

%While large  Open Data are shared today as RDF graphs, 
%users need help finding out the
%most interesting facets of a graph they are not familiar
%with. We consider the problem \emph{
%  automatically identifying the $k$ most interesting aggregate
%  queries} which can be evaluated over the content of an RDF graph, given a user-specified integer $k$ and a
%user-chosen \emph{interestingness function}. 
%The problem, recalling  relational data warehouses (DWs), differs from
%it since: ($i$)~in an RDF graph we are not \emph{given} but must  \emph{identify}
%the facts, dimensions, measures, and aggregate functions;
%($ii$)~facts may have zero, one or more values for dimensions and measures.
%We propose an \emph{extensible end-to-end framework} for selecting and
%evaluating interesting aggregates, based on a \emph{one-pass
%  algorithm for efficiently evaluating  a lattice of RDF aggregates}
%and an \emph{early-stop technique} (with probabilistic guarantees) that
%can reduce the aggregate evaluation effort. 
%
%We describe experiments on real and synthetic graphs demonstrating the
%ability of our framework to find interesting aggregates, as well as
%its scalability wrt the graph size and complexity.
\end{abstract}
	
	
	
	
	\maketitle

	%\IM{For uniformity, please let's stick the ``informal enumeration'' style to ($i$)~sth, ($ii$)~sth else, etc. (the number in math mode, unbreakable white space after it)}
\section{Introduction}
RDF graphs are increasingly being published and shared as part of the Linked Open Data \cut{(LOD) }movement. Given the size, heterogeneity, and complexity of these graphs, their information content is hard to grasp, in particular for non-expert users.
In this work, we explore automatic insight extraction from RDF graphs~\cite{dagger,daggerplus,diao:hal-02152844}. %More specifically,
Given a graph and an integer $k$, we seek to automatically identify the $k$ most \emph{interesting insights} in the graph. An insight is an \emph{RDF analytical \cut{(aggregate) }query} that results in aggregated \emph{measures} over the data, grouped by a set of \emph{dimensions}. The query can be expressed in a language such as SPARQL 1.1, the W3C's standard RDF query language~\cite{sparql11}, and evaluated by any RDF query engine. The \emph{interestingness} of an insight is assessed based on a statistical measure of the query result.

\rev{\textbf{Motivating application.}
Computational Lead Finding (CLF) \cite{belfodil:hal-02383776,clf-query-perturbations} is one of the target applications of our work. For journalists, a ``lead” is an idea based on which they may write an interesting article. Given a dataset, CLF aims to automatically identify the interesting leads from the data. Below, we outline our approach to RDF insight extraction using examples from statistical lead discovery.
%Available data may contain multiple interesting leads, and CLF aims to identify them automatically. \sys{} achieves that on RDF graphs.
%For instance, on our CEOs real dataset, top-k results returned by \sys{} show insights such as: ``American philanthropists and stakeholders are the wealthiest CEOs overall''; ``there is a big difference in the number and wealth of male and female CEOs''; ``the number of Chinese CEOs is not very high but the sum of their wealth stands out compared to the sums in other nationalities''.
%The above insights can be used as ``leads'' if the journalist deems them interesting. In CLF, an interesting aggregate is one that surprises the journalist, i.e., deviating from their prior knowledge, which in turn is likely to surprise the audience. If the prior knowledge indicates the aggregate results (a) to be uniform across different groups, then variance as a statistical measure will identify the aggregates with results far from the average value; or (b) to follow a normal distribution (over a sorted dimension), then the third and fourth moments can capture the deviation from that.
}

\begin{figure*}[t!]
	\centering
	\includegraphics[width=\textwidth,height=50mm]{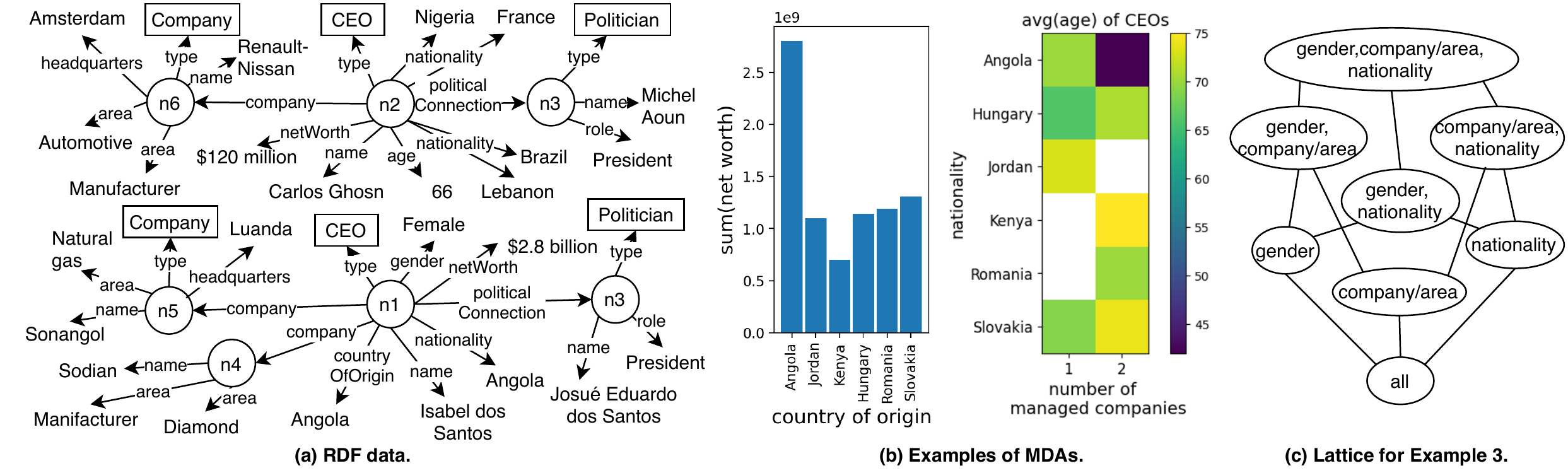}
	\veat{-8mm}
	\caption{Running examples using the CEOs dataset.\label{fig:ceos-running-example}}
	\Description{Running examples using the CEOs dataset.}
	\veat{-3mm}
\end{figure*}

\textit{Running examples.} Consider an RDF graph comprising politicians, CEOs, and connections between them. We can extract such a graph, for instance, from the WikiData open-source RDF repository. \cut{To do that, we identify RDF nodes of the types \href{https://www.wikidata.org/wiki/Q82955}{CEO} and \href{https://www.wikidata.org/wiki/Q484876}{politician}, respectively, and follow the edges in the neighborhood of these RDF nodes up to a certain distance.}
Figure~\ref{fig:ceos-running-example}(a) shows an example RDF graph where CEOs are linked with politicians, e.g., \href{https://www.wikidata.org/wiki/Q456034}{Isabel dos Santos}, a wealthy Angolan CEO (at the heart of the \href{https://www.icij.org/investigations/luanda-leaks/}{Luanda Leaks} scandal), is the daughter of a former president of Angola.
Starting from the graph, we aim to \emph{automatically identify} a small set of aggregate queries 
\rev{that are \emph{statistically interesting}. Here, interestingness is a statistical measure that indicates deviation from the prior knowledge of the journalists. For example, the interesting aggregate results may deviate from a uniform distribution of values over different aggregate groups, or a normal distribution over numeric dimensions such as \textit{age}.}

Table~\ref{tab:example-aggregates} shows three example aggregates, whose
dimensions and measures  are either properties in the RDF graph or properties that we derive  to enrich the scope of the analysis. In Example~1, \textit{CEO}s, \textit{politicalConnection}s, \textit{countryOfOrigin}, and \textit{netWorth} are either types or properties in the RDF graph in Figure~\ref{fig:ceos-running-example}(a). Example~2 analyzes the CEOs along \emph{the number of managed companies}, which is not a property in the graph: we \emph{derive} it by \emph{counting} the properties of  each CEO. This enables us to discover, e.g., that the average age of Angolan CEOs that manage two companies is low compared to other nationalities. Example~3 analyzes CEOs by \emph{areas of companies}; we derive this from the graph by following a path from the CEOs to the companies they manage, then to their areas. Similar path examples include \textit{company/headquarters}, \textit{politicalConnection/role}; longer paths produce a larger number of novel angles for the analysis.

Among all possible aggregate queries that we can generate, \rev{the above three examples are selected because their results show significant deviation from uniform values (having outliers)}. %, e.g., trend or unexpected spikes/dips.
%For Example~1, 
%Figure~\ref{fig:ceos-running-example}(b) exhibits a very high value for Angola due to Dos Santos. 
\rev{For Examples~1 and 2, Figure~\ref{fig:ceos-running-example}(b) shows respectively a histogram that exhibits an outlier in  $sum$(\textit{netWorth}) for Angola,  and a heat map where 
the dark color reflects a  low value of $avg$(\textit{age}) of CEOs, both due to Dos Santos. }
We can show to the user such interesting insights as ($i$)~histograms (if one-dimensional), ($ii$)~heat maps (if two-dimensional), or ($iii$)~tables (for high-dimensional aggregates). 
%See examples in Figure~\ref{fig:ceos-running-example}(b).

\begin{table}[t!]
	\begin{small}
		\begin{tabular}{@{}p{14mm}@{\hskip 2.5mm}p{50mm}@{}}%|p{1.5cm}|p{6cm}|
			\toprule
			%\noalign{\vskip -0.4mm}
			%\hline
			{\tt Example 1} & Sum of the net worth of CEOs with political connections grouped by country of origin.\\
			%\hline
			%\noalign{\vskip -0.4mm}
			{\tt Example 2} & Average age of CEOs grouped by nationality and number of managed companies.\\
			%\hline
			%\noalign{\vskip -0.4mm}
			{\tt Example 3} & Number of CEOs grouped by nationality, gender, and area of the companies they manage.\\
			%\noalign{\vskip -0.4mm}
			\bottomrule
			%\hline
		\end{tabular}
	\end{small}
	\caption{Examples of interesting aggregates.\label{tab:example-aggregates}}
	\veat{-9mm}
\end{table}

Our goal to discover the $k$ most interesting aggregates from an RDF graph poses two unique challenges:

\textbf{Challenge C1 - Aggregate identification.}  Automatic extraction of interesting aggregates is one among many existing techniques for data exploration and visualization recommendation. %in particular for multidimensional data. 
Yet, most prior works assume a fixed relational schema~\cite{botang,seedb}. In contrast, in RDF graphs, facts, dimensions, and measures are not specified but must be identified therein. %In Example~2 we automatically identified as facts \textit{CEO}s, as dimensions \textit{nationality} and \textit{``number of managed companies''}, and as measure \textit{age}. The measure is aggregated through the avg function. 
To address this challenge, given an RDF graph, we provide a variety of strategies to create new dimensions and measures, which enable us to examine a rich space of candidate aggregates and to discover the most interesting ones. We further develop a modular framework for aggregate identification, which can be extended or customized as needs arise.

%Data exploration through interesting aggregate extraction has been intensively studied for \emph{relational data warehouses} (DWs)~\cite{botang,seedb,qagview,deepeye}. However, in a DW, the facts, measures and dimensions are clearly identified through the analytical schema built by the warehouse designers.  Thus, a \emph{first challenge} is that 

\textbf{Challenge C2 - Efficient and correct aggregate evaluation.} Since we define interestingness on an aggregate \emph{result}, we must evaluate candidate aggregates to determine if they are among the $k$ most interesting ones. A key feature of our work is that we look for \emph{multidimensional aggregates} (MDAs), such as Examples~2 and~3. A set of $N$ dimensions, among which we enumerate candidate aggregates, leads to a lattice~\cite{ullman96} of $2^N$ nodes, each of which is an MDA (see Figure~\ref{fig:ceos-running-example}(c)). We may have many such %dimension sets
lattices to consider %simultaneously. Evaluating such large numbers of aggregates
at once, and efficiently evaluating them all
poses a salient challenge.
%However, it also leads to an \emph{explosion of our search space} as, the more new dimensions we create, the greater the number of candidate queries to consider. Such explosion is further emphasized by the fact that we look for  An MDA with $N$ dimensions determines a lattice~\cite{ullman96} of $2^N$ nodes (that is, MDAs) from the top (which groups by the $N$ dimensions) to intermediary levels (group by $N-1$, $N-2$ dimensions, etc.) to the bottom node that does not group at all (see Figure~\ref{fig:ceos-running-example}(c)). Once the dimensions at the top are identified, each lattice is a candidate MDA.

To address the challenge, first, we revisit a classical framework for \emph{lattice-based MDA computation} in \emph{relational data warehouses} (DWs).
Efficient algorithms, such as ArrayCube~\cite{molap}, compute an aggregate in the lattice from the result of one of its parents, and  
% this shares computation across all the nodes and 
compute all aggregates in the lattice in a single pass over the data.
However, a crucial observation we make in this work is that the classical one-pass approach to lattice computation is \emph{incorrect} for RDF data, due to a phenomenon called \emph{multi-valued dimensions}, that is, \emph{an RDF node (fact) may have multiple values along a given dimension}.
To tackle the issue, while retaining the benefits of one-pass algorithms, we provide a theoretical analysis of how the classical approach produces errors. Furthermore, we develop a new RDF-compatible one-pass algorithm that ($i$)~correctly and efficiently handles  lattice-based MDA computation where the aggregates use multi-valued dimensions; ($ii$)~for each node in the lattice (with a given set of dimensions), simultaneously handles many aggregates that differ in the measure (among many possible ones) and the  aggregate function in use; ($iii$)~saves computation cost by sharing measures across all lattices that analyze the same set of facts.

Second, to further improve efficiency, %aggregate evaluation performance, 
%\textbf{Challenge C3 - Aggregate evaluation and pruning.} 
%This is the most challenging step in our approach. The interestingness of an aggregate is computed by means of the statistic moments (variance, skewness, kurtosis). Thus, we cannot rely on the interestingness function to be monotonic. That is, if we know that aggregate $A_1$: \textit{``number of CEOs by nationality''} is not interesting, we cannot use this knowledge to determine if the aggregates whose dimensions overlap with $A_1$ (e.g., \textit{``number of CEOs by nationality and gender''}) are interesting or not. Thus, we cannot easily prune our huge search space by relying on apriori-like pruning strategies.
% \IM{I think this delves into too much details - monotonicity, kurtosis... I don't think this belongs here. It could be good to recycle it somewhere else}.
%
%To face \textbf{C3}, 
we develop a new technique to stop the evaluation of an MDA as soon as we can determine (with high probability) that it will not be among the top $k$. %based on an interestingness function of choice. 
Our technique builds on the work in~\cite{DBLP:conf/sigmod/HellersteinHW97}, which provides \emph{confidence-interval} (CI) bounds on an \emph{approximate} aggregate result.
Our problem is harder because we want to approximate \emph{the interestingness score computed over the aggregate result}, which amounts to estimating the result of a \emph{nested} aggregate query, whereas the prior work does not support such nested queries. Using advanced statistical tools, 
%such as the Delta Method\footnote{\url{https://en.wikipedia.org/wiki/Delta_method}}, 
we construct CIs for the interestingness function including \emph{variance}, \emph{skewness}, and \emph{kurtosis} over estimated results of candidate aggregates,  enabling early pruning of %nonviable
uninteresting aggregates.
%This allows us to focus  evaluation on the interesting aggregates, reducing the overall evaluation time. 

In summary, the contributions we make in this work include: 
%\begin{itemize}
%[nosep,leftmargin=1em,labelwidth=*,align=left]
%	\item 

$\bullet$ \sys{}, a new \emph{RDF-oriented end-to-end framework} that automatically identifies, enumerates, and efficiently evaluates RDF MDAs to determine the most interesting ones (Section~\ref{sec:overview});
	
%\item 
$\bullet$ \ouralgo{}, the first correct and efficient algorithm for \emph{one-pass lattice-based computation of RDF MDAs} (Section~\ref{sec:lattice-based-computation});
	
%\item 
$\bullet$ A novel \emph{early-stop technique} that stops the evaluation of MDAs that, with a high probability, will not be in the top-$k$ list (Section~\ref{sec:early-stop});
	
%\item  
$\bullet$ \emph{Experimental results} validating ($i$)~the ability of \sys{} to extract insights from a large space of candidate aggregates; ($ii$)~the frequent, and potentially high errors that existing algorithms introduce on real-life, heterogeneous RDF graphs; ($iii$)~the efficiency of our one-pass algorithm, which is faster than PostgreSQL's GROUP BY CUBE implementation by 20\% to 80\%; ($iv$)~the extra speedup of 10\% to 43\% achieved by our early-stop technique, and ($v$)~the scalability of \sys{} as the size and complexity of the graphs increase (Section~\ref{sec:experiments}). 
%\YD{Need to cite numbers. These results are too vague.}\IM{I added the main ones}
%\end{itemize}

%An additional difficulty, due to RDF graph heterogeneity, is that \emph{a given fact may have zero, one, or more values for a dimension}. % e.g., the nationality of a CEO can be unknown, or they may have multiple nationalities; the same may be true about measures. 
%This has a big impact on the lattice-based computation of MDAs as the following example shows. Consider \href{https://m.wikidata.org/wiki/Q356719}{Carlos Ghosn}, for many years CEO of Renault-Nissan. Ghosn is a citizen of Lebanon, France, Brazil and Nigeria. In a \emph{relational data warehouses} (DWs), such a CEO would be stored as a tuple in the fact table and their multiple nationalities would be modeled by means of a four-tuple relation. Then, we could find the result of Example~2 with a query $q$ that joins the two relations, groups the data by the dimensions and finally aggregates the measure. 

	\section{Problem statement and notation}

We consider RDF data defined over three pairwise disjoint sets: the set of URIs $\mathcal{U}$, the set of literals $\mathcal{L}$, and the set of blank nodes $\mathcal{B}$. An RDF graph $\graph$ is a finite set of triples of the form $(s, p, o)$, called subject, property, and object, such that $s \in (\mathcal{U} \cup \mathcal{B})$, $p \in \mathcal{U}$, and $o \in (\mathcal{U} \cup \mathcal{B} \cup \mathcal{L})$. The RDF property \textit{rdf:type} is used to attach types % (i.e. classes)
to an RDF node, which may have zero, one or several types.
\rev{Such an RDF graph may have an ontology stating relationships among its types and properties, e.g., any \textit{CEO} is a \textit{BusinessPerson}. An ontology leads to implicit triples that together with the triples explicitly present in $\graph$ are the graph's semantics. All the implicit triples can be materialized via \emph{saturation}, iteratively deriving new ones from $\graph$ and the rules; we consider ontologies for which this process is finite as in~\cite{goasdoue:hal-00804503}, and apply it prior to our analysis.}

A \textbf{candidate fact set} (CFS) is a set of RDF nodes that we build an \cut{interesting }aggregate on; we call a member of the set a \textbf{candidate fact} (CF).

An \textbf{attribute} is either a (direct) \textbf{property} (P) of a CF in the original RDF data, or a \textbf{derived property} (DP), which we create from the data and attach to a CF to enrich the analysis. For instance, one may attach to each CEO the number of companies \rev{they manage} (the full set of derivation strategies is discussed %. We support several derivation strategies (counting, language detection, keyword extraction, path derivations, as we detail
in Section~\ref{sec:overview}). %and envision that others can be flexibly added.
An attribute can be used as a \textbf{dimension}, to group CFs by value, or as a \textbf{measure}, to be aggregated within each group of CFs.

We employ an \textbf{aggregate function}, $f$, that ranges over the common set $\Omega=\{count, min, max, sum, avg\}$.

A \textbf{multidimensional aggregate} (MDA), $A=\langle CFS, \mathcal{D}, M, f\rangle$, is determined by: a CFS, a set $\mathcal{D}=\{D_1, D_2,\dots, D_N\}$ of dimensions (which are attributes), a measure $M$ (also an attribute), and an aggregate function $f$. The semantics of $A$  is that of a SPARQL 1.1 aggregate query~\cite{sparql11}, which also agrees with that of the RDF analytical queries introduced in~\cite{akbariazirani:hal-01187448,colazzo:hal-00960609}.
The result of $A$ on an RDF graph $\graph$, denoted $A(\graph)$, is the set of tuples, one per each distinct combination of dimension values (aggregate group) in the data:
\begin{align*}
A(\graph)=\{(  & d_1, d_2,\dots, d_N, f\{m_j\ |\ \exists\, CF_i\in CFS, CF_i.D_1=d_1, \\
& CF_i.D_2=d_2,\dots, CF_i.D_N=d_N, CF_i.M=m_j\})\}
\end{align*}

\noindent where $CF_i$ has (at least) the values $d_1,d_2, \ldots, d_N$ along the dimensions $D_1,D_2, \ldots, D_N$, and  $m_j$ iterates over the set of values of the measure $M$ on $CF_i$. Finally, $f\{\cdot\}$ is the result of running the aggregate function $f$ over the measure values from a given set.

Our semantics, unlike that of relational DWs, does account for \emph{heterogeneity in RDF data}: ($i$)~Some CFs may miss dimensions and/or measures, and thus they do not contribute to the result. For the graph in Figure~\ref{fig:ceos-running-example}, the result for Example~1 is \{(Angola, \$2.8B)\}, due to $n_1$, 
%Although it belongs to the CFS, 
whereas $n_2$ does not contribute to the result as it lacks the \textit{countryOfOrigin} dimension;
($ii$)~A CF may contribute to multiple groups in $A$ (if it has multiple values for a dimension), and/or multiple times to the aggregated value (if it has several values for the measure). The result for Example~2 is \{(Nigeria, 1, 65), (France, 1, 65), (Lebanon, 1, 65), (Brazil, 1, 65)\}, all obtained from $n_2$ given its four distinct values of \textit{nationality}. Although $n_1$ has both dimensions, it does not contribute to the result as it misses the \textit{age} measure.

An \textbf{interestingness function}, $h$, is applied over the result of an aggregate $A$. Let $W$ be the number of tuples in $A(\graph)$ and, for each tuple $t_i\in A(\graph)$, let $t_i.v$ be the aggregated value computed by $f$. Then $h$ takes the set $\{t_1.v, t_2.v, \dots, t_W.v\}$ and returns a score, i.e., a positive real number, reflecting a measure of interestingness of $A$. The user chooses the function to be used during the analysis. %\sys{} allows users to choose the function to be used during the analysis and it natively supports variance, skewness, and kurtosis. \PG{We repeat the same sentence in the next section while decribing the approach and features of \sys{}. In this section we didn't introduce \sys{} yet.}

Finally, the problem we address is stated as follows: 
\begin{problemst}
	\sloppy
	Given an RDF graph $\graph$, a positive integer $k$, and an interestingness function $h$ of choice, find the aggregates $A_1(\graph),\ldots, A_k(\graph)$ whose interestingness on $\graph$ is the highest. %\IM{I changed in A(G) because only this (evaluated, leading to concrete results) has a well-defined interestingness.
\end{problemst}

	\section{Overview of the approach\label{sec:overview}}

\begin{figure}[t!]
	\centering
	\includegraphics[width=0.95\columnwidth, height=50mm]{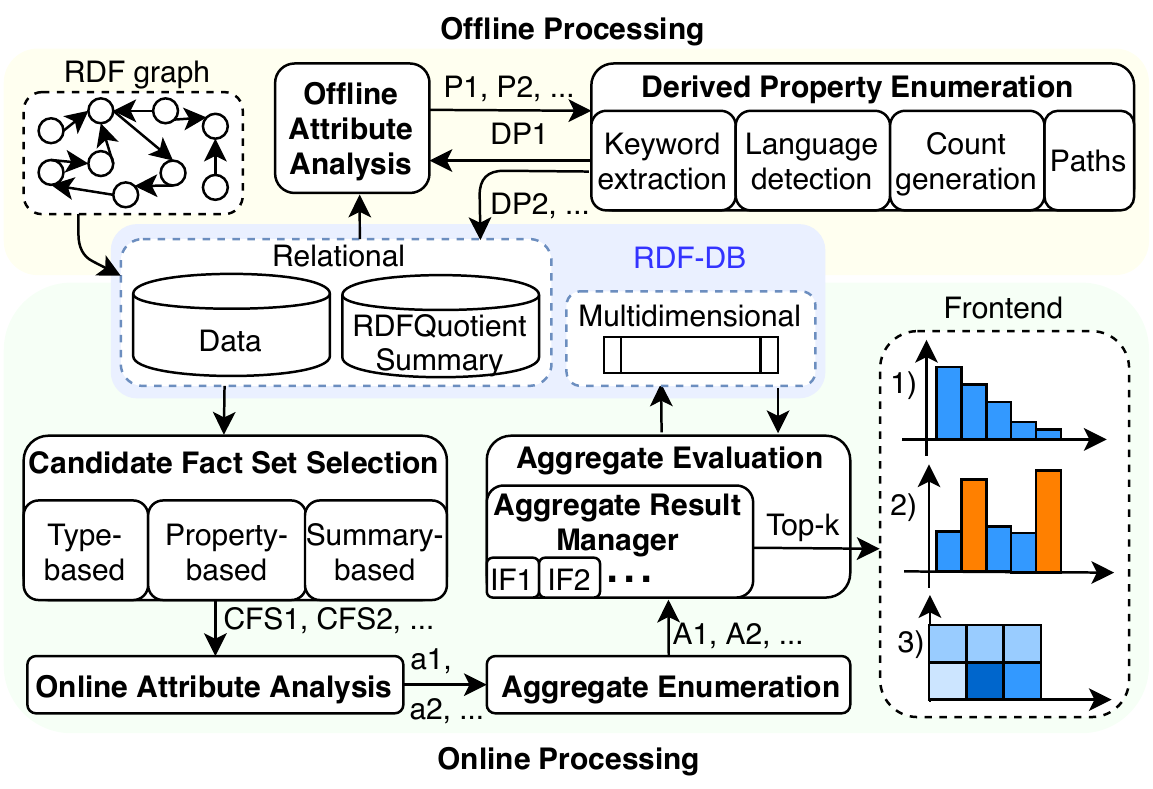}
	\veat{-4.5mm}
	\caption{The architecture of \sys{}.\label{fig:framework}}
	\Description{The architecture of \sys{}.}
	\veat{-5.5mm}
\end{figure}

In this section, we describe the system design of \sys{}, a new \emph{RDF-oriented end-to-end framework} that automatically identifies, enumerates, and efficiently evaluates MDAs to determine the most interesting ones. Figure~\ref{fig:framework} shows \sys{}'s analytics pipeline; it comprises an \emph{offline} phase, where an RDF graph is loaded and pre-processed, and an \emph{online} phase, where  user-specific analysis is performed.

\textbf{Offline Processing.} Upon loading an RDF graph, we first build a structural summary thereof, using the open-source RDFQuotient tool~\cite{VLDBJournal2020}. The summary captures all the properties occurring in the graph and proposes a set of RDF node groups such that the RDF nodes in each group are considered equivalent. \sys{} uses the summary to expedite several steps of the analysis, e.g., the enumeration of RDF types and properties, as described below. 

Next, we perform \textbf{Offline Attribute Analysis} with three main purposes: ($i$)~to gather a set of \emph{statistics} for each property in the graph, ($ii$)~to determine if derivations should be generated for a given property, and ($iii$)~to decide if pre-aggregated values of some properties should be computed and stored in the database.

Derived properties are the key to a rich search space and to effectively addressing challenge \textbf{C1}. With this aim, we compute statistics \rev{including} the type of property values (e.g., String, Integer, Date) and, if they are multi-valued, their number of distinct values, the lowest and highest values, etc. Based on these results, \textbf{Derived Property Enumeration} generates: ($i$)~\textit{property counts} for multi-valued properties, e.g., how many companies a CEO manages; ($ii$)~\textit{keywords occurring in property values}, e.g., if a company's description is  ``Sonangol oversees petroleum production'',  we attach to the company the multi-valued attribute \textit{kwInDescription} with the values  ``Petroleum'' and ``Production''; ($iii$)~the \textit{language} of a text property, e.g., a company may gain the attribute \textit{langOfDescription} with the value ``English''; ($iv$)~\textit{paths}, e.g., a CEO politically connected to a ``President'' gains the attribute \textit{politicalConnection/role} with the value ``President''.
Finally, each derived property is also analyzed and stored along with its statistics in the database. 

In addition, for each multi-valued attribute, we create a table in the database storing its values, pre-aggregated on the RDF nodes that have it. More specifically, for each RDF node, we compute and store the aggregated value for each (attribute, aggregate function) pair, e.g., the sum of $a_1$, the count of $a_1$, the minimum of $a_2$.
This allows \sys{} to account for facts with multiple measure values and improve Aggregate Evaluation during Online Processing. 

%LONGER VERSION
%In addition, for each multi-valued attribute a table is created in the database storing its values, pre-aggregated on the RDF nodes that have it. This allows \sys{} to account for facts with multiple measure values and improve Aggregate Evaluation during Online Processing. First, depending on the type of its values, each attribute is assigned the set of possible aggregate functions (e.g., only numeric attributes are assigned min, max, sum and avg). Then, \emph{for each RDF node}, the aggregated value for each (attribute, aggregate function) pair, e.g., the average of $a_1$, the count of $a_1$, the minimum of $a_2$, is computed and stored.

\textbf{Online Processing.} The analysis of RDF graphs suits the specific needs of users and proceeds in the following steps. 

\textbf{Step~1} is \textbf{Candidate Fact Set Selection}. To address challenge \textbf{C1}, \sys{} identifies CFSs in three ways: ($i$)~\textit{type-based}: for each type $T$ in the graph, the set of RDF nodes of type $T$; ($ii$)~\textit{property-based}: for a (user-specified) set of properties, all the RDF nodes having those outgoing properties; ($iii$)~\textit{summary-based}: each set of RDF nodes identified as equivalent by the RDFQuotient summary; RDF nodes in the same equivalence class tend to have many common properties, making them interesting candidates to be analyzed together. 

\textbf{Step~2} is \textbf{Online Attribute Analysis}. In this step, for each CFS, we first enumerate all direct and derived properties. Then, we enrich the offline-analysis results by adding CFS-dependent statistics, e.g., the support of an attribute among all the facts in the CFS, the number of CFs that have such an attribute more than once, and the number of distinct values. \sys{} exploits the gathered statistics  in different steps, e.g., to guide the choice of dimensions, measures, and aggregate functions and to improve Aggregate Evaluation. 
%For example, dimensions and measures must be \emph{frequent} (i.e., have support greater than a defined threshold); dimensions should not have \emph{too many distinct values when compared to the number of facts to examine} (e.g., we do consider counting the number of CEOs by their birthday).

%\YD{Derived properties are key to a rich  search space, and address \textbf{C1}. Should be mentioned somewhere. In addition, shall we say how we determine if a given property is allowed to be dimension? }

\textbf{Step~3} is \textbf{Aggregate Enumeration}. \sys{} uses the pool of analyzed attributes to generate candidate MDAs. To address challenge \textbf{C1}, we 
%develop an (extensible) set of primitives to 
generate a rich space of candidate aggregates while applying rule-based pruning to avoid meaningless candidates. 
%a set of rules ensure that while we enrich the search space, only meaningful aggregates are enumerated, as outlined below. 

($a$)~\textit{Identifying dimensions and measures from (derived) properties}: 
We first enumerate all the (derived) properties and consider them for dimensions or measures, subject to the following rules: 
($i$)~Dimensions and measures must be \emph{frequent}, i.e., having a support greater than a defined threshold; ($ii$)~Dimensions should \emph{not have too many distinct values} when compared to the number of facts to examine (e.g., we do not consider counting the number of CEOs by their birthday as there are too many distinct values\cut{ for the birthday}).

($b$)~\textit{Identifying the dimension set of each lattice}:
We compute the Maximal Frequent Sets of attributes~\cite{maximalfrequent} in the CFS. Each of the found sets is the root of one lattice. We further filter them so that  each lattice: ($i$)~has at most $N$ attributes, and ($ii$)~does not contain attributes that are derived one from the other, e.g., \textit{nationality} and \textit{numOfNationalities} are not allowed as dimensions of the same lattice.\cut{ Each lattice provides a smaller analysis problem with dimensionality up to $N$.}
%we support the customization of its dimensionality. 
Although we aim to offer a general approach, we also note that the readability of MDAs by human users is maximized at levels of relatively low dimensionality, i.e., $N\in \{1, 2, 3, 4\}$. 

($c$)~\textit{Identifying the measures in each lattice}:
Once a lattice acquires dimensions $\mathcal{D}_i$, we assign it a measure set $\mathcal{M}_i$ that comprises all the analyzed attributes of the CFS except those in $\mathcal{D}_i$, and those that are derived from a dimension in $\mathcal{D}_i$, e.g., \textit{numOfNationalities} cannot be a measure in an aggregate whose dimension is \textit{nationality}.

%To address challenge \textbf{C1}, a set of rules ensures that only meaningful aggregates are enumerated. For example, we restrict the measure to differ from each dimension, choose the aggregate function based on the measure type (e.g., only average numeric measures), etc. 
\sloppy
Several lattices may be found for a CFS, e.g., for CEOs, we have three: \{\textit{countryOfOrigin}\}, \{\textit{nationality}, \textit{numOfCompanies}$\}$, and \{\textit{nationality}, \textit{gender}, \textit{company/area}$\}$ (Examples~1-3). They might partially overlap in dimensions and/or measures; e.g., Examples~2~and~3 share \textit{nationality}. %During their evaluation,
\sys{} ensures that the results of evaluated MDAs are reused (not recomputed) in the other lattices where they appear. 

\textbf{Step~4}  is \textbf{Aggregate Evaluation}. This step triggers the actual evaluation of the enumerated MDAs. To address challenge \textbf{C2}, we combine: ($i$)~our novel early-stop technique to quickly prune the unpromising MDAs, and ($ii$)~our \ouralgo{} algorithm to efficiently compute the remaining MDAs in a single pass. The final results are produced in an incremental fashion 
%Indeed, our one-pass algorithm, \ouralgo{}, produces aggregate results for some groups at a time. 
and handled by the \textbf{Aggregate Result Manager} (ARM). The ARM stores them and incrementally updates statistics such as minimum and maximum values, as we explain in Section~\ref{sec:lattice-based-computation}. These are used to determine the interestingness of the computed MDAs (by applying $h$) in one pass over their results.

\textbf{Step~5} finally performs \textbf{Top-$k$ Computation}. Once the evaluation is complete, the ARM retrieves all the evaluated MDAs, computes their \emph{interestingness} score by applying $h$, and returns the $k$ best aggregates. \rev{\sys{} natively supports three interestingness functions, from which the user can choose to suit their preferences: \rev{($i$)~\emph{variance}, ($ii$)~skewness, and ($iii$)~kurtosis, where variance can detect deviation from uniform aggregate values, whereas the latter two can detect deviation from a normal distribution of aggregated values over numeric dimensions}.}
\cut{For example, Figure~\ref{fig:ceos-running-example} shows a histogram that exhibits a peak (outlier) in $sum$(\textit{netWorth}) for Angola, due to Dos Santos, and a heat map where 
the dark color reflects a particularly low value of $avg$(\textit{age}) of CEOs, again due to Dos Santos. 
%the color reflects $avg$(\textit{age}) of CEOs and a dark cell shows a particularly low value, again due to Dos Santos. 
The two aggregates are interesting because of their high variance scores. }
Figure~\ref{fig:ceos-running-example}(b) shows two example aggregates with high variance scores.
More sophisticated interestingness functions for insight detection can be applied on the Step~4 results via the ARM; \rev{we discuss early-stop extensions in Section~\ref{sec:estimating-interestingness}.}

	\section{Lattice-based computation\label{sec:lattice-based-computation}}
We first recall a classical optimized method of computing all aggregates in a lattice. We then explain its limitations and the errors it makes in our setting.  Finally, we present our new algorithm to compute lattices of RDF aggregates correctly and efficiently.

\subsection{Classical one-pass lattice computation\label{sec:arraycube}}
In relational DWs, nodes in a multidimensional lattice are often computed from one of their parents to reuse computation and limit the number of passes over the data. Among the existing algorithms~\cite{DBLP:journals/csur/MorfoniosKIK07}, ArrayCube~\cite{molap} computes the whole lattice in a single pass\cut{ over the data}. Given a set of $N$ dimensions, a measure, and an aggregate function, it relies on an \emph{array representation} of data and evaluates $2^N$ nodes through a \emph{Minimum Memory Spanning Tree}, as we recall below.

\begin{figure}[t!]
	\centering
	\veat{-2.5mm}
	\includegraphics[width=\columnwidth,height=37mm]{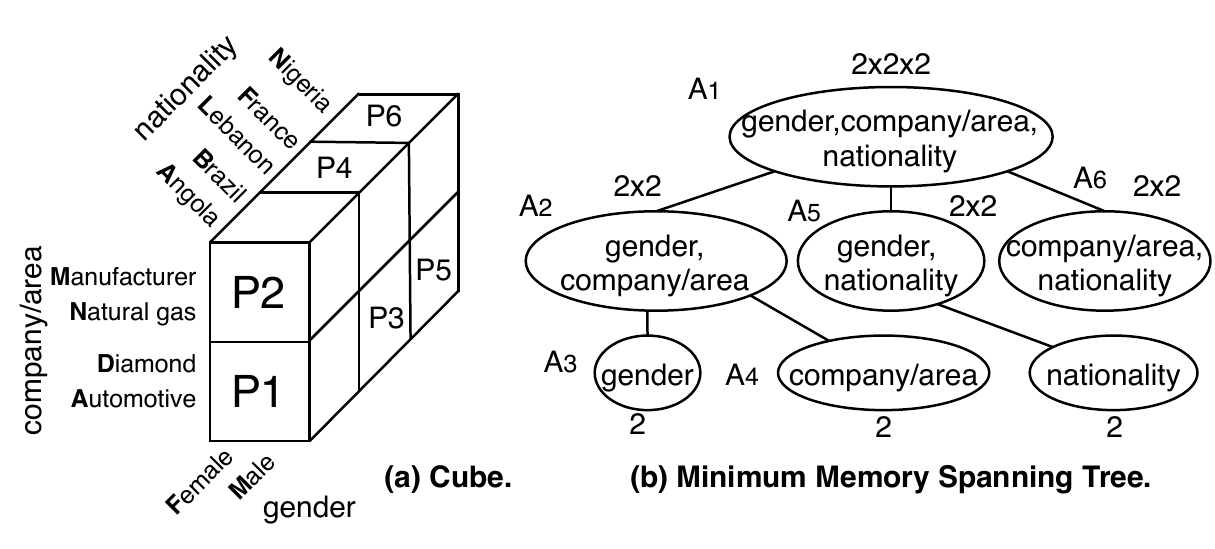}
	\veat{-9mm}
	\caption{Multidimensional space and MMST for Example~3.\label{fig:multiDimSpaceArrayCube}}
	\Description{Multidimensional space and MMST for Example~3.}
	\veat{-4mm}
\end{figure}

\begin{figure}[t!]
	\centering
	\includegraphics[width=0.93\columnwidth,height=37mm]{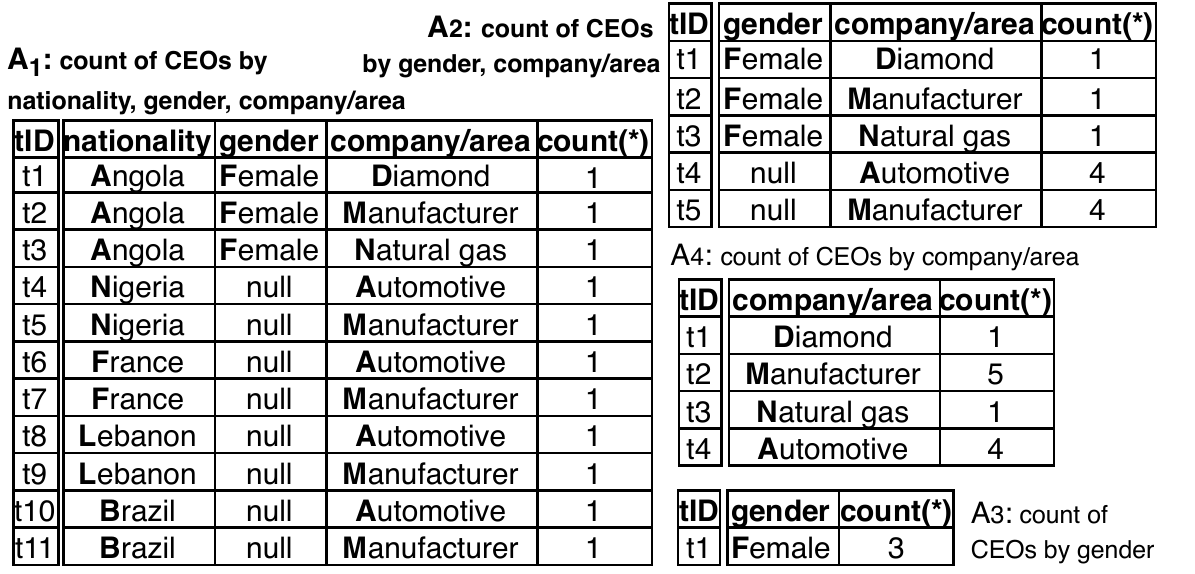}
	\veat{-4mm}
	\caption{Relational aggregation.\label{fig:aggregationOfCEOs}}
	\Description{Relational aggregation.}
	\veat{-5mm}
\end{figure}

\sloppy\textbf{Array representation of data.}
The distinct values of each dimension are \emph{ordered}, leading to a set of \emph{cells}, each corresponding to a unique combination of indices of values along the $N$ dimensions (axes). In Example~3, assuming \textit{nationality} $\in$ \rev{\{A, B, F, L, N\}}, \textit{gender} $\in$ \rev{\{F, M\}} and \textit{company/area} $\in$ \rev{\{A, D, M, N\} (we denote initials of the respective values in Figure~\ref{fig:multiDimSpaceArrayCube})}, the multidimensional space has 40 cells, e.g., in cell 0, \textit{nationality}=\rev{A}, \textit{gender}=\rev{F} and \textit{company/area}=\rev{A}; in cell 1, \textit{nationality}=\rev{B}, \textit{gender}=\rev{F} and \textit{company/area}=\rev{A}. Each cell of the $N$-dimensional array contains the value of the aggregated measure over all facts in that cell; in Example~3, this is the count of CEOs. Further, cells are grouped in \emph{partitions}: each partition is a contiguous part of the array, containing the cells corresponding to a predefined number of distinct values along each dimension, e.g., if this is 2, the 40-cell array has 6 partitions. Figure~\ref{fig:multiDimSpaceArrayCube}(a) shows the array. Note that an initial pass over the data is required to bring it from the relational to the array representation.

\textbf{Minimum Memory Spanning Tree (MMST).} The dimensions in Example~3 determine the lattice in Figure~\ref{fig:multiDimSpaceArrayCube}(b). 
To evaluate all nodes, ArrayCube chooses, for each non-root node $A$, a parent node \cut{in the lattice }to compute \cut{the aggregate in }$A$ from, hence forming a spanning tree of %covering all the nodes in
the lattice. The memory needed to evaluate all the aggregates in one pass over the data depends on the ordering of dimensions, their numbers of distinct values, and the partition size. ArrayCube chooses the tree that minimizes the overall memory needed; it is called the MMST.
%The total amount of memory required by the MMST is bounded as follows.
%
%Given a chunked multidimensional array $A$ with size $\prod_{i=1}^{N}|D_i|$ where $|D_i|=d$ for all $i$, and each array chunk has the size $\prod_{i=1}^{J}|C_i|$ where $|C_i|=c$ for all $i$, the total amount of memory $M_T$ to compute the cube of the array in one scan of $A$ is less than $c^N + (d+1+c)^{N-1}$.

\textbf{Lattice computation} proceeds as follows. The MMST is instantiated, allocating to each node the required memory. Partitions are loaded from the array representation of data, one at a time, into the root of the MMST. The content of each cell in the root is propagated to the children and used to incrementally update the aggregated measures of all the nodes in the MMST. Once a partition is evaluated, \emph{each node checks if it is time to store its memory content to disk}. For instance, after scanning partition P1 in Figure~\ref{fig:multiDimSpaceArrayCube}(a), the subarray with \textit{nationality} $\in$ \rev{\{A, B\}} and \textit{company/area} $\in$ \rev{\{A, D\}} is exhausted. Thus, the counts of CEOs with \rev{either of the two nationalities, and A or D company area are computed.} %($nationality$=Angola or $nationality$=Brazil) and ($company/area$=Automotive or $company/area$=Diamond) have been computed.
Now, $A_6$ (Figure~\ref{fig:multiDimSpaceArrayCube}(b)) can store its result to disk and reuse the memory in the subsequent computation. Similarly, once processed, the two subarrays of P2, ($i$)~\textit{nationality} $\in$ \rev{\{A, B\}}, and \textit{company/area} $\in$ \rev{\{M, N\}}; ($ii$)~\textit{nationality} $\in$ \rev{\{A, B\}}, are exhausted, and both $A_6$ and $A_5$ can store their results to disk. $A_6$ stores its result \cut{to disk }after every partition, $A_5$ after every two.

%\YD{(1) Do we need the memory analysis here? (2) One-pass: what about the chunking method that takes another pass? }
\subsection{Incorrectness in the RDF setting\label{sec:theory}}
%To compute all MDAs in the lattice determined by the nationality and number of managed companies, efficient relational DW algorithms compute an aggregate in the lattice from one of its parents: this reuses aggregation effort and computes all aggregates in the latice in a single pass over the data~\cite{molap}. Thus, we would use the results of $q$ to find the \textit{``average age of CEOs  by nationality''} and the \textit{``average age of CEOs  by number of managed companies''}. However, the result of $q$ contains \emph{four} tuples describing  (one for each of their nationalities): if we aggregate these results, e.g., to compute the average CEO age, we obtain wrong results, as Ghosn's age would count four times instead of once. 

%To correctly compute the whole lattice from the root aggregate, we can: ($i$)~ensure that each CEO fact is represented by at most one tuple, by ignoring  all but one of Ghosn's nationalities: this would clearly miss an interesting part of the data; ($ii$)~compute each of the $2^N$ aggregate in the lattice separately, missing the benefits of efficient one-pass algorithms; this would entail a high run time overhead. 
%Instead, \emph{one of our main objectives is to provide an efficient one-pass algorithm that will correctly compute the whole lattice in the presence of multi-valued dimensions}.

Results computed by ArrayCube may be \emph{incorrect in the presence of multi-valued dimensions}. %(a fact may have multiple values along a given dimension), 
%which are frequent in RDF data. %is prevalent in RDF data. 
Consider our running examples that show CEOs with \emph{various} nationalities and \emph{at most one} gender who manage companies in \emph{several} areas. In a relational DW, \rev{each such CEO} would be stored as a tuple in the fact table, and \rev{their} multiple nationalities (respectively, company areas) would be modeled as a dimension table associating \rev{them} with each of \rev{their} nationalities (company areas).
We could then find the result for Example~3 with a query $q$ that joins all the relations, groups the data by the dimensions, and finally aggregates the measure. To evaluate all MDAs in the lattice determined by the dimensions\cut{ in Example~3}, ArrayCube would use the MMST in Figure~\ref{fig:multiDimSpaceArrayCube}(b) and compute the aggregate $A_1$ by means of $q$, using its result to compute the rest of the lattice.

Figure~\ref{fig:aggregationOfCEOs} shows the result of $A_1$ when applied to the two CEOs in Figure~\ref{fig:ceos-running-example}. The tuples $t_1$ to $t_3$ are derived from Dos Santos (the RDF node $n_1$), whereas $t_4$ to $t_{11}$ are due to Carlos Ghosn (the RDF node $n_2$). Since $n_2$ lacks gender information, the tuples $t_4$ to $t_{11}$ have \textit{gender}=\texttt{null}. We need to keep them to compute the rest of the lattice correctly. %Indeed, 
Since $n_2$ has valid values for \textit{nationality} and \textit{company/area}, we must count this CEO when computing aggregates over one or both of these dimensions, e.g., $A_4$ in Figure~\ref{fig:aggregationOfCEOs}.
We obtain the result of $A_2$ by aggregating $A_1$'s result to project away the \textit{nationality} dimension. For instance, the tuples $t_4$, $t_6$, $t_8$, and $t_{10}$, which are all associated with $n_2$, collapse into the tuple $t_4$ in $A_2$ where now this CEO counts as \emph{four}.
Then, $A_2$ is further aggregated by projecting away \textit{company/area} to compute $A_3$ and separately \textit{gender} to compute $A_4$.
The cardinality ``bug'' introduced in $A_2$ \emph{propagates} down the lattice. In $A_4$'s result, we find \emph{five} CEOs managing Manufacturer companies, whereas there are only \emph{two}. A similar error occurs in $A_3$ where we count \emph{three} female CEOs\cut{ (which is wrong)} because the tuples $t_1$ to $t_3$ of $A_2$ are aggregated into the same tuple and are, thus, counted three times (although they all represent $n_1$). 
\cut{Since $A_3$ will not be further aggregated, we can remove the tuple having \textit{gender}=\texttt{null}.}

The above example shows that multiple values for a dimension may lead to errors \emph{when an aggregate is computed from one of its parents}. To correctly compute the whole lattice from the root aggregate, na\"ive solutions may: ($i$)~require that each CEO fact be represented by at most one tuple, e.g., in our example, by ignoring all but one of Ghosn's nationalities (company areas); this would clearly miss an interesting part of the data; ($ii$)~compute each of the $2^N$ aggregates in the lattice separately, missing the benefits of efficient one-pass algorithms; this would entail a high run time overhead. 

%The error is avoided if we compute $A_2$ from the initial data:  we would first join the CFS with the dimensions \textit{gender} and \textit{company/area}, not with \textit{nationality} (as it is not a dimension of $A_2$), and thus no ``wrong duplicate'' is introduced. However, computing each lattice aggregate from the base data is very inefficient. 

Interestingly, if we alter the query $q$ to \emph{count \underline{distinct} CEOs in each group} (in lieu of $count(*)$), no errors occur in the result of Example~3. By design, ArrayCube cannot compute aggregates including \emph{distinct}: instead, it computes all aggregates from the result of the lattice root, where information about individual facts is no longer present. Other one-pass algorithms for lattice-based aggregate computation, such as PostgreSQL's GROUP BY CUBE implementation~\cite{pgcube} (\pgcube{}), do support the counting of distinct values and can thus be used to obtain the correct result for Example~3. \cut{Thus, one could get correct result for Example~3 by asking $count$(distinct \textit{CEO}) rather than $count(*)$. }However, in the presence of multi-valued dimensions, computing aggregates from the result of one of their parents in the lattice \emph{may still lead to wrong results}, as illustrated in the following variation\techreport{s}{} of Example~3.

\textit{Variation\techreport{~1}{}.} Consider the aggregate \textit{``sum of the net worth of CEOs by nationality, gender, and area of the companies they manage''}. We first augment the data in the root aggregate $A_1$ with the sum of \textit{netWorth} ($NW$). The tuples $t_1$ to $t_3$ contain the $NW$ of Dos Santos: \$2.8 \rev{billion}. The tuples $t_4$ to $t_{11}$ contain the $NW$ of Ghosn: \$120 \rev{million}. We then compute the sum of $NW$ by \textit{company/area}. The tuples $t_2$, $t_5$, $t_7$, $t_9$, and $t_{11}$ (all having \textit{company/area}=\rev{M}) sum up into one tuple, and result in the sums of \$2.8B of Dos Santos\cut{ (from $t_2$)}, and $4\,\times\, $\$120M of Ghosn\cut{ (from the other tuples)}, whereas both CEOs should have contributed exactly once.
Moreover, we cannot solve this issue with the $sum$(distinct \textit{NW}) aggregate. If both CEOs had the same $NW$, a $sum$(distinct) would sum $NW$ once, instead of (correctly) summing it twice.% (once per each of them).

\techreport{Similarly, the following variation illustrates another scenario leading to wrong results. 

\textit{Variation~2.} Consider the aggregate \textit{``average age of CEOs by nationality, gender, and area of the companies they manage''}. We obtain it as $sum$(\textit{age})/$count$(\textit{age}), i.e., the sum in Variation~1 is divided by 5. Instead, the correct value is sum of ages of Dos Santos and Ghosn divided by 2. As in Variation~1, we cannot solve this issue by using $avg$(distinct \textit{age}).}{}

%\YD{The second variation does not bring much new information. It can be summarized in a single sentence. Save the space for other sections.}

\techreport{\veat{-0.11mm}}{}
As our experiments show (Section~\ref{sec:comparisonWithCube}), the number of incorrectly computed aggregates, and the magnitude of the error itself, can be quite significant. This is because of the flexible RDF model, which allows multi-valued dimensions. Conversely, in a relational DW, once a fact table is joined with dimension tables, ArrayCube assumes that each fact has exactly one value for a dimension (for instance, due to a functional dependency). Below, we formally characterize the situations when ArrayCube introduces errors on RDF data.

\textbf{Analysis of ArrayCube errors on RDF.}
%\subsection{Analysis of ArrayCube errors on RDF\label{sec:theory}}
\techreport{}{\veat{-0.11mm}}
Consider an RDF graph $\graph$ and a lattice of $N$ dimensions ($2^N$ nodes) on $\graph$.
Whether a lattice node can be computed correctly from one of its parents, depends on the presence of \emph{multi-valued dimensions} in the lattice:

\begin{lemma}\label{theo:fddoesnothold}
Let $\graph$ be an RDF graph. Let $P=\langle CFS, \mathcal{D}_P, M, f\rangle$, $C=\langle CFS, \mathcal{D}_C, M, f\rangle$ be two aggregates in a lattice on $\graph$ such that $P$ is a parent of $C$,  $\mathcal{D}_P=\mathcal{D}_C\cup \{D\}$ where $D$ is a dimension,  $f\in\{count(*), count(M), sum(M), avg(M)\}$, 
%differ by dimension $D$. Let $D$ be a multi-valued dimension such that $D\in\mathcal{D}_P$ and $D\not\in\mathcal{D}_C$.
%\IM{In the absence of functional dependencies (which hold in general, at an abstract level, even if one has no graph in mind, we cannot say ``multi-valued dimension'': we neeed to tie this to a graph.}
and there exists a fact $n\in CFS$ with more than one value along the dimension $D$. Then, computing $C(\graph)$ from the result of $P(\graph)$ may lead to wrong results. 
\end{lemma}

\techreport{
\begin{proof}
Let the fact $n \in CFS$ have the values $n.D=\{ a, b\}$ and, for each $D_j\in \mathcal{D}_P$, $D_j\neq D$, $n.D_j=d_j$ and $d_j$ is not \texttt{null}.
By definition of $P$, there exist tuples $t_1, t_2\in P(\graph)$ such that $t_1=(d_1,\dots, a, \dots, d_N, v_1)$ and $t_2=(d_1,\dots, b, \dots, d_N, v_2)$, to both of which $n$ contributes. Hence, there exists a tuple $t_3\in C(\graph)$ such that $t_3=(d_1,\dots, d_N, v_3)$, in which\cut{, by hypothesis,} the dimension $D$ does not appear.

When computing $C(\graph)$ from $P(\graph)$,  the aggregated value $v_3$ is obtained from $t_1.v_1$ and $t_2.v_2$ based on the  function $f$. For instance,  if $f$ is $count(*)$, the fact $n$ will be counted twice, instead of just once. If  $f$ is $sum(M)$, the $M$ value(s) of $n$ will be summed twice, which falsifies the result (except for the particular case where their sum is 0). Computing the $avg$ may similarly lead to wrong results.
\end{proof}
}{}

%\IM{The Theorem has moved from \emph{1 aggregate} to \emph{1 parent-child agg pair}, in other words, from \emph{being about a node} to \emph{being about an edge} in the lattice. Observe that it just states \emph{which edges introduce errors} - without questioning \emph{if the parent was correctly computed, to start with}. However, our real interest here is in a \emph{complete lattice computation}, where we don't care if an edge introduces an error, but rather, if the path from the root to each aggregate introduces one. Thus: }

How does Lemma~\ref{theo:fddoesnothold} impact the one-pass lattice-based computation for a given graph $\graph$? We show the following result: 

\begin{theorem}\label{theo:incorrectness}
Given an RDF graph $\graph$ and a lattice on $\graph$, let $\mathcal{MD}\subseteq \mathcal{D}$ be the set of all the dimensions for which some fact(s) $n\in CFS$ have more than one value, and let $K>0$ be the size of $\mathcal{MD}$.
($i$)~A one-pass algorithm \emph{cannot} compute correctly all the lattice aggregates.
($ii$)~The maximum number of MDAs (lattice nodes) that can be computed correctly (depending on the choice of the MMST) is  $2^{N-K}$.
\end{theorem}

\techreport{
\begin{proof}
($i$)~Among the $N\cdot 2^{N-1}$ lattice edges, $K\cdot 2^{N-1}$
% \IM{I think this is the correct number (not $K*2^{K-1}$) because: the lattice has $N*2^{N-1}$ edges; all edge labels (all dimensions) appear with equal frequency; thus, the K properties will be on N * (K/N) *2 ^{N-1} edges, which is $K*2^{N-1}$.}
are labeled with a dimension from $\mathcal{MD}$, meaning that the dimension is projected away when computation follows this edge. As Lemma~\ref{theo:fddoesnothold} shows, if the MMST contains one such edge, the result of the child node of that edge may contain errors.
However, no spanning tree, thus, no MMST, can avoid \emph{all} edges labeled with  a dimension in  $\mathcal{MD}$. This is because to go from the root, whose dimensions are $\mathcal{D}$, to a node lacking one dimension $D\in \mathcal{MD}$, by the construction of the lattice, the MMST \emph{must} traverse an edge labeled $D$.

($ii$)~The lattice nodes that can be computed correctly in one pass (starting from the root's result) are exactly those having all the $\mathcal{MD}$ dimensions: a node lacking one such dimension would be obtained by aggregating a parent's result along that dimension, and thus, by Lemma~\ref{theo:fddoesnothold}, be wrongly computed. The lattice has $2^{N-K}$ such nodes. \emph{Fewer} nodes may be computed correctly if the MMST picks a ``wrong'' edge, even if it could have avoided doing so. 
\end{proof}
}{Due to space constraints, we delegate the proofs of Lemma~\ref{theo:fddoesnothold} and Theorem~\ref{theo:incorrectness} our technical report~\cite{TR}.}

\subsection{\ouralgo{} Algorithm\label{sec:mvdcube}}
We now present \textbf{Multi-Valued Data Cube} (\ouralgo{}), our new one-pass MDA evaluation method. Going beyond existing algorithms~\cite{molap}, \ouralgo{}: ($i$)~\emph{produces correct results} even in the presence of missing or multi-valued dimensions and/or measures, ($ii$)~computes  \emph{several aggregate functions} over \emph{a large set of measures} in the same lattice, and ($iii$)~saves computation cost by sharing measures across all lattices from a given CFS.

%\IM{GraphCube? RDFCube? GCube?}

\begin{figure*}
	\centering
	\includegraphics[width=\textwidth]{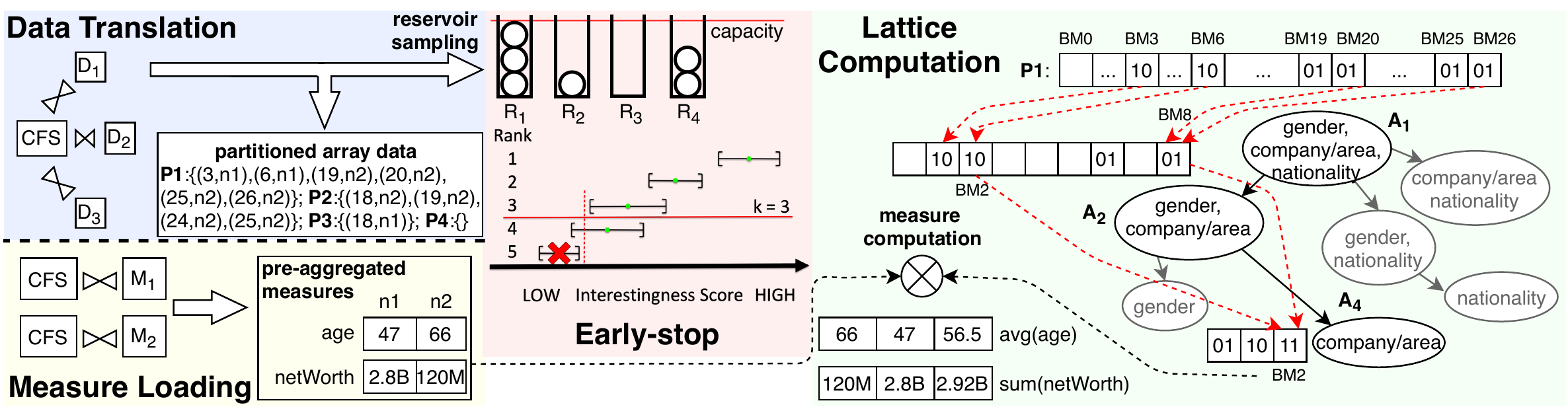}
	\veat{-7.5mm}
	\caption{Aggregate evaluation using \ouralgo{} and early-stop.\label{fig:mvdcubewithes}}
	\Description{Aggregate evaluation using \ouralgo{} and early-stop.}
	\veat{-3.5mm}
\end{figure*}

Before we move forward, we clarify that our RDF database uses the following storage: a CFS is represented by a single-column table storing the identifiers (IDs) of the facts; for each attribute $a$, a table $t_a$ stores ($s$, $o$) pairs for each $(s,a, o)$ triple in the RDF graph.

Figure~\ref{fig:mvdcubewithes} depicts the main features of \ouralgo{}.
Our \ouralgo{} evaluation method proceeds in the following steps, with the pseudo-code of its core functions shown in Algorithm~\ref{alg:molapforrdf}.

\textbf{Building MMSTs.} 
Given a CFS and a set of lattices (identified in Step 3 of \sys{}'s pipeline), each with the dimensions $\mathcal{D}_i$ and the measures $\mathcal{M}_i$, we construct one MMST per lattice as in~\cite{molap}. 
% and requires the same number of memory cells. However, a cell does not store aggregated measure values, but a (bitmap-compressed) set of RDF nodes. \MM{maybe discussion about lifting optimility here? cells have variable length}

\textbf{Data Translation.} 
For each lattice, we process the root node by sending a join query to the database to obtain all the CFs that have a value for at least one of the dimensions in $\mathcal{D}_i$. We then translate the join result to lay the data in a \emph{partitioned array representation} of cells.
A partition is a set of pairs (cell index, CF). We assign each RDF node a cell index based on its dimensions' values; in the case of multiple values for a dimension, we assign indexes of all corresponding cells. % e.g., in Figure~\ref{fig:ceos-running-example}, CEO $n_1$ is related to three company areas, for any aggregate on the dimension \textit{company/area}, $n_1$ will contribute for each of the three values.
%This scheme treats RDF nodes with missing dimensions such that these RDF nodes  will be absent from the cells aggregating along that dimension. On the other hand, a
We add the special value \texttt{null} in the domain of each dimension to account for missing values.
Therefore, each cell is associated with \emph{the set of RDF nodes that correspond to the combination of dimension values that this cell represents}. %The cells contain, thus, \emph{base (non-aggregated) data}.
Like ArrayCube, we take an initial pass over the data to bring it into the array representation, where the (conceptual) multidimensional array is stored as a serialized one-dimensional array. If the data does not fit into the available memory, we partition it, store to disk, and later read back, one partition at a time; otherwise, \ouralgo{} accesses the array directly from the main memory, in a single pass, in subsequent steps.
%Moreover, if the data does not fit into available memory, it is chunked and stored to disk. Otherwise, one chunk containing all the data is created and accessed directly from main memory. 

\textbf{Measure Loading} is performed in parallel to the Data Translation step. For each measure $M$ in $\mathcal{M}_i$, we query the database to retrieve, for each CF, the pre-aggregated values of $M$ (which were computed and stored offline). We load the values \emph{ordered by the IDs of the CFs}, and share them among all MMSTs in a given CFS. As they are stored at the granularity of a CF, they can be used to compute aggregate results for all cells, as we describe below. 

\textbf{Lattice Computation} is then carried out in one pass over the data using the MMST. \ouralgo{} associates an MMST node with a (large) set of aggregates; we denote such a node as $A_i =\langle CFS, \mathcal{D}_j\rangle$. Each node then \emph{represents all the MDAs that have dimensions $\mathcal{D}_j$} (but might differ in their measure and aggregate function). Suppose that we want to compute the lattice with $\mathcal{D}$=\{\textit{gender}, \textit{company/area}, \textit{nationality}\}, $\mathcal{M}$=\{\textit{age}, \textit{netWorth}\} and that \textit{age} is associated with $avg$, and \textit{netWorth} is associated with $sum$. Node $A_2$ in Figure~\ref{fig:mvdcubewithes} represents the two MDAs: ($i$)~average age of CEOs, and ($ii$)~sum of \textit{netWorth} of CEOs, both grouped by \textit{gender} and \textit{company/area}.

In the MMST, we allocate, for each node, the needed memory. We load partitions successively into the root. In Figure~\ref{fig:mvdcubewithes}, we assume that each partition contains 3 distinct values of each dimension, hence 27 cells. 
%(line~\ref{mvdcubeline:loadchunk} in Algorithm~\ref{alg:molapforrdf}). 
For compactness, we encode each set of RDF nodes in a cell using a Roaring Bitmap%\footnote{\url{https://roaringbitmap.org/}}
~\cite{roaring}
%\IM{I shortened here as I didn't want to spend readers' attention with more bitmap references. If really needed, put them in the related work}
%tend to outperform conventional compressed bitmaps such as WAH~\cite{wah}, EWAH~\cite{ewah} or Concise~\cite{concise}. They are 
(also adopted in Spark because of the strong compression and lookup performance).
%they can be hundreds of times faster and they often offer significantly better compression. 
In Figure~\ref{fig:mvdcubewithes}, each cell stores a set of CEOs (a subset of the facts $n_1$ and $n_2$). The \emph{bitmaps follow the same ordering of the CFs applied during Measure Loading}. The cell of index 3 in $A_1$ contains a bitmap of size 2, BM$_3=10$, representing that $n_1$ is in the set, whereas $n_2$ is not.

\textit{(a) Projection and bitmap propagation.} We scan the bitmaps in the cells of the root node and immediately propagate them to the child nodes in the MMST as dimensions are projected away (line~\ref{mvdcubeline:updatesubtree} in Algorithm~\ref{alg:molapforrdf}). We union (OR) the bitmap in each cell in a child node with each bitmap received from the parent (line~\ref{updatesubtreeline:updatebitmap}): this models \emph{the contribution of all facts in a parent node to the corresponding cell in the child node}. In particular, as we project away a multi-valued dimension from a parent node to a child node, if a fact has  multiple values of the dimension, it belongs to  different cells in the parent node, but will be consolidated in the same cell in the child node. 

Red arrows in Figure~\ref{fig:mvdcubewithes} show propagations. For example, the bitmap of cell 2 in node $A_4$, BM$_2$, is initially empty (i.e., 00). Then it is updated to 01 when \rev{BM$_8$} from $A_2$ is propagated, and later  to 11 when BM$_2$ from $A_2$ is also propagated.

Once a partition is evaluated, we apply the ArrayCube check (Section~\ref{sec:arraycube}) in the nodes to learn if it is time to write results to disk (line~\ref{updatesubtreeline:storetodisk}). If so, we first propagate their memory content to their child nodes (line~\ref{updatesubtreeline:updatesubtree}), and then we compute the values of the aggregated measures and store them (line~\ref{updatesubtreeline:computemeasure}). 

\textit{(b) Measure computation} (denoted as $\otimes$). %at line~\ref{line:combine} 
When a node is ready to write to disk, we scan its memory one cell at a time. For each cell: ($i$)~we identify the pre-aggregated measures of each RDF node in the cell's bitmap, and ($ii$)~we apply the relevant aggregate functions to them.
Note that measure computation is very fast as both the bitmaps and the pre-aggregated measures are ordered by the fact ID, and can aggregate different measures simultaneously. 

Revisit $A_4$ in Figure~\ref{fig:mvdcubewithes}. 
Once P1 and P2 are evaluated, $A_4$ is ready to write current results to disk. We scan the three cell bitmaps, and for each bitmap: ($i$)~identify the age and the net worth of each CEO in the bitmap by accessing the pre-aggregated measures, ($ii$)~aggregate the respective measures by applying $avg$ on the age and $sum$ on the net worth. For example, for BM$_2$, we identify the ages (respectively, net worth) of $n_1$: 47 (\$2.8B) and $n_2$: 66 (\$120M) because they are both present in the bitmap, then compute their average (respectively, sum).
The  Aggregate Result Manager (line~\ref{line:store-res-manager}) receives the computed measures, and the %true 
values of the dimensions obtained from the cell index.
%by reverting the data translation. 
\cut{For BM$_2$, the ARM will receive ($i$)~57.5 and 2.92B, ($ii$)~\textit{company/area}=M.}
Finally, we empty $A_4$'s memory in the MMST and reuse it to evaluate the aggregate on the next partition\cut{ of data} (line~\ref{updatesubtreeline:emptymemory}). 

\veat{-2mm}
\begin{algorithm}
	\caption{\ouralgo{}(root, partitions)\label{alg:molapforrdf}}
	\small
	\SetKwFunction{Main}{Main}
	\SetKwFunction{updateSubtree}{updateSubtree}
	\SetKwFunction{computeAndStoreAggregatedMeasures}{computeAndStoreAggregatedMeasures}
	
	\SetKwProg{Fn}{Function}{:}{}
	\Fn{\Main{root, partitions}}{
		\ForEach{P $\in$ partitions}{
			root.loadPartition(P)\;\label{mvdcubeline:loadchunk}
			root.updateSubtree()\;\label{mvdcubeline:updatesubtree}
			root.computeAndStoreAggregatedMeasures()\;\label{mvdcubeline:computemeasure}
		}
	}
	\Fn{\updateSubtree{}}{
		\ForEach{child $\in$ children}{
			\ForEach{pair (partition, offset) $\in$ memory}{
				child.updateBitmap(partition, offset)\;\label{updatesubtreeline:updatebitmap}
				\If{timeToStoreToDisk()}{\label{updatesubtreeline:storetodisk}
					child.updateSubtree()\;\label{updatesubtreeline:updatesubtree}
					child.computeAndStoreAggregatedMeasures()\;\label{updatesubtreeline:computemeasure}
					child.emptyMemory()\;\label{updatesubtreeline:emptymemory}
				}
			}
		}
	}
	\Fn{\computeAndStoreAggregatedMeasures{}}{
		\ForEach{pair (partition, offset) $\in$ memory}{
			currentBitmap = getBitmap(partition, offset)\;
			\ForEach{pair (measure, aggFunction)}{
				aggregatedMeasure = currentBitmap $\otimes$ preAggregatedMeasure(measure,aggFunction)\;\label{line:combine}
				resultManager.add(partition,offset,aggregatedMeasure);\label{line:store-res-manager}
			}
		}
	}
\end{algorithm}
\veat{-2mm}

\textbf{Memory usage.}
Our memory analysis builds on \rev{the corresponding ArrayCube study~\cite{molap}}. % the upper bound of the working set of ArrayCube.
Assuming $N$ dimensions with $d$ distinct values each and $c$ distinct values per partition, the MMST uses at most $M_T=c^N + (d+1+c)^{N-1}$ array cells to compute \emph{one} aggregated measure.
%Assuming $N$ dimensions, each has $d$ distinct values, and each partition contains $c$ distinct values,~\cite{molap} provides the upper bound of the working set of ArrayCube as $M_T=c^N + (d+1+c)^{N-1}$. This is the number of array cells needed by the MMST to compute \emph{one} aggregated measure.
In \ouralgo{}, the memory for an MMST is also upper bounded by $M_T$ cells. However, cells have a variable size as each of them contains a Roaring Bitmap (RB). For this reason, we 
provide a worst-case estimation of \ouralgo{}'s memory needs for the MMST and the pre-aggregated measures.
%comprising ($a$)~the memory for the MMST and ($b$)~the storage of pre-aggregated measures.

% given a set of measures $\mathcal{M}=\{M_1, M_2, \dots M_m\}$, and denoting by $S_{M_i}$ the set of aggregate functions which can be computed over $M_i$\footnote{If $M_i$ is numeric, this includes $min$, $max$, $avg$, $count$, possibly also using distinct; otherwise, only $count$ with or without distinct.}, the memory to compute $m$ aggregated measures over the lattice is bounded by $M_T* \sum_{i=1}^{m} |S_{M_i}|$.

($a$) %Bitmaps have an impact on the MMST memory needs. \IM{9/2/2021: the previous proposition looks redundant wrt prior paragraph}
The size of an RB used to store $Z$ integers in the interval $[0,u)$ is bound in~\cite{roaring} to $M_{RB} = 2 \cdot Z +  9 \cdot (u/65535+1) + 8$, that is, beyond a fixed overhead for $u$, the universe size, RBs never use more than 2 bytes per integer. In the worst case, we could have $|CFS|$ facts in each cell, occupying a total of $M_T \cdot M_{RB}$ bytes.

($b$)~For $m$ measures, \ouralgo{} needs $|CFS| \cdot\sum\limits_{i=1}^{m} |S_{M_i}|$ float numbers in the worst case, where $M_i$ refers to each measure and $S_{M_i}$ is the set of aggregate functions assigned to the measure. As an optimization, we detect, offline, the \emph{numeric} properties having \emph{at most one value} for all their RDF nodes, e.g., the age of CEOs. To save memory, we allocate a single float number for all pre-aggregated results ($min$, $max$, and $sum$) for such properties. 
% Note that the memory required by the MMST is independent of the number of measures assigned to a lattice. \IM{I'm not sure why this was needed/why we come back to the MMST memory needs here}

	\section{Early-stop aggregate pruning\label{sec:early-stop}}
%\PG{I think this section's name shouldn't contain the word ``computation'': we only estimate the interestingness of aggregates; we don't evaluate them entirely. To me it can be misleading. I would replace it with ``pruning''.}

To reduce the effort required to compute lattices of aggregates, we have developed a novel technique called \textbf{early-stop} (ES).

% problem statement
\subsection{The early-stop principle}
Given an aggregate $A=\langle CFS, \mathcal{D}, M, f\rangle$ and an interestingness function $h$, finding, how interesting $A$ is, amounts to evaluating a query of the form: 

\begin{tabular}{l}
\textsf{\small \textbf{SELECT} $h$(aggregated) \textbf{FROM}}\\
\textsf{\small $\quad$ (\textbf{SELECT} $D_1,D_2,\ldots D_N, f(M)$ \textbf{AS} aggregated}\\
\textsf{\small $\quad$ \textbf{FROM} $CFS^{{\mathcal D},M}$ \textbf{GROUP BY} $D_1, D_2, \ldots, D_N$) \textbf{AS} inner;}\\
\end{tabular}
%\begin{lstlisting}
%SELECT $h(\texttt{aggregated})$
%FROM (
 %   SELECT $D_1, D_2, \ldots, D_n, f(M)$ AS $\texttt{aggregated}$
  %  FROM $A$
  %  GROUP BY $D_1, D_2, \ldots, D_n$
%) AS $\texttt{inner}$;
%\end{lstlisting}

% problem reduction: n dimensions -> 1 dimension
\noindent where $CFS^{{\mathcal D},M}$ is $CFS$ joined with dimensions $\mathcal{D}$ and the (pre-aggregated) measure $M$. Note that we only need to present the result of the inner query to the user, if $A$ ends up in the top-$k$. This leads to the following idea:  
% estimation, confidence intervals
we could reduce the effort to compute some aggregates \emph{if we can determine (with high probability) that they will not be among the $k$ most interesting ones}.

The literature~\cite{DBLP:conf/sigmod/HellersteinHW97,LargeSample} introduced conservative and large-sample confidence intervals as means of estimating \emph{the result} of a query such as \textsf{\small inner} but not the result of the full nested query, i.e., the \emph{interestingness} score that we aim to obtain.
%We follow the track open in~\cite{DBLP:conf/sigmod/HellersteinHW97}, which shows how to approximate the results, and continued in~
Recent work on visualization recommendation~\cite{seedb} shows how to \rev{stop} the evaluation of low-utility one-dimensional aggregates \rev{early} on relational data. \rev{In doing} so, it relies on a \emph{worst-case} (\emph{conservative}) confidence-interval-based pruning.
%\PG{I think I found a compelling, and distinguishing characterization of our confidence intervals. We construct large-sample intervals, as opposed to worst-case (also known as conservative) intervals offered in SeeDB (Section 4.2). These two terms (conservative and large-sample) appear in Online Aggregation paper in the introduction on p. 2. In fact, our technique abstracts away the problems of multi-valued dimensions/measures. We focus on one specific aggregate (the one in the query), and in 5.3 we make sure the data is fetched correctly, i.e., the sample is propagated using bitmaps, measures are pre-aggregated, etc.}
In contrast, we extend the line of research on  aggregate pruning by constructing a \emph{large-sample confidence interval} around the interestingness score estimator.
%, but developing novel probabilistic guarantees (Section~\ref{sec:es-formula}).
We provide our novel approach and formalize its probabilistic guarantees below.

%In the regular evaluation \PG{Maybe regular evaluation needs some rewording: I refer to P1}, we have to evaluate the inner query before computing the interestingness score $h$ in the outer query. Instead, 

To enable early-stop pruning, we estimate the interestingness of the aggregate $A$ using an estimator $\widehat{H}_r$, and bound this approximate score within our large-sample confidence interval. (We derive the formula for the interval in Section~\ref{sec:estimating-interestingness}.) %Moreover, each such estimation can be bounded using a confidence interval. Once we are able to determine the estimated scores, we can stop some unpromising aggregates early.
We draw from each aggregate group a sample containing the same number of facts. For the sake of efficiency, our sampling procedure proceeds in batches of a given size. After scanning a batch, we update the estimate of the aggregate's interestingness based on the (pre-aggregated) measure values of the facts in the batch.
%We draw a stratified sample of subjects, i.e., a sample with the same number of values from each group with the same amount of values. We fetch the sample in batches, updating the estimates after each such fetching round. Observe that we can rank all aggregates by their point estimate $\widehat{h}_r$. Further, given our goal to compute the top-$k$, an aggregate can be stopped applying the early-stopping criterion: \emph{if its upper bound is smaller than the lower bound of the $k$-th best aggregate so far}.
To prune some aggregates,  if we find that the \emph{upper-bound on the estimate of $A$'s interestingness is lower than the current lower-bound of the $k$-th best aggregate}, we can give up evaluating $A$, and thus obtain the top-$k$  aggregates more quickly. 
\rev{The} central part of Figure~\ref{fig:mvdcubewithes} %\ref{fig:early-stopping-condition} %shows a snapshot of on-going early-stopping with 
illustrates this with five aggregates and $k=3$: the fifth aggregate can be stopped after the current batch, whereas the estimation of the fourth aggregate will continue in the next batch. 
This procedure terminates once the sample is exhausted or no aggregates have been pruned in a given number of batches.

\subsection{Estimating the interestingness score\label{sec:estimating-interestingness}}
% estimation using confidence intervals: variance, skewness and curtosis; count, avg, sum

% basic notation
\textbf{Notation recall.} 
A \textbf{simple random sample} of size $r$ is a vector $[v_1,\ldots,v_r]$ of values drawn uniformly without replacement from a population $V$ of size $R$; the sample is modeled by a set of independent, identically distributed (i.i.d.) random variables $X_1,\ldots,X_r$.

An \textbf{estimator} is a random variable equal to a linear or nonlinear combination of $X_1,\ldots,X_r$ (typically modeling a simple random sample).
Evaluating the estimator on a vector $[v_1,\ldots,v_r]$ of concrete values taken by these random variables yields an \textbf{estimation}.

Let $S$ be a statistic of $V$, $\widehat{S}_r$ be an estimator of the true value of $S$ based on a sample of size $r$, and $(1-\alpha)$ be a confidence level for $0\leq\alpha\leq 1$. Then, a \textbf{$(1-\alpha)$-confidence interval} (CI) is a random interval such that for each $1\leq r\leq R$, $P(\widehat{S}_r-\underline{\varepsilon}_r\leq S\leq\widehat{S}_r+\bar{\varepsilon}_r)=1-\alpha$.
%Let $S$ be a statistic of $V$, $\widehat{S}_r$ be an estimator of the true value of $S$ based on a sample of size $r$, and $(1-\alpha)$ be a confidence level for $0\!\leq\!\alpha\!\leq\! 1$. A \textbf{$(1-\alpha)$-confidence interval} (CI) is a random interval such that for each $1\!\leq\! r\!\leq\! R$, $P(\widehat{S}_r-\underline{\varepsilon}_r\!\leq\! S\!\leq\!\widehat{S}_r+\bar{\varepsilon}_r)=1-\alpha$.
%\footnote{Often $\underline{\varepsilon}_r=\bar{\varepsilon}_r$, if we aim to have a symmetric interval}
One interval is derived deterministically from one sample; the probability is taken over all such intervals.
%\footnote{That is, if we were to draw samples of size $r$ many times, on average $(1-\alpha)$ of their corresponding %$[l_r, u_r]$
%intervals would contain the true value of $S$.}. 
We denote $L_r=\widehat{S}_r-\underline{\varepsilon}_r$ and $U_r=\widehat{S}_r+\bar{\varepsilon}_r$, respectively, the lower and the upper bounds at $(1-\alpha)$ confidence level on $\widehat{S}_r$. As in~\cite{DBLP:conf/sigmod/HellersteinHW97}, the \emph{large-sample confidence interval} contains the true value with the probability approximately equal to $1-\alpha$.

\sloppy
%Early-stop is capable of estimating the answer to query $Q$ for all the three statistical moments we use as interestingness functions over either avg, sum, min, or max as aggregate function. 
\textbf{Constructing the estimator.} We begin by developing formulas for the point estimator $\widehat{H}_r$ of the query's result when the aggregate function ($f$) in use is $count$, $sum$, or $avg$ and the interestingness function ($h$) is \emph{variance}, \emph{skewness}, or \emph{kurtosis}. We first detail this for $avg$, and variance and then discuss extensions to other functions.
% in case of variance over average in groups. \PG{add citations to Online Aggregation and mention related works: here or in the state-of-the-art section?}

Let $g_1,g_2,\ldots,g_G$ be the aggregate groups of $A$ and $\bm{\mu}=(\mu_1,\mu_2,\ldots,\mu_G)^\intercal$ be the true result of $A$,
that is, the vector containing, for each group, the average of the pre-aggregated values of $M$ for facts from that group. 
Further, for each group $g_i$, 
%$\bm{\bar{Y}}_{(g_i)}=\frac{1}{r}\sum\limits_{j=1}^{r} X_j$ 
let $\bm{\bar{Y}}_i=\frac{1}{r}\sum\limits_{j=1}^{r} X_j$ be the sample mean estimator, where the variable $X_j$ has mean $\mu_i$ and variance $\sigma_i^2$ and models the (pre-aggregated) measure value of the $j$-th fact of the sample of size $r$, drawn from the facts in $g_i$.
Note that, from the Central Limit Theorem (Theorem~5.5.14 in~\cite{CaseBerg:01}), each $\bm{\bar{Y}}_i\sim\mathcal{N}(\mu_i,\frac{\sigma_i^2}{r})$ as ${r\to\infty}$, where $\mathcal{N}(\mu_i,\sigma_i^2)$ is the normal distribution centered in $\mu_i$ with standard error $\sigma_i$.
%\PG{I fixed the variance of \bar{Y}_i}, indeed it should be divided by $r$.

We estimate $\widehat{H}_r(\bm{\mu})$ with $\widehat{H}_r(\bm{\bar{Y}})$, where
%$\bm{\bar{Y}}=\left(\bm{\bar{Y}}_{(g_1)},\bm{\bar{Y}}_{(g_2)},\ldots,\bm{\bar{Y}}_{(g_G)}\right)^\intercal$
$\bm{\bar{Y}}=\left(\bar{Y}_1,\bar{Y}_2,\ldots,\bar{Y}_G\right)^\intercal$ is the vector of all the group estimators. We thus obtain the (unbiased) estimator of the variance of a vector $\bm{y}=(y_1,y_2,\dots,y_G)^\intercal$:
%$\widehat{H}_r(\bm{y})=\frac{1}{G-1}\sum\limits_{i=1}^G\left(y_i-\frac{1}{G}\sum\limits_{j=1}^G y_j\right)^2$ 
%$\widehat{H}_r:\mathbb{R}^G\rightarrow\mathbb{R}$
\begin{equation}\label{eq:variance}
\widehat{H}_r(\bm{y})=\frac{1}{G-1}\sum_{i=1}^G\left(y_i-\frac{1}{G}\sum_{j=1}^G y_j\right)^2
\end{equation}
%We are thus interested in the distribution of $\widehat{H}_r(\bm{\bar{Y}})$.

\textbf{Deriving CI bounds.} We aim at providing a large-sample confidence interval around $\widehat{H}_r(\bm{\bar{Y}})$. Our formal result is as follows:

\begin{theorem}\label{theorem:early-stop}
Let $\widehat{H}_r$ be the estimator of variance. There exists an error $\varepsilon_r>0$ such that $\widehat{H}_r(\bm{\mu})\in[\widehat{H}_r(\bm{\bar{Y}})-\varepsilon_r,\widehat{H}_r(\bm{\bar{Y}})+\varepsilon_r]$ with the probability approximately equal to $1-\alpha$.
\end{theorem}

\begin{proof}
We prove Theorem~\ref{theorem:early-stop} constructively, thus exhibiting a concrete formula for $\varepsilon_r$.
To derive the confidence interval, first, we approximate $\widehat{H}_r(\bm{\bar{Y}})$ around $\bm{\mu}$ using the first two terms of its Taylor series expansion: $\widehat{H}_r(\bm{\bar{Y}})\approx \widehat{H}_r(\bm{\mu})+\nabla \widehat{H}_r(\bm{\mu})\cdot(\bm{\bar{Y}}-\bm{\mu})$.
Then, we apply the Multivariate Delta Method (Theorem~5.5.28 in~\cite{CaseBerg:01}) to state that
\begin{align}\label{eq:delta-method}
\sqrt{r}\left[\widehat{H}_r(\bm{\bar{Y}})-\widehat{H}_r(\bm{\mu})\right]\xrightarrow{D}\mathcal{N}(0,\tau^2)
\end{align}
where $\xrightarrow{D}$ denotes convergence in distribution,   $\tau^2=\sum\limits_{s=1}^G\sum\limits_{t=1}^G\sigma_{s,t}\frac{\partial \widehat{H}_r(\bm{\mu})}{\partial y_s}\frac{\partial \widehat{H}_r(\bm{\mu})}{\partial y_t}$,
$\sigma_{s,t}=\Cov(\bm{\bar{Y}}_s, \bm{\bar{Y}}_t)$ for $1\leq s,t\leq G$.
In other words, the difference between the correct value of interestingness, $\widehat{H}_r(\bm{\mu})$, and that on the %sample-based
estimator, $\widehat{H}_r(\bm{\bar{Y}})$, converges in distribution to a $0$-centered normal distribution. %, or, equivalently: as the number of sample grows, the sample-based interestingness estimator converges to the true value.

To apply this theorem, we must show that 
(\textbf{1})~$\widehat{H}_r$ has continuous first partial derivatives and that (\textbf{2})~$\tau^2>0$.
Condition (\textbf{1}) can be easily shown by applying basic calculus on Eq.~\ref{eq:variance}.
% %\begin{align*}
% 	$\frac{\partial \widehat{H}_r(x)}{\partial x_k}
% 	% &=\frac{\partial}{\partial x_k}\left[\frac{1}{G-1}\sum\limits_{i=1}^G\left(x_i-\frac{1}{G}\sum\limits_{j=1}^G x_j\right)^2\right]\\
% 	% &=\frac{1}{G-1}\left[\frac{\partial}{\partial x_k}\sum\limits_{i=1, i\neq k}^G\left(x_i-\frac{1}{G}\sum\limits_{j=1}^G x_j\right)^2+\frac{\partial}{\partial x_k}\left(x_k-\frac{1}{G}\sum\limits_{j=1}^G x_j\right)^2\right]\\
% 	% &=\frac{2}{G-1}\left[\sum\limits_{i=1, i\neq k}^G\left(x_i-\frac{1}{G}\sum\limits_{j=1}^G x_j\right)\cdot\left(-\frac{1}{G}\right)+\left(x_k-\frac{1}{G}\sum\limits_{j=1}^G x_j\right)\cdot\left(1-\frac{1}{G}\right)\right]\\
% 	% &=\frac{2}{G-1}\left[-\frac{1}{G}\sum\limits_{i=1}^G\left(x_i-\frac{1}{G}\sum\limits_{j=1}^G x_j\right)+x_k-\frac{1}{G}\sum\limits_{j=1}^G x_j\right]\\
% 	% &=\frac{2}{G-1}\left[x_k-\frac{1}{G}\left(\sum\limits_{i=1}^G\left(x_i-\frac{1}{G}\sum\limits_{j=1}^G x_j\right)+\sum\limits_{i=j}^G x_j\right)\right]\\
% 	% &=\frac{2}{G-1}\left[x_k-\frac{1}{G}\left(\sum\limits_{i=1}^G x_i-\frac{G}{G}\sum\limits_{j=1}^G x_j+\sum\limits_{j=1}^G x_j\right)\right]\\
% 	% &
% 	=\frac{2}{G-1}\left(x_k-\frac{1}{G}\sum\limits_{i=1}^G x_i\right)$ for $1\leq k\leq G$.
% %\end{align*}
% Thus, $\frac{\partial \widehat{H}_r(x)}{\partial x_k}$, as a sum of continuous functions, is also continuous.
For (\textbf{2}), we assume that $\bm{\bar{Y}}_{1},\bm{\bar{Y}}_{2},\ldots,\bm{\bar{Y}}_{G}$ are independent random variables. 
%\IM{Maybe ES doesn't work if there are \emph{correlations} among the attributes whose averages these estimate?} 
Hence, for $1\leq s,t\leq G$, if $s\neq t$, then $\Cov(\bm{\bar{Y}}_s, \bm{\bar{Y}}_t)=0$, else $\Cov(\bm{\bar{Y}}_s, \bm{\bar{Y}}_t)=\Var(\bm{\bar{Y}}_s)=\frac{\sigma^2_s}{r}$, and $\tau^2
%=\sum\limits_{s=1}^G\sigma^2_s\left(\frac{\partial \widehat{H}_r(\bm{\mu})}{\partial y_s}\right)^2
=\sum\limits_{s=1}^G\frac{\sigma^2_s}{r}\left(\frac{2}{G-1}\left(\bm{\mu}_s-\frac{1}{G}\sum\limits_{i=1}^G \bm{\mu}_i\right)\right)^2$ is positive. 
%\end{assumption}

We now move toward a formula for the confidence interval based on the samples in the groups. We derive it by ``standardizing'' the distribution of the difference obtained in Eq. \ref{eq:delta-method}, and taking quantiles of the standard normal distribution, $\mathcal{N}(0,1)$, as the interval's ends.
%From the Central Limit Theorem~(5.5.14 in~\cite{CaseBerg:01}) it follows that 
%$$\frac{\sqrt{r}\left[\widehat{H}_r(\bm{\bar{Y}})-\widehat{H}_r(\bm{\mu})\right]}{\sqrt{\tau^2}}\xrightarrow{D}\mathcal{N}(0,1)$$
%\PG{The passage through the CLT is not necessary, I commented it out.}

Let $\widehat{\tau^2}=\sum\limits_{s=1}^G\frac{\widehat{\sigma^2_s}}{r}\left(\frac{2}{G-1}\left(\bm{\bar{Y}}_s-\frac{1}{G}\sum\limits_{i=1}^G \bm{\bar{Y}}_i\right)\right)^2$, where $\widehat{\sigma^2_s}$ are (unbiased) estimators of variances %\footnote{Not to be confused with the interestingness function which is also variance.} 
in all the $G$ groups.
From the Strong Law of Large Numbers (Theorem~5.5.9 in~\cite{CaseBerg:01}), we have that $\lim\limits_{r\to\infty}\widehat{\tau^2}=\tau^2$ almost surely. % (or shortly a.s.).
Then, applying Slutsky's theorem (Theorem~5.5.17 in~\cite{CaseBerg:01}), we get %$\frac{\sqrt{r}\left[\widehat{H}_r(\bm{\bar{Y}})-\widehat{H}_r(\bm{\mu})\right]}{\sqrt{\widehat{\tau^2}}}\xrightarrow{D}\mathcal{N}(0,1)$.
$\sqrt{r}\left[\widehat{H}_r(\bm{\bar{Y}})-\widehat{H}_r(\bm{\mu})\right]/\sqrt{\widehat{\tau^2}}\xrightarrow{D}\mathcal{N}(0,1)$.
In turn, for large $r$, we obtain:
%\begin{small}
%\begin{align*}
	$P\left(\left|\widehat{H}_r(\bm{\bar{Y}})-\widehat{H}_r(\bm{\mu})\right|\leq\varepsilon_r\right)=P\left(\frac{\sqrt{r}\left|\widehat{H}_r(\bm{\bar{Y}})-\widehat{H}_r(\bm{\mu})\right|}{\sqrt{\widehat{\tau^2}}}\leq\frac{\varepsilon_r\sqrt{r}}{\sqrt{\widehat{\tau^2}}}\right) \approx 2\Phi\left(\frac{\varepsilon_r\sqrt{r}}{\sqrt{\widehat{\tau^2}}}\right)-1$
%\end{align*}
%\end{small}
, where $\Phi$ denotes the cumulative distribution function of a normally distributed variable. 

Let $z_p$ be the $\frac{p+1}{2}$ quantile of $\Phi$. Solving $z_p=\frac{\varepsilon_r\sqrt{r}}{\sqrt{\widehat{\tau^2}}}$ for $\varepsilon_r$, gives us $\varepsilon_r=\sqrt{\frac{z_p^2 \widehat{\tau^2}}{r}}$. Finally, choosing $z_p=z_{1-\alpha}$ we obtain the approximation at the desired confidence level:

\begin{center}
$P\left(\left|\widehat{H}_r(\bm{\bar{Y}})-\widehat{H}_r(\bm{\mu})\right|\leq\sqrt{\frac{z_{1-\alpha}^2 \widehat{\tau^2}}{r}}\right)\approx (1-\alpha)$
\end{center}
\veat{-3mm}
\end{proof}

\textbf{Other interestingness functions.} To derive confidence intervals for skewness and kurtosis, we follow similar derivations by replacing the definition of $\widehat{H}_r$ (Eq.~\ref{eq:variance}) with 
%$g(x)=\frac{\frac{1}{G}\sum\limits_{i=1}^G\left(x_i-\frac{1}{G}\sum\limits_{j=1}^G x_j\right)^3}{\left(\frac{1}{G-1}\sum\limits_{i=1}^G\left(x_i-\frac{1}{G}\sum\limits_{j=1}^G x_j\right)^2\right)^\frac{3}{2}}$, and 
%$h(x)=\frac{\frac{1}{G}\sum\limits_{i=1}^G\left(x_i-\frac{1}{G}\sum\limits_{j=1}^G x_j\right)^4}{\left(\frac{1}{G}\sum\limits_{i=1}^G\left(x_i-\frac{1}{G}\sum\limits_{j=1}^G x_j\right)^2\right)^2}-3$, respectively.
their respective formulas. We derive the CIs based on the Delta Method --
both cases exhibit continuous first partial derivatives%
%\YD{positive first partial derivatives}\PG{This was a mistake, thanks for catching it: fixed now}
\techreport{; see Appendix \ref{app:skewness-kurtosis}.}{. %Due to space constraints,
We delegate the details \rev{to our technical report~\cite{TR}}.}
%\YD{Cite a TR?} \IM{SIGMOD answer pending}
%A TR would be good, the question is: \emph{should we make it available}, and \emph{when}? Some people interpret the CfP (see ``Availability'' under \url{https://2021.sigmod.org/calls_papers_sigmod_research.shtml}) as ``We can make the TR available 2 weeks after the deadline''. What do you think?}
In general, one can derive similar formulas for any interestingness function that meets conditions (\textbf{1}) and (\textbf{2}). %examined in the proof.
%has \YD{continuous first partial derivatives}. \PG{Worth double-checking, but I think it's true.}

\textbf{Other aggregate functions.} For $sum$, we estimate the group sizes while sampling and compute the estimate as a product of the $avg$ and $count$ estimates. %\PG{Minor notation comment: somebody put math mode around aggregate functions names, I put it in rest of the occurrences in this section. Should we leave it like this?}\IM{Yes}
For $min$ and $max$, we use the sample min and the sample max, respectively, as point estimates; we apply Popoviciu's and Sz\H{o}kefalvi-Nagy's inequalities~\cite{popoviciu} for the upper and lower bounds, respectively\techreport{. See Appendices~\ref{app:sum}, and \ref{app:min-max} for details.}{. \rev{See~\cite{TR} for details}.}
%\YD{Check understanding.} \PG{In fact, the bounds aren't approximate, but still useful in practice: we can mention that.}

\subsection{Plugging early-stop into \ouralgo{}\label{sec:es-in-molap}}
We integrate early-stop into \ouralgo{} to speed up Aggregate Evaluation, and thus address challenge \textbf{C2}.
The evaluation of an MMST begins with the Data Translation step, run \emph{in parallel} with Measure Loading (recall Section~\ref{sec:mvdcube}).
%At this step of the pipeline, \sys{} has identified a set of lattices of aggregates to be computed, and has associated them with their MMSTs. For each MMST the system proceeds by: ($i$)~loading the measure values into memory and, \emph{in parallel}, ($ii$)~translating the data into the multidimensional representation (based on cells and chunks, recall Section~\ref{sec:lattice-based-computation}) to be used by \ouralgo{}.
%Given an MMST with $\mathcal{D}$ dimensions at its root, we collect tuples $(d_1, d_2,\ldots, d_N, CF)$ where $d_i$ is the value of $D_i$ and $CF$ identifies each fact in CFS. For each such tuple, we compute its address in the multidimensional space, i.e., the ID of the chunk it belongs to and its cell address inside this chunk.
We exploit the data translation to create a stratified sample of facts for the early-stop pruning. Given the MMST, each address in the multidimensional space in the root corresponds to a unique group of facts.
We allocate empty reservoirs $R_1,R_2,\ldots, R_G$, one per aggregate group, each with a capacity equal to the sample size: this way we ensure stratification. While reading each tuple, we determine its group, hence also the reservoir, and either put the fact in or not with some probability. If the reservoir is full, we discard one of the previously inserted facts. This strategy is known as reservoir sampling and guarantees a choice of a simple random sample~\cite{reservoir-sampling}. Figure~\ref{fig:mvdcubewithes} shows an on-going sampling process with four reservoirs $R_1$ to $R_4$, each of size 3.
\cut{; should $R_1$ accept a new fact, it will overwrite one of its three facts to avoid an overflow.}

The sample thus obtained is used by early-stop as follows. %in two steps: ($i$)~propagation of the sample down the MMST; ($ii$)~early-stopping.
Once the translation is finished, we propagate the facts sampled from the MMST's root down the tree using Roaring Bitmaps as in \ouralgo{} (see \cut{bitmap operations in }Figure~\ref{fig:mvdcubewithes})%
%Indeed the sampling is performed at the root level and by propagating them through the MMST \IM{Up to here this was repeating the previous phrase}
% we ensure that the remaining nodes in the tree have their own sample. \IM{This is different (talks about each node's sample) but let's clarify it even more:}
: each node in the MMST receives its own sample. 
Then, we perform the early-stop pruning based on these samples. All the aggregates that have not been pruned (deemed sufficiently interesting) by early-stop are subsequently evaluated by \ouralgo{}.

	\section{Experimental evaluation\label{sec:experiments}}

\begin{table}[t!]
	\ra{1.1}
	\resizebox{\columnwidth}{!}{
		\begin{tabular}{@{}l@{}rrrrrrrrr@{}} 
			\toprule
			\textbf{Dataset} & \textbf{\#triples} & \textbf{\#CFSs} & \textbf{\#P} & \textbf{\#A} & \multicolumn{4}{c}{\textbf{\#DP}} & \textbf{\#A} \\
			\cmidrule{6-9} 
			& & & & \textbf{woD} & \textbf{kw} & \textbf{lang} & \textbf{count} & \textbf{path} & \textbf{wD}\\
			\midrule
			%	Airline delays\footnotemark & 56M & 1 & 30 & 5923 & 0 & 0 & 0 & 0 & 5923\\
			%	CEOs\footnotemark& 85k & 237 & 61 & 159 & 1 & 1 & 37 & 462 & 27860 \\
			%	DBLP\footnotemark & 33M & 1 & 21 & 1 & 5 & 3 & 8 & 19 & 961 \\
			%	Foodista\footnotemark & 1M & 5 & 13 & 0 & 1 & 1 & 6 & 38 & 14 \\
			%	NASA\footnotemark & 99k & 10 & 37 & 19 & 3 & 15 & 3 & 87 & 1449 \\
			%	Nobel Prizes\footnotemark  & 87k & 15 & 39 & 58 & 3 & 3 & 18 & 87 & 30658 \\
			Airline~\cite{airline-delays-dataset} & 56M & 1 & 30 & 5\rev{,}923 & 0 & 0 & 0 & 0 & 5\rev{,}923\\
			CEOs~\cite{ceos-dataset} & 85k & 237 & 61 & 159 & 1 & 1 & 37 & 462 & 27\rev{,}860 \\
			DBLP~\cite{dblp-dataset} & 33M & 1 & 21 & 1 & 5 & 3 & 8 & 19 & 961 \\
			Foodista~\cite{foodista-dataset} & 1M & 5 & 13 & 0 & 1 & 1 & 6 & 38 & 14 \\
			NASA~\cite{nasa-dataset} & 99k & 10 & 37 & 19 & 3 & 15 & 3 & 87 & 1\rev{,}449 \\
			Nobel~\cite{nobel-prizes-dataset}  & 87k & 15 & 39 & 58 & 3 & 3 & 18 & 87 & 30\rev{,}658 \\
			\bottomrule
		\end{tabular}
	}
	\caption{\small Real datasets used for testing.\label{tab:realdatasets}}% \IM{Align dataset order between this and the next tables. Maybe use alphabetic order?}}\PG{Sorted}
	\veat{-8mm}
\end{table}

\begin{figure}[t!]
	\begin{subfigure}{0.3\columnwidth}
		\includegraphics[scale=0.25]{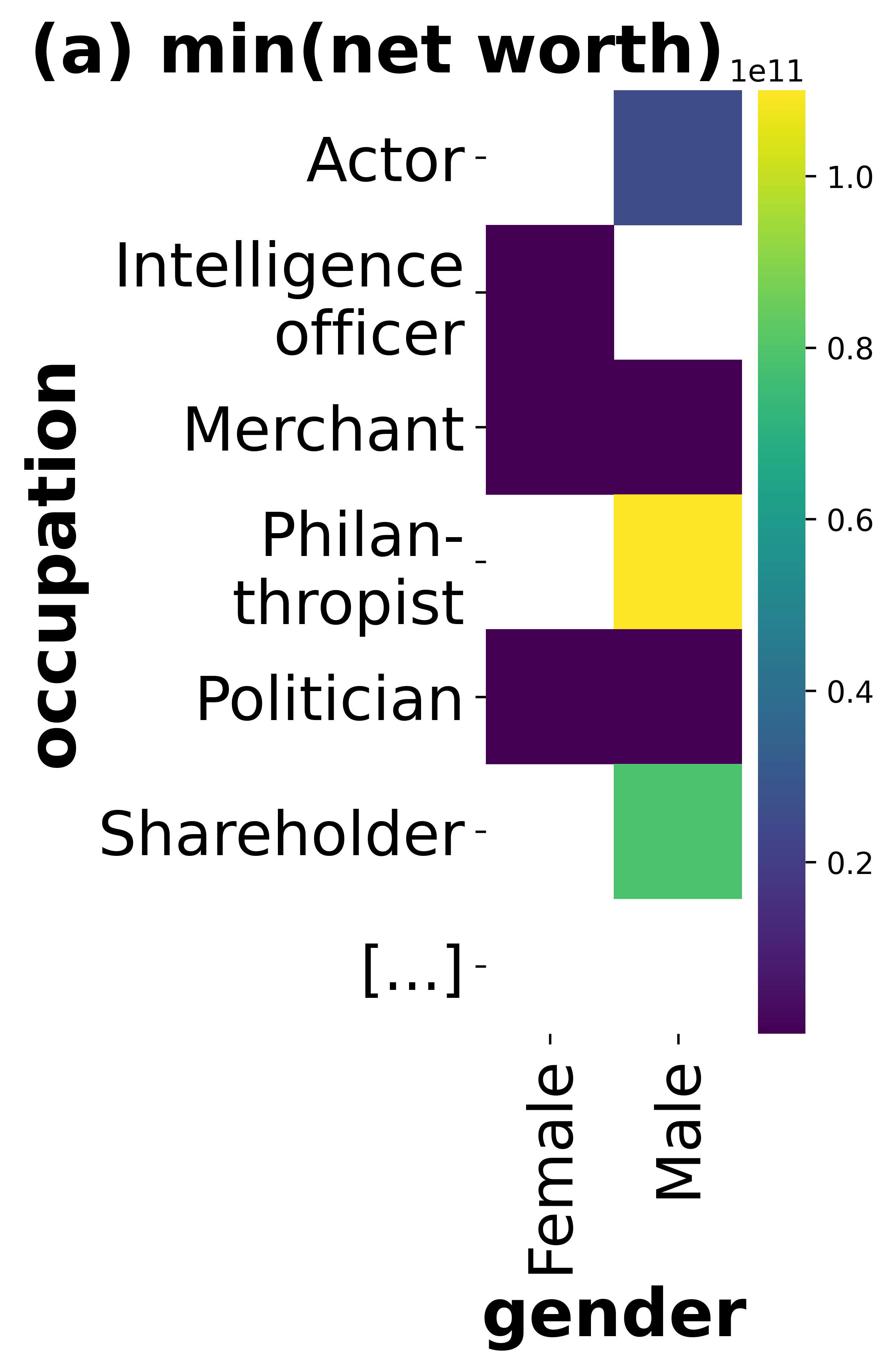}
	\end{subfigure}%
	%\hfill{}
	%\begin{subfigure}{0.24\columnwidth}
	%	\includegraphics[width=\columnwidth]{figures/MDAsFromRealRDFs/selected_aggregates/minimum_net_worth_of_CEOs_by_nationality_and_occupation.png}
	%\end{subfigure}%
	\hspace{1mm}
	\begin{subfigure}{0.3\columnwidth}
		\includegraphics[scale=0.26]{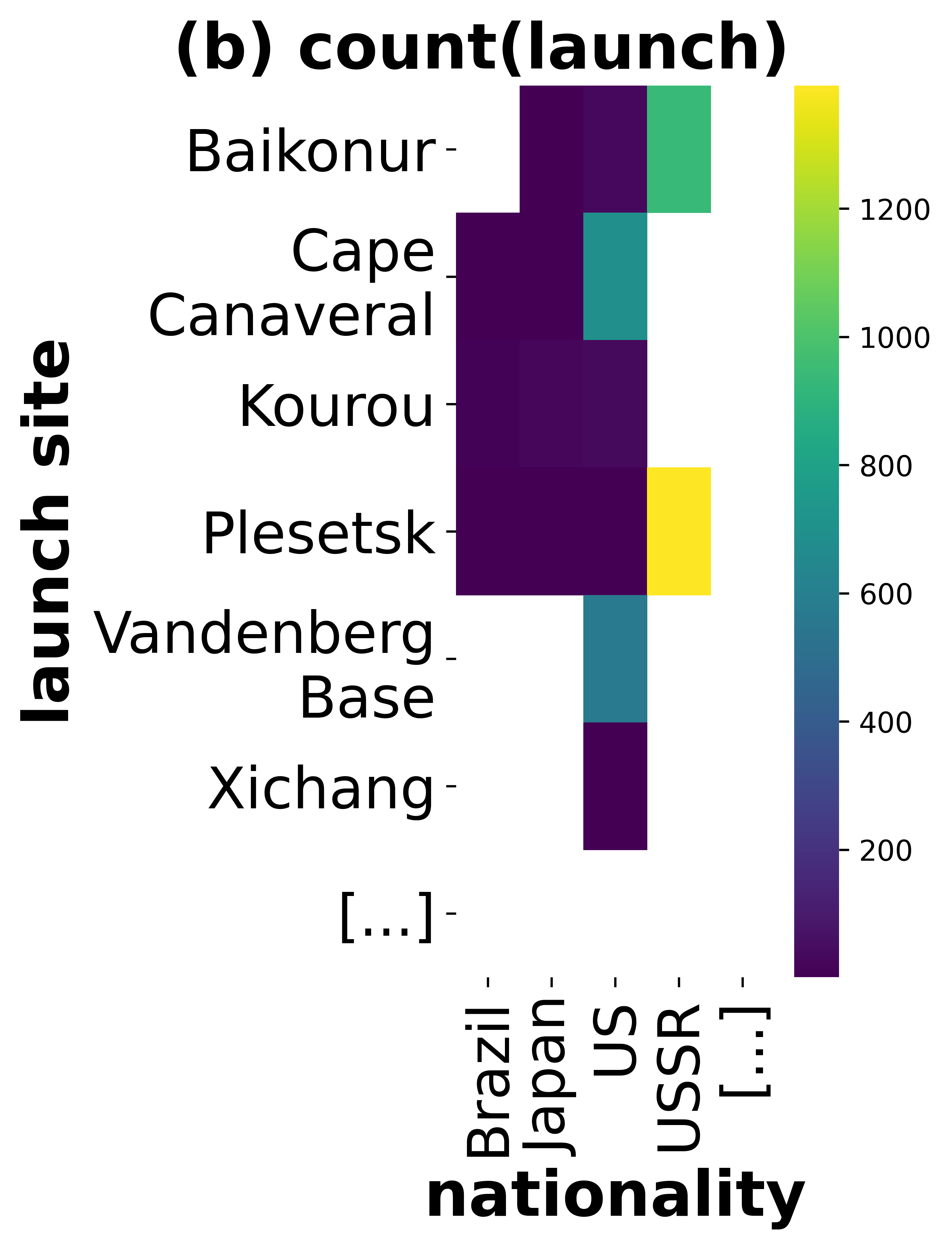}
	\end{subfigure}%
	\hspace{4.5mm}
	\begin{subfigure}{0.3\columnwidth}
		\includegraphics[scale=0.27]{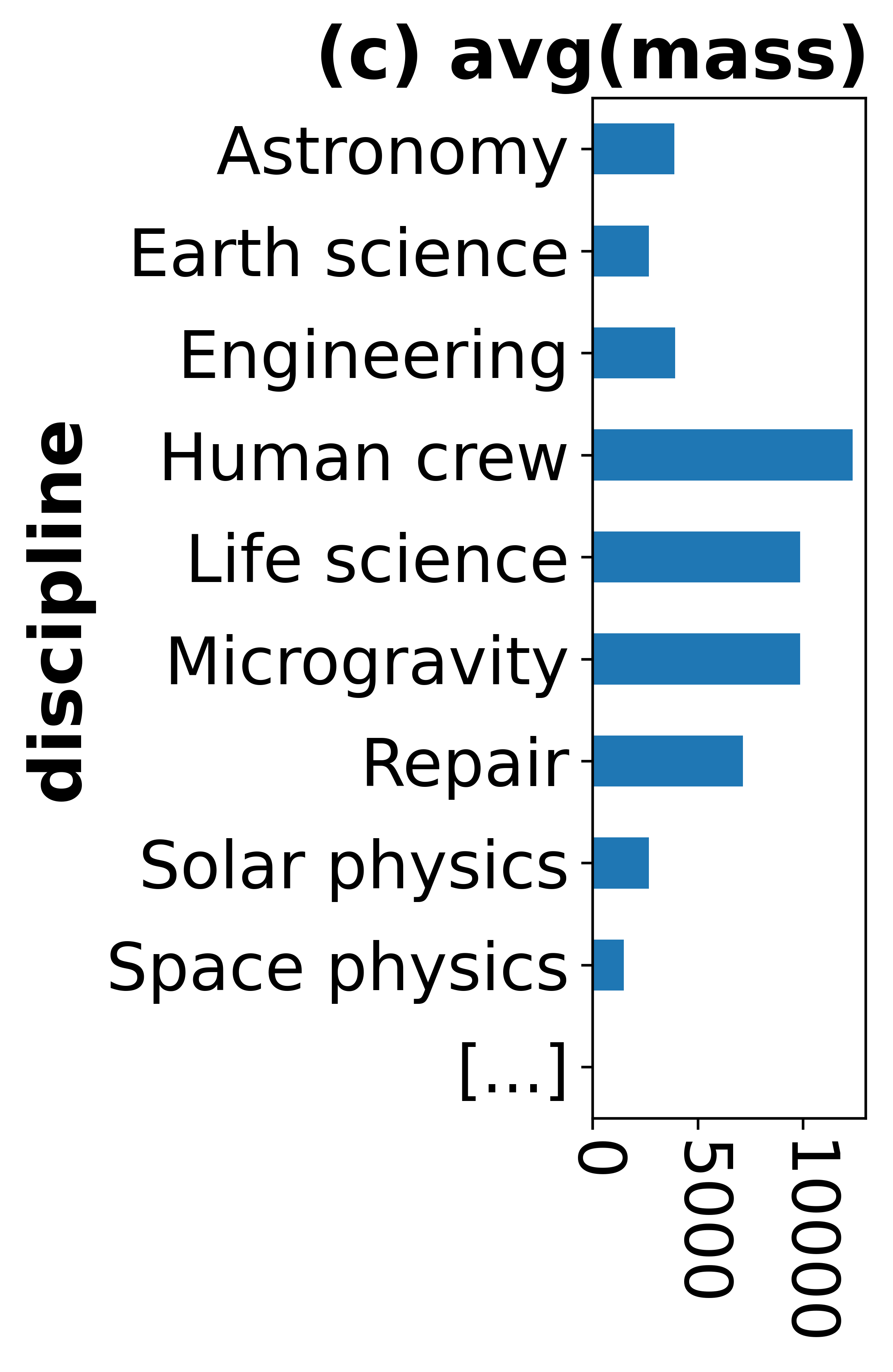}
	\end{subfigure}%
	\veat{-4.5mm}
	\caption{\small \rev{Examples of interesting aggregates found by \sys{}.}\label{fig:interesting-aggregates}}
	\Description{Examples of interesting aggregates found by \sys{}.}
	\veat{-4mm}
\end{figure}

% PG: I put the figures from experimental section here to force the LaTeX compiler to put them eariler on (a better placement).
%\input{example-aggregates-and-interestingness-profiles}

\textbf{Computational environment.} We ran all experiments on an Intel Xeon CPU E5-2640 v4 @ 2.40GHz, 40 cores (2 sockets with 10 physical cores each, hyper-threading enabled), running CentOS 7 with 90GB for JVM \rev{(OpenJDK 1.8)} and 30GB for PostgreSQL 12. 

\textbf{Systems.} We implemented \sys{} in Java 1.8 (18k lines of code); it relies on OntoSQL \rev{1.0.12}, an efficient RDF storage and query answering platform on top of an RDBMS~\cite{vldb2016,eswc2019,edbt2020} (PostgreSQL in our case).%; we used PostgreSQL 12 as the RDBMS.
%OntoSQL adopts the common practice of encoding space-consuming URIs and literals into compact integers, together with a dictionary table which allows going from one to the other. For a given class $c$, all triples of the form $x$ $type$ $c$ are stored in a single-column table $t_c$ holding the codes of the subjects $x$; for each property $p$ other than $type$, a table $t_p$ stores ($s_{code}$, $o_{code}$) pairs for each $(s,p, o)$ triple in the graph $G$.
%
\cut{Recall from Section~\ref{sec:mvdcube} that for a given RDF node $n$, all triples of the form $x$ \textit{rdf:type} $n$ are stored in a single-column table $t_n$ holding the subjects $x$; for each property $p$ other than \textit{rdf:type}, a table $t_p$ stores ($s$, $o$) pairs for each $(s,p,o)$ triple in the RDF graph.}
%
%\YD{If space is needed, details of OntoSQL such as URL encoding, etc., can be left to the appendix, and later moved to a TR online.}
%After loading an RDF graph, \sys{} builds an RDFQuotient~\cite{VLDBJournal2020} summary thereof; it also finds CFS, enumerates and analyzes their properties, creates and stores derived properties directly as (subject, object) tables in Postgres12.
%
%\YD{Here, we have to explain why we choose \pgcube{} as a baseline}%, instead of many other algorithms in the literature -- reviewers would definitely raise the question.} 
% \IM{Made a try below}
We compare the performance of our \emph{aggregate evaluation} method against the best-effort \textbf{baseline}, which 
%, similarly to \ouralgo{}, starts with the data stored in an RDBMSs. Specifically, we used
uses PostgreSQL's GROUP BY CUBE implementation, since 2016 based on an efficient one-pass computation
%\footnote{\url{https://tinyurl.com/y5f9wv75}}%https://git.postgresql.org/gitweb/?p=postgresql.git;a=commitdiff;h=f3d3118532175541a9a96ed78881a3b04a057128
%}} 
 of all aggregates in a lattice~\cite{postgres-one-pass}, that supports additional features such as $count$(distinct), which were not available in ArrayCube~\cite{molap}. We denote this by \pgcube{}. As discussed in Section~\ref{sec:theory}, \pgcube{} may fail to compute correct results in the presence of multi-valued dimensions.
%We compare our \ouralgo{} algorithm, against the best-effort implementation baseline using existing techniques: \pgcube{}. In the presence of multi-valued dimensions, \pgcube{} might fail to compute the semantically correct results as it first builds a set of GiST indexes over the data, then uses them to compute the lattice \emph{top-down in one pass}, i.e, using the results of a node to compute the results of its children.
% \IM{As we noted in rdfanalytics/SPADE/postgres_cube/infos.txt, the GiST is related to the CUBE data type and UNRELATED to how Postgres evaluates group by cube.}
However, the support for counting of distinct values may help \pgcube{} correct some wrong results. Thus, we consider two variants: ($i$)~\pgcube{} computing counts using $count(*)$, denoted \pgstar{}, and ($ii$)~\pgcube{} computing counts using $count$(distinct), denoted \pgdist{}. In both cases, \emph{our Java code is at a disadvantage against a C/C++ engine}.

\textbf{Real-world graphs.} Our experiments involve a set of real-application RDF graphs, %
%We used a set of real application graphs
 for which Table~\ref{tab:realdatasets} shows: the number of triples, the number of CFSs, the number of (direct) properties and derived properties (\#P and \#DP, respectively) in the graph, and the number of aggregates without and with derivations (\#A$_{woD}$ and \#A$_{wD}$, respectively). The graph sizes in this work are similar to the real-world dataset sizes used in comparable relational works, e.g., 20k tuples in~\cite{botang}, and up to 60M tuples in~\cite{seedb}. Airline was originally a relational dataset on flight delays used in prior work~\cite{seedb}; we converted it into RDF (each tuple becomes a CF with a fixed set of properties), whereas the others are natively RDF. We discuss differences between this and the other graphs shortly. 
%\IM{These graph sizes are similar to the dataset sizes used in comparable relational works, e.g., up to ? tuples in~\cite{seedb}, ? tuples in~\cite{...}, etc. (If the others make a simplification, e.g., ``which only considers one measure'', also add that info.)}\PG{I think both SeeDB and Bo Tang's works focus on a large number of dimensions and measures. In Bo Tang's some dimensions are derived.}

%\MM{should we add the synthetic datasets to the table as well?}\IM{No}

%\YD{We should explain why we include a relational dataset.}\IM{Done}

\begin{figure}[t!]
	\veat{-1mm}
	\includegraphics[width=0.93\columnwidth]{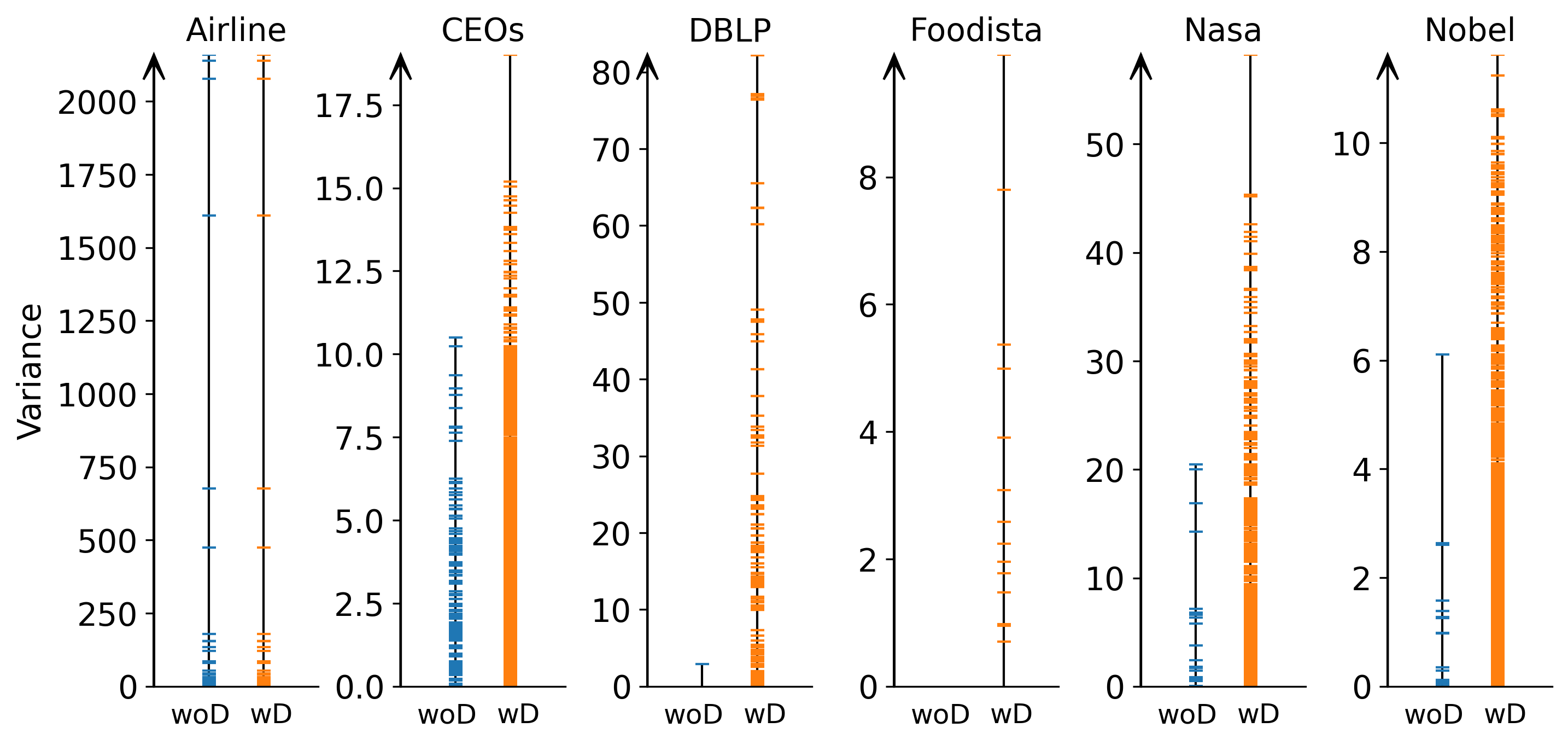}
	\veat{-4.5mm}
	\caption{\small Interestingness of MDAs due to derivations.\label{fig:profiles}}
	\Description{Examples of interesting aggregates found by \sys{}, and interestingness of MDAs due to derivations.}
	\veat{-5mm}
\end{figure}

\subsection{Analysis of example results}
\rev{We begin by showing, in Figure~\ref{fig:interesting-aggregates}, example interesting aggregates found by \sys{} when using variance as an interestingness score:}% \PG{I'm not sure if we want to stress that since we don't provide the rank of these aggregates.}

\rev{($a$)~\textit{``Minimum net worth of CEOs by gender and occupation''}: ($i$)~there are two \cut{cells clearly indicating }outliers, male philanthropists and male shareholders: their  minimum net worth is much higher than others'; ($ii$)~the net worth value is known for all but one occupation for male CEOs, but only in a half of them for female CEOs\cut{ (e.g., there are no female philanthropists)}; ($iii$)~the minimum net worth of female CEOs is nearly the same across occupations.}

\cut{\rev{($b$)~\textit{``Minimum net worth of CEOs by nationality and occupation''}: ($i$)~there are two outliers, American philanthropists and American stakeholders; they exhibit the highest minimum net worth among all; ($ii$)~behaviors also vary by country: the highest net worth is associated with the two occupations above for American CEOs, but with business people, instead, for Chinese CEOs.}}

\rev{($b$)~\textit{``Number of launches by launch site and spacecraft/agency''} in the NASA graph: ($i$)~very high values for USSR spacecrafts launched from Plesetsk and Bajkonur; ($ii$)~the \cut{two }most used USA launch sites are Cape Canaveral and Vandenberg Base.
	%\textit{``number of launches by launch site and spacecraft agency''} in the NASA graph: ($i$)~very high values for USSR spacecrafts launched from Plesetsk and Bajkonur; ($ii$)~the two most used USA launch sites are Cape Canaveral and Vandenberg Base. 
	These interesting insights were discovered thanks to our path derivations.}

\rev{($c$)~\textit{``Average mass of spacecrafts by discipline''}: here 4 disciplines, i.e., Human crew, Microgravity, Life sciences and Repair stand out with the average spacecraft mass significantly higher than others'.}

\rev{Nonetheless, many candidate MDAs are uninteresting: \techreport{Figure~\ref{fig:uninteresting-aggregates} shows}{e.g.,} the aggregate \textit{``minimum number of occupations of CEOs by gender and number of companies''}\techreport{ in the CEOs dataset}{}, where all aggregated values are uniformly equal to 1; or \textit{``average number of launched vehicles by launch site''}\techreport{ in the NASA dataset}{}, where most values are equal to 1, and only 8 out of 35 bars are slightly higher but still less than 1.05\techreport{. These aggregates don't exhibit any significant outliers and were therefore ranked low by \sys{}.}{ (see~\cite{TR} for more details).}}%
\techreport{
\begin{figure}[t!]
	\begin{minipage}{\columnwidth}
		\begin{subfigure}{\columnwidth}
			\centering
			\includegraphics[width=0.3\columnwidth]{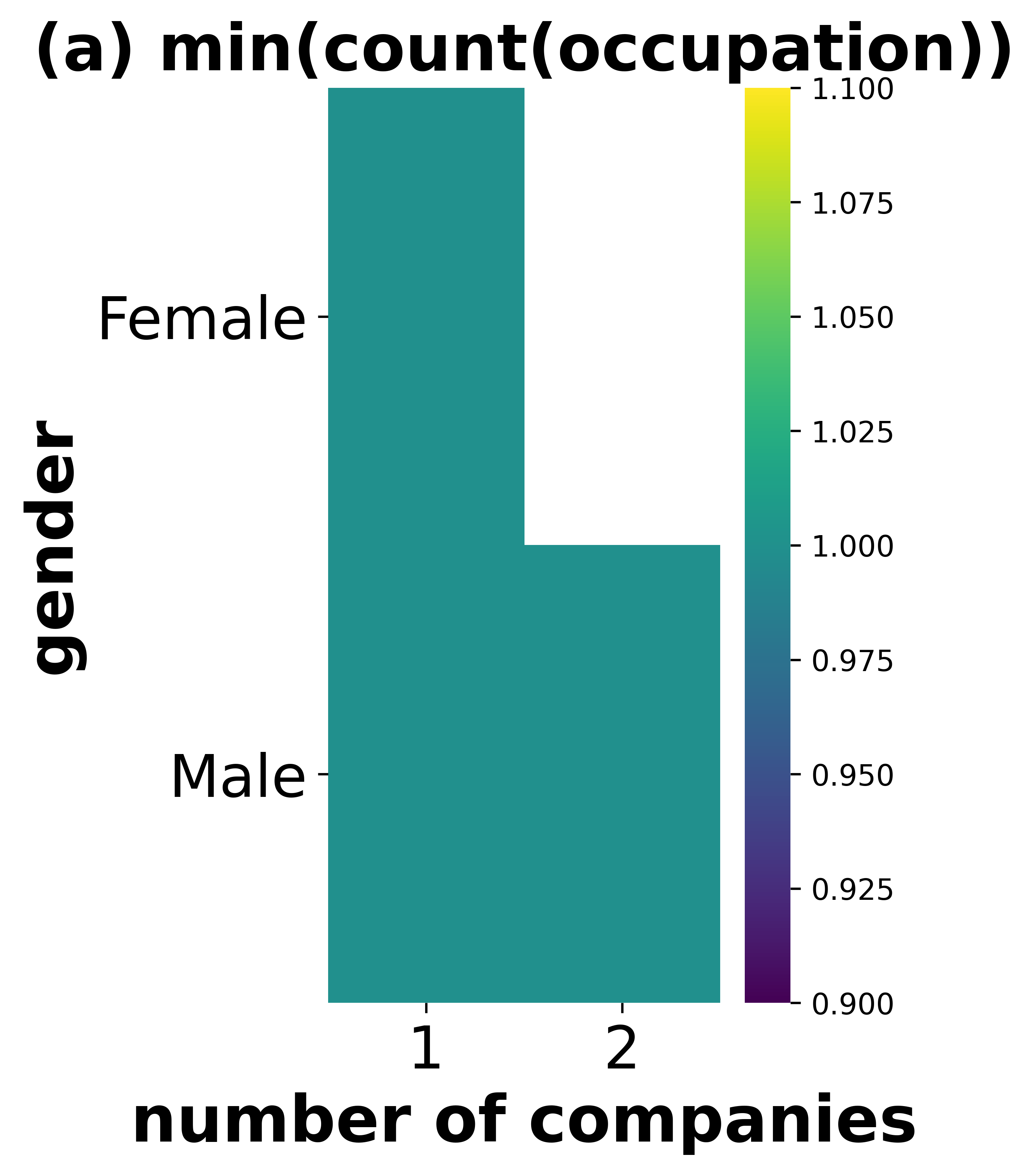}
		\end{subfigure}
		\begin{subfigure}{\columnwidth}
			\includegraphics[width=\columnwidth, height=65mm]{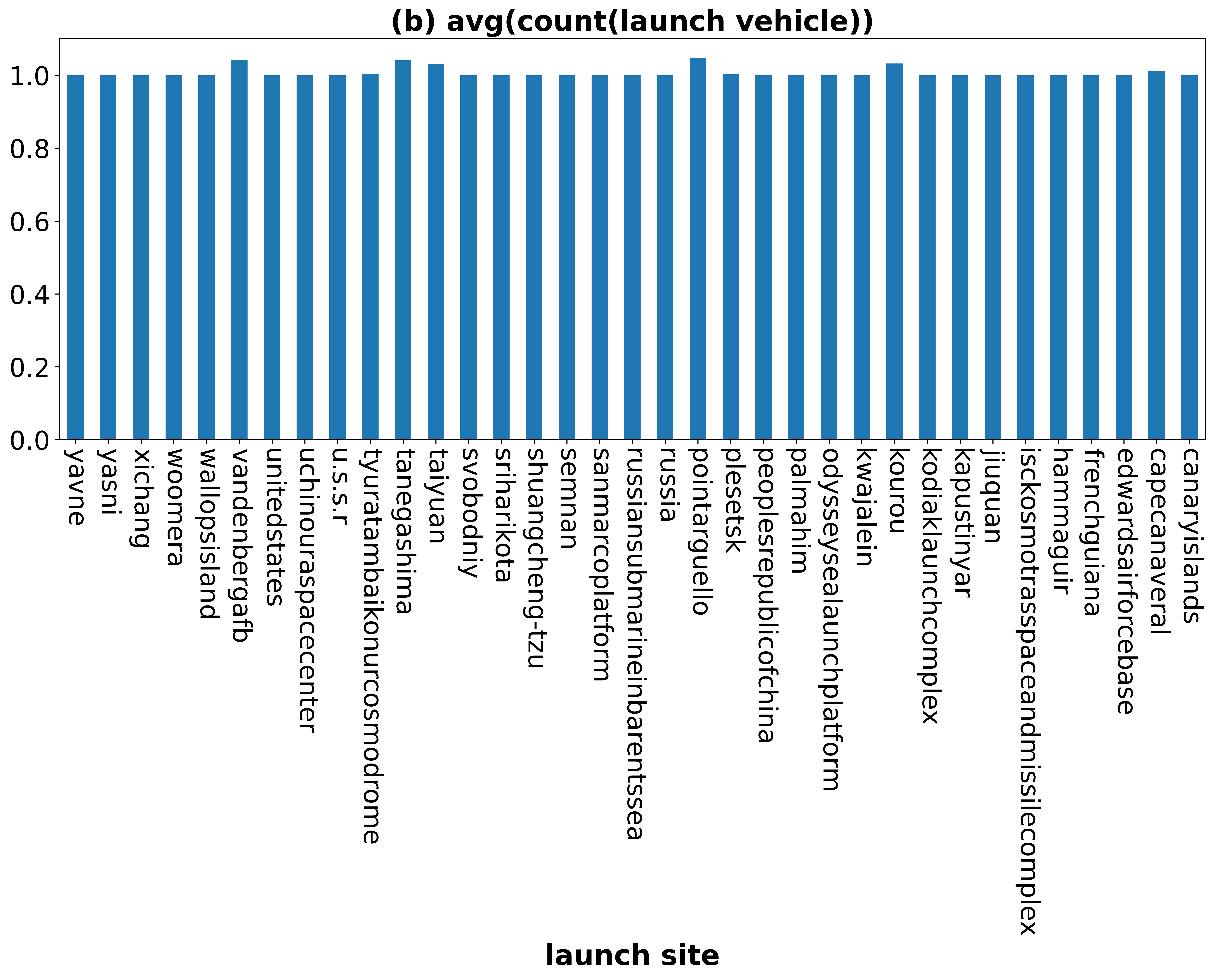}
		\end{subfigure}
		\veat{-5mm}
		\caption{\small \rev{Example uninteresting aggregates found by \sys{}.\label{fig:uninteresting-aggregates}}}
		\veat{-5mm}
	\end{minipage}
\end{figure}
}{}
%\rev{Figure~\ref{fig:uninteresting-aggregate} shows a simplified result of the aggregate \textit{``maximum number of death places by gender and country of affiliation''} in Nobel Prizes graph, which is uniform across all the aggregate groups, and was ranked 30,627.}
\rev{This confirms the need for using early-stop to  prune such MDAs.}

\techreport{\rev{It could have been in principle envisioned to compare the interestingness of the aggregates found by our system with that of some manually chosen aggregates. However, doing so is hampered by the lack of feasible selection methods available to human users. For this reason, the starting point of our work is precisely the observation that it is very hard to select aggregates manually. There are several reasons for this: ($i$)~The sheer size of the graph impedes human understanding, and it is hard to induce human users to attempt solving such a computationally expensive task at all. Even if they did try to solve it, typically, such users would use a simple SPARQL engine that can evaluate aggregates, and hence they would have to formulate the queries themselves, which requires expertise in writing such complex aggregate queries.
($ii$)~Even if we reduce a graph to a modest size, e.g., through summarization~\cite{graph-summarization-methods-survey,cebiric:hal-01925496} or sampling, the reduced graph may not reflect ($a$)~all possible combinations of facts, dimensions, and measures in the original data; ($b$)~the graph values, e.g., the frequent values and value distributions; or ($c$)~any derived properties. Even under strong (unrealistic) assumptions, e.g., ($b$) and ($c$) are both known for a simple, regular RDF graph, users would still not know which aggregates are interesting (e.g., deviating from a uniform distribution) before enumerating and evaluating them all at least partially.
($iii$)~Supporting interactions with the system leads users inevitably to inject some information about their preferences in the aggregate selection process. For example, in the NASA dataset, the users may prefer to investigate launches grouped by the launch site rather than the discipline of the spacecraft staff.
In contrast, \sys{} is a fully automated approach to discovering \emph{statistically interesting} aggregates, with no user input required. It defines and enumerates a large set of candidate aggregates by applying heuristics to generate potentially interesting dimensions and measures and evaluates them efficiently.}
		
\rev{As examples in Figure~\ref{fig:interesting-aggregates} show, our highly-ranked results returned from the six real datasets reveal interesting insights. Due to the automatic nature of \sys{}, in some datasets, there may be a small fraction of aggregates that, despite being statistically sound, are unlikely to be chosen by the user. For example, the aggregate \textit{minimum net worth of CEOs by nationality/image} uses a derived property, \textit{nationality/image}, which is statistically similar to other meaningful dimensions, e.g., \textit{nationality/label}, but the user is unlikely to choose it. This indicates that a ``human-in-the-loop'' approach can further improve the effectiveness of our automated approach. While for the above example, the user can simply add \textit{nationality/image} to a stop list for dimensions, a full design of ``human-in-the-loop'' data exploration will be a focus of our future research.}}{}

%Section~\ref{sec:XXX} we show how our early-stop technique allows us to successfully prune up to 90\% of them to focus on the evaluation of the promising ones.

\cut{\textbf{Outline.} We establish the interest of derived properties in Section~\ref{sec:exp-derive}. Next, we compare \ouralgo{} against the baseline in Section~\ref{sec:comparisonWithCube}, then we demonstrate the interest of the early-stop technique in Section~\ref{sec:exp-early-stop} and the scalability of \ouralgo{} in Section~\ref{sec:exp-scale}.}

%\begin{figure}[t!]
%	\begin{subfigure}{0.66\columnwidth}
%		\includegraphics[width=\columnwidth]{figures/MDAsFromRealRDFs/selected_aggregates/interesting_aggregate.png}
%		\caption{\rev{Interesting.}\label{fig:interesting-aggregate}}
%	\end{subfigure}%
%	\hfill{}
%	\begin{subfigure}{0.3\columnwidth}
%		\includegraphics[width=\columnwidth]{figures/MDAsFromRealRDFs/selected_aggregates/uninteresting_aggregate.png}
%		\caption{\rev{Uninteresting.}\label{fig:uninteresting-aggregate}}
%	\end{subfigure}%
%	\caption{\rev{Example aggregates found by \sys{}.}}
%\end{figure}

\subsection{The benefits of derived properties\label{sec:exp-derive}}

We begin our evaluation by validating the benefits of Derived Property Enumeration (Section~\ref{sec:overview}). This step is crucial to address challenge \textbf{C1}. We show that it allows us to increase the pool of attributes and to generate a large and rich space of interesting aggregates.
% is not needed in a relational DW, where dimensions are known in advance. In contrast, in our setting, it drastically enlarges the space of interesting aggregates, as we show below.

\textbf{Experiment 1.} We compare the results of our analytical strategy when: ($i$)~only RDF graph properties were used for the analysis ({\em woD}), and ($ii$)~derived properties were also considered ({\em wD}). As Table~\ref{tab:realdatasets} shows, the Airline dataset (originally relational) leads to no derivations: tuples are not linked to each other, and thus no paths can be derived; it lacks multi-valued attributes, thus no count derivation applies; the data is mostly numeric, so keyword or language attributes are not derived. 
The other (native RDF) graphs differ drastically: they feature several CFSs, multi-valued properties, links among RDF nodes leading to many path derivations (Table~\ref{tab:realdatasets} shows counts of path derivations of length 1, as they are the most numerous); textual attributes are also quite frequent.
%We show that the search space in ($ii$)~is much richer as we generate not only a bigger pool of candidate MDAs but also candidates with higher interestingness. To this purpose we analyze the real graphs and, in both cases (no/yes derivations), exhaustively compute the whole space of enumerated MDAs. In
Figure~\ref{fig:profiles} further shows, for each graph, the interestingness of its MDAs (measured with variance) in woD and wD settings (left and right lines, respectively);
%in the two settings. From the top down, there is a pair of horizontal bands for each graph, first without, then with derivations. A vertical tick in a band depicts an MDA.
a horizontal tick in a line depicts an MDA.

Our first main observation, denoted as \textbf{remark (R1)}, is that ($i$)~\emph{derivations increase the total number of enumerated MDAs}: for instance, on Foodista, no MDA exists without derivations, whereas we find several by deriving the recipe language, the count of ingredients, etc.; on DBLP, only \textit{year} is a good dimension, whereas through derivations we obtain, e.g., $keyword(title)$; ($ii$)~\emph{derivations increase the interestingness of the best aggregates}.

Henceforth, we enable derivations in our experimental analysis.
% we focus our experimental analysis on the setting where derivations are enabled.
% raises consistently\PG{maybe we replace ``raises consistently'' with something else?},

%Notice that derivations broadened analysis needs. In DBLP articles without derivations, only $year$ is a good dimension. Derivations allow us to significantly raise the number of candidate MDAs, we add, e.g., $keywords(title)$. This is also very visible in Foodista where, without derivations, we are unable to detect any MDA, whereas, by adding, e.g., the language of the recipe's description, the count of ingredients for each recipe, we can detect 14.

%\MM{Putting here the footnotes to move them on the same page where the table is}
%\footnotetext[6]{\url{https://www.kaggle.com/giovamata/airlinedelaycauses}}
%\footnotetext[7]{\url{https://tinyurl.com/y3mtws5h}}%https://www.dropbox.com/s/af8kzjwesz1vs2y/CEOsWithAllTheirData_Plus2hops.nt}}
%\footnotetext[8]{\url{http://www.rdfhdt.org/datasets/}}
%\footnotetext[9]{\url{https://old.datahub.io/dataset/foodista}}
%\footnotetext[10]{\url{https://data.nasa.gov/}}
%\footnotetext[11]{\url{http://data.nobelprize.org/}}

\subsection{Analysis of \ouralgo{} against \pgcube{}\label{sec:comparisonWithCube}}
Our next set of experiments focuses on Aggregate Evaluation, the last step of our online pipeline, where most computation takes place. Since \pgcube{} is not able to prune unpromising aggregates, for fairness, in this section, early-stop is disabled.

\textbf{Experiment 2.} We compare \ouralgo{} with \pgcube{} in \emph{run time} and \emph{quality} (\emph{correctness}). Recall that \pgcube{}'s results may be erroneous (Section~\ref{sec:theory}). We use the six real graphs with derivations.

\begin{figure}[t!]
	\includegraphics[width=\columnwidth]{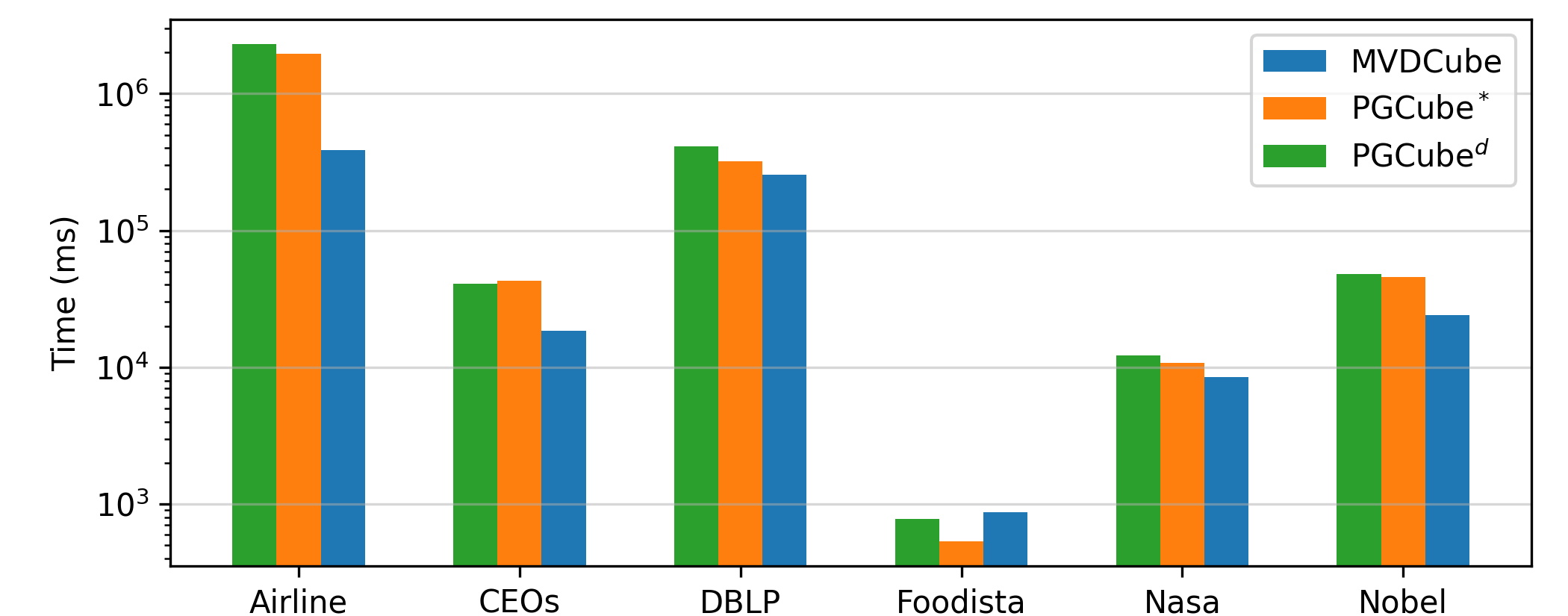}
	\veat{-7mm}
	\captionsetup{justification=centering}
	\caption{\small \rev{Run times (on log scale) of \ouralgo{} and \pgcube{}.\label{fig:mvd-cube-vs-pg-cube-run-times}}}
	\Description{Run times (on log scale) of \ouralgo{} and \pgcube{}.}
	\veat{-4mm}
\end{figure}

\begin{table}[t!]\ra{1.1}
	\begin{minipage}{0.6\columnwidth}
		\resizebox{0.88\columnwidth}{!}{
			\begin{tabular}{@{}lr@{}r@{}r@{}}
				\toprule
				\textbf{Dataset}& \textbf{\pgstar{}} & \phantom{a} & \textbf{\pgdist{}} \\
				\cmidrule{2-2} \cmidrule{3-4}
				& \textbf{\#wrong aggs} & \phantom{a} & \textbf{\#wrong aggs} \\
				\midrule
				Airline & 0 & & 0 \\
				CEOs & 4\rev{,}723 & & 3\rev{,}998 \\
				DBLP & 102 & & 87 \\
				Foodista  & 2 & & 0 \\
				NASA  & 378  & & 312 \\
				Nobel  & 4\rev{,}154  & & 3\rev{,}821  \\
				\bottomrule
			\end{tabular}
		}
		\caption{\small \rev{\pgstar{} and \pgdist{}\\errors on real-graph aggregates.\label{tab:erroneousaggregates}}}
	\end{minipage}%
	\begin{minipage}{0.4\columnwidth}
		\veat{-4.75mm}
		\includegraphics[width=0.85\columnwidth]{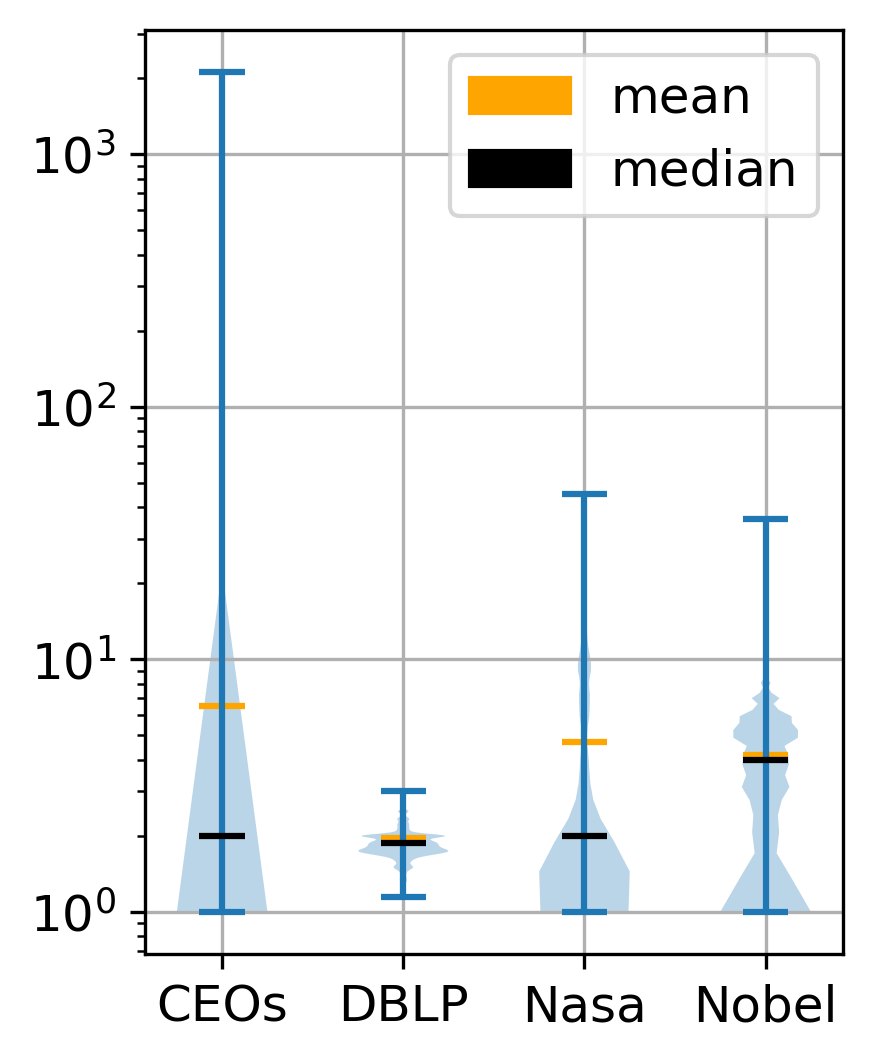}
		\veat{-5mm}
		\captionof{figure}{\small Distribution\\ of \pgdist{} errors.\label{fig:errorsincube}}
	\end{minipage}
	\veat{-8mm}
\end{table}

\begin{figure}[t!]
	\captionsetup{justification=centering}
	\centering
	\includegraphics[width=0.8\columnwidth]{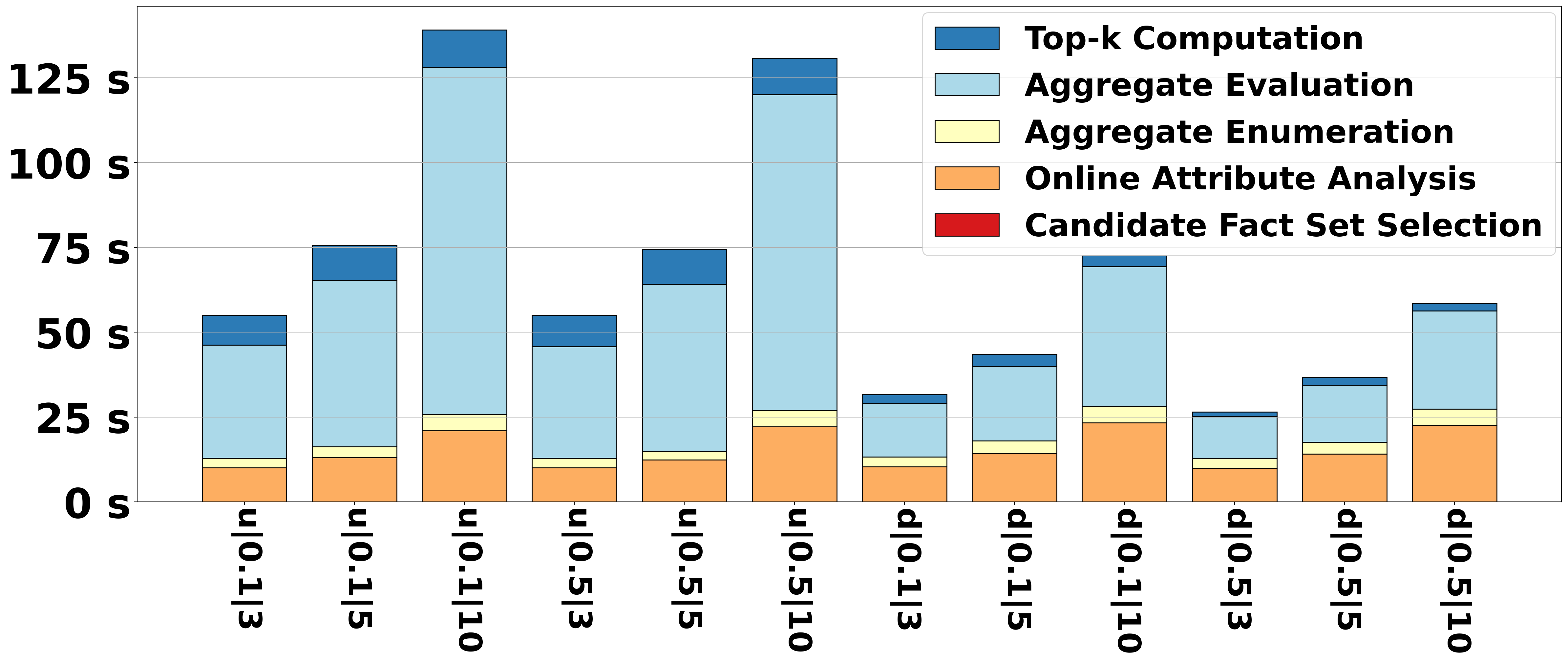}
	\veat{-4.5mm}
	\caption{\small Run times of the steps in \sys{}'s online pipeline.\label{fig:scalabilityStudyVaryingNumberOfMeasures}}
	%\caption{Scalability study varying the number of measures.\label{fig:scalabilityStudyVaryingNumberOfMeasures}}
	\Description{Run times of the steps in \sys{}'s online pipeline.}
	\veat{-4mm}
\end{figure}

\rev{Regarding the \emph{run time}, Figure~\ref{fig:mvd-cube-vs-pg-cube-run-times} shows \ouralgo{} against \pgstar{}  and \pgdist{} on our real datasets.} We observe that 
%We observe that, \underline{time-wise}, \ouralgo{} performs \emph{significantly better} in 83\% of the configurations. Our results show that \textbf{(R2)} 
\emph{\ouralgo{} achieves a time gain of 20\% to 80\% over \pgstar{} and of 30\% to 83\% over \pgdist{}} on most datasets (\textbf{R2}). 
Specifically,  \emph{\ouralgo{} outperforms \pgcube{} when there are many (more than 15) aggregates to  evaluate}  \textbf{(R3)}. This is because \ouralgo{}: ($i$)~shares measures across all the  aggregates from the same CFS, and ($ii$)~computes each aggregate only once, even if it appears in several lattices. %avoiding re-computing aggregates. In \ouralgo{}, an initial effort is needed to load all measures in memory; they are then used by all the lattices that need them. On the contrary,
In contrast, \pgcube{} evaluates each lattice in a separate query, each of which joins the facts with the measures. 
%When there are few aggregates and sharing is limited, \pgcube{} performs better (because it needs to join few times). This is the case on Foodista. The difference is not dramatic given that the times on Foodista are short: \ouralgo{} takes 99ms more than \pgdist{}. 
%On the contrary, on CEOs, NASA and Nobel Prizes, where many MDAs are evaluated,
%\ouralgo{} gains 40\% over \pgcube{}. Similarly, Airline delays leads to almost 6k MDAs, the dataset is rather large (6M facts), and the repeated joins are expensive: \pgstar{} takes 5 times \ouralgo{}'s time. 
Except for the Foodista dataset, which has a small number of aggregates and both methods run under a second,
\ouralgo{} shows significant gains on CEOs, NASA and Nobel Prizes graphs, where many MDAs are evaluated,
% we witness an explosion of the number of MDAs in the presence of derivations. This is mostly due to many path derivations (see Table~\ref{tab:realdatasets}) that are deemed good dimensions. 
\ouralgo{}  gains 40\% over \pgcube{}.
Similarly, Airline leads to almost 6k MDAs, the dataset is rather large (6M facts), and the repeated joins are expensive: \pgstar{} takes 5 times \ouralgo{}'s time. 

%We observe that, \underline{time-wise}, \ouralgo{} performs significantly better than \pgcube{} when many aggregates are to be evaluated. In nobelprizes with derivations it takes almost half of the time needed by \pgstar{}, in CEOs with derivations it takes even less than half of \pgstar{}'s time. Derivations increase the number of aggregates but the time is not strictly dependent on them. Indeed, in airlinedelays without derivations \emph{\ouralgo{} takes one fifth of the time needed by \pgstar{}}. In general, 

%\YD{Is the better performance of PG due to some algorithmic aspects, or the simple fact that it is C (against Java).}

\rev{Regarding the \emph{errors}, Table~\ref{tab:erroneousaggregates} shows, for each graph, the number of aggregates with incorrect results (\#wrong aggs) for \pgstar{} and \pgdist{}.} We observe that \pgstar{} and \pgdist{} produce errors in, respectively, 14\% and 12\% of all computed aggregates \textbf{(R4)}. 
%\underline{Error-wise} we can see that \pgstar{} produces errors in 58\% of the configurations. This percentage is slightly reduced (to 50\%) by \pgdist{}. Moreover, we remark 
%The amount of aggregates that contain errors is significant. Indeed, 
\pgdist{}, \pgcube{}'s best effort to generate correct results, still produces errors in 9\% to 21\% of the computed aggregates across different datasets. As shown in Section~\ref{sec:theory}, errors are related to multi-valued attributes in the data. Indeed, CEOs, NASA, and Nobel Prizes datasets have the greatest number of multi-valued attributes and the highest error, ranging from 12\% to 21\%.

\textbf{Experiment 3.} We now \emph{quantify the error} in those aggregates that are computed wrongly by  \pgdist{}. Given an aggregate $A$, we denote $m_j^A$ the value of the aggregated measure of the $j$-th group in $A$, as computed by \ouralgo{}.
We denote by $p_j^A$ the value that \pgdist{} computes for the same group. %\IM{I simplified the notation a bit: maybe we can do without a second aggregate here...}
%We compare the result found by \ouralgo{} ($A_{M}$) and that found by \pgdist{} ($A_{C}$) by computing, for each group in $A$, the ratio $m_j^{A_{C}}/m_j^{A_{M}}$. Since $m_j^{A_{C}}$ can only be \emph{higher than or equal to} the correct value $m_j^{A_{M}}$, 
As $p_j^{A}$ can only be \emph{higher than or equal to} the correct value $m_j^{A}$,
ideally, this ratio should be $1$. 
When an aggregate is shared by two lattices, %e.g., both Example~2 and Example~3 have dimension \textit{nationality}, 
it can be computed from either lattice, leading to different error ratios. %: in particular, $A$ may be computed \emph{correctly} in one lattice and \emph{incorrectly} in the other. 
% This might produce two different results: consider Example~3 and the variation of Example~2: \textit{``number of CEOs by nationality and number of managed companies''}. Aggregate $A_5$: \textit{``number of CEOs by nationality''} can be computed from both. Example~3 has two multi-valued dimensions but $A_5$ contains only one, thus, according to Theorem~\ref{theo:fddoesnothold}, it might contain errors. Instead, Example~2 has one multi-valued dimension which is present in $A_5$, thus, it is computed correctly.
When this happens, we record the maximum error, to measure the ``worst-case risk'' incurred by evaluating the lattice through \pgcube{}. Each aggregate thus leads to a set of error ratios, one per group. 
Figure~\ref{fig:errorsincube} shows their distribution, for $count$ and $sum$ aggregates, for the four datasets from Table~\ref{tab:erroneousaggregates} where errors were detected.
We note that \emph{errors can easily exceed one order of magnitude} \textbf{(R5)}: in 3 out of 4 cases, \pgdist{} produces at least 1 tuple whose value is more than 30 times the true value. In CEOs, one group records an error ratio greater than $10^3$; it comes from a three-dimensional lattice where  all dimensions were multi-valued. Such  incorrect values would severely falsify  the selection of the $k$ most interesting aggregates. 

%The presence of errors has a big impact on the whole pipeline of \sys{}. Indeed, they invalidate the interestingness of aggregates and, thus, prevent us from finding the true top-K.

%\YD{Again this discussion should lead to clear conclusions, which  we can cite  with numbers in Abstract, Introduction, conclusions, etc.}

\subsection{Impact of early-stop on \ouralgo{}\label{sec:exp-early-stop}}
%by pruning unpromising aggregates, and thus, reducing the evaluation effort.
%In addition to the previously shown results, from Experiment~3 we also learned that evaluation is a costly step. We want to reduce the evaluation effort using early-stop to prune unpromising aggregates before the complete evaluation. Therefore, here we study the performance of early-stop, P3, compared to P1. 

% \ouralgo{} already outperforms \pgcube{}, but we can do better
%Let's now consider our main objective: finding the $k$ most interesting aggregates. Even though only \ouralgo{} can compute them correctly, while still remaining efficient compared to \pgcube{}, we learned from Table~\ref{tab:erroneousaggregates} that such evaluation can be costly.
%From Figure~\ref{} we observe that, across all the datasets, evaluation took XX time on average.
%To address this practical concern, we study the performance gains thanks to early-stop.

%\textbf{Experiment 3.} We focus on early-stop effectiveness, that is: ($i$)~we compare the results obtained from P1 (where all aggregates are computed) and P3 (where unpromising aggregates are pruned online) and ($ii$)~we report the accuracy of P3. Given \{$MDA_{P1}$\} and \{$MDA_{P3}$\}, the set of aggregates with the highest score found from P1 and P3 respectively, the accuracy is defined, as in~\cite{seedb}, as the fraction of true positives in \{$MDA_{P3}$\}: $\frac{|\{MDA_{P3}\} \cap \{MDA_{P1}\} |}{|\{MDA_{P1}\} |}$

\textbf{Experiment 4.} We next study the effectiveness of our early-stop technique (ES).  For our real graphs, Table~\ref{tab:early-stop} shows: ($i$)~the evaluation time taken by \ouralgo{} alone, ($ii$)~the time with ES enabled, as described in Section~\ref{sec:es-in-molap}, ($iii$)~the time gain due to ES, ($iv$)~the fraction of aggregates pruned and ($v$)~the accuracy of ES. Following~\cite{seedb}, if $T_k^{w/o}$ and $T_k^{w}$ are the sets of the top-$k$ aggregates returned by \ouralgo{} without and with ES, the accuracy is computed as the fraction of true positives in $T_k^{w}$: $|T_k^{w/o} \cap T_k^{w}|/|T_k^{w/o}|$. %\IM{Removed \frac as it occupies a lot of space}
We show this for $k\in\{3,5,10\}$, in keeping with comparable works in a relational DW setting~\cite{seedb} %\IM{Also in Bo Tang's papers?} \PG{They range up to top-100. However, I don't think it makes sense: no one is going to browse this much.}, 
and using a sample size of $60$ with 2 batches, a configuration we found empirically to work well.
Table~\ref{tab:early-stop} leads to two observations. First,  \emph{ES can bring significant evaluation time gains}, from 10\% to 43\% in our experiments; and it \emph{aggressively prunes uninteresting aggregates}, frequently as much as 70\% \textbf{(R6)}.
ES is especially beneficial on graphs with more than 100 aggregates, except for DBLP, where  translating the data into an array representation is much more expensive than evaluation, and thus, the saved evaluation effort appears small.
In some cases, the impact of ES was negative (and very small), due to a sampling overhead. 
Second, \emph{\ouralgo{} with ES is often quite accurate} \textbf{(R7)}: 100\% accuracy is attained in the majority of cases, %although there are a few negative cases in particular for Nobel Prizes and Foodista.
except for Nobel Prizes, where, e.g., the true top-$10$ contains aggregates with interestingness score greater than $10.49$, whereas ES returns those greater than $9.45$.%\PG{Maybe we can somehow connect it to Figure~\ref{fig:profiles}}.
%\YD{What is the problem with Nobel Prizes? Need to say a bit more about Foodista.}

%\PG{I put the table corresponding to sample size = 60 and k = 1, 3, 5, 10, no derivations, and evaluation time only. We can say that we found empirically sample size = 60 to work well for us.}
%We focus on early-stop effectiveness. To measure the impact on \sys{}, we show that early-stop \textbf{(R4.1)}~achieves performance gains from 10\% up to 43\% for $k$ ranging in ${1, 3, 5, 10}$; \textbf{(R4.2)}~aggressively prunes uninteresting aggregates, frequently as much as 70\% of the total number; and \textbf{(R4.3)}~excels in accuracy of the top-$k$, scarcely below 100\%. 
%\PG{Agree on the numbering of results \textbf{R}} \IM{Reading from the beginning, it appeared better to call these \emph{remarks}.} 
%\PG{Remarks don't sound very impressive... These are our results.}\IM{We also have algorithmic and theoretical results, so these are \emph{some} of our results. Maybe call them Experiment Conclusions (\textbf{EC}$_i$)?}

\begin{table*}[t]
	\resizebox{\textwidth}{!}{
		\begin{tabular}{l|rrrrr|rrrrr|rrrrr}
			\ & TOP 3 & \ & \ & \ & \ & TOP 5 & \ & \ & \ & \ & TOP 10 & \ & \ & \ & \ \\
			\toprule
			dataset & \ouralgoshort{} &\ouralgoshort{}+ES & gain\% & pruned\% & acc\% & \ouralgoshort{} &\ouralgoshort{}+ES & gain\% & pruned\% & acc\% & \ouralgoshort{} &\ouralgoshort{}+ES & gain\% & pruned\% & acc\% \\
			\midrule
			Airline & 381\rev{,}710 & 316\rev{,}168 & \textbf{17.17} & \textbf{96.13} & \textbf{100.00} & 369\rev{,}369 & 316\rev{,}885 & \textbf{14.21} & \textbf{93.5}2 & \textbf{100.00} & 373\rev{,}660 & 330\rev{,}467 & \textbf{11.56} & \textbf{88.10} & 90.00 \\

			CEOs & 18\rev{,}114 & 14\rev{,}624 & \textbf{19.27} & \textbf{79.21} & 33.33 & 18\rev{,}685 & 14\rev{,}108 & \textbf{24.50} & \textbf{72.86} & \textbf{100.00} & 18\rev{,}047 & 15\rev{,}108 & \textbf{16.29} & 66.86 & \textbf{100.00} \\

			DBLP & 256\rev{,}832 & 255\rev{,}918 & 0.36 & \textbf{88.03} & \textbf{100.00} & 250\rev{,}916 & 248\rev{,}982 & 0.77 & \textbf{85.33} & \textbf{100.00} & 249\rev{,}463 & 256\rev{,}325 & -2.75 & \textbf{80.85} & \textbf{100.00} \\

			Foodista & 855 & 917 & -7.25 & 0.00 & \textbf{100.00} & 1\rev{,}173 & 893 & 23.87 & 0.00 & \textbf{100.00} & 886 & 920 & -3.84 & 0.00 & \textbf{100.00} \\

			NASA & 8\rev{,}633 & 7\rev{,}366 & \textbf{14.68} & \textbf{82.40} & \textbf{100.00} & 8\rev{,}581 & 7\rev{,}750 & 9.68 & \textbf{76.54} & 80.00 & 8\rev{,}458 & 8\rev{,}151 & 3.63 & \textbf{59.01} & \textbf{100.00} \\

			Nobel & 24\rev{,}633 & 13\rev{,}897 & \textbf{43.58} & \textbf{95.94} & 0.00 & 24\rev{,}453 & 13\rev{,}829 & \textbf{43.45} & \textbf{95.59} & 20.00 & 23\rev{,}848 & 14\rev{,}267 & \textbf{40.18} & \textbf{94.70} & 30.00 \\
			\bottomrule
		\end{tabular}
	}
	\caption{\small Early-stop effectiveness on real datasets. All times in ms; in bold: gain\% > 10\%, pruned\% > 70\%, and acc\% = 100\%.\label{tab:early-stop}}
	\veat{-7.5mm}
\end{table*}

\subsection{Scalability study\label{sec:exp-scale}}
% Points we want to make in this section
% \ouralgo{} with or without early-stop scales well: see curves
% P1 is better than P2
% P3 is better than P1
% ergo P3 is what we want to use
We finally analyze the scalability of our approach and compare it with \pgcube{}, when varying different data characteristics. To be able to fully control them, we designed a \textbf{synthetic benchmark} (a set of graphs) with fixed numbers of facts $|CFS|$, $N$ dimensions and $M$ measures. All property values are numeric. 
We ensure that a single CFS is found and that each dimension $D_i$, $1\leq i \leq N$, takes at most $100$ values (so that they are considered good dimensions, recall Step 2 in Section~\ref{sec:overview}). We denote each graph by $|D_1|$ : $|D_2|$ : $\ldots$ : $|D_N|$, the maximum number of distinct values along each dimension. To obtain realistic distributions of the facts in this multidimensional space, we randomly assign dimension values as in~\cite{DBLP:conf/vldb/AgarwalADGNRS96}, controlled by a sparsity parameter $s\in [0,1]$. To ensure \pgcube{} correctness, each fact has only one value for each dimension. 

%As in~\cite{DBLP:conf/vldb/AgarwalADGNRS96}, we control the ratio among the numbers of distinct values by a formula of the form $d_1$:$d_2$:$\ldots$:$d_N$, as well as the sparsity $s\in[0,1]$ of the values chosen for each dimension. For the $i$-th dimension, we draw random values between $1$ and $(|CFS|/s)^\frac{1}{N} d_i / (d_1\cdot d_2\cdot\ldots\cdot d_N)^\frac{1}{N}$. 

%\YD{This figure is hard to read for several reasons: (1) the steps do not match those in Figure 2; (2) the x-labels are too complex and too small. Please reconsider the main messages that we want to deliver and re-organize results to highlight those messages. } \PG{I tried to answer these comments.}

% Experiment 5: vary the number of measures and the sparisty of dimensions: times using P1
\textbf{Experiment 5.} We analyze the performance of \emph{the entire online pipeline of \sys{}} on benchmark datasets.
%We vary the number of measures, the number of distinct values, and the sparsity of dimensions.
We use 12 configurations, each having $|CFS|$=$1M$, 3 dimensions, and 3, 5, or 10 measures. We also use ($i$)~two different combinations of %the maximum numbers of 
distinct values for dimensions, 100:100:100 \rev{(uniform}) and 100:5:2 \rev{(decreasing)}, and ($ii$)~two different sparsity coefficients, $0.1$ and $0.5$.
In Figure~\ref{fig:scalabilityStudyVaryingNumberOfMeasures}, each bar represents one configuration 
\rev{(``u'' or ``d'' for value distribution | sparsity coefficient | number of measures)}
and reports the total execution time of \sys{} using \ouralgo{} without early-stop. Each segment of a bar covers one computation step (recall Figure~\ref{fig:framework}).
In the pipeline order of steps, we observe that:
($i$)~Candidate Fact Set Selection is too fast to be visible; although there is only one CFS here, \emph{in all our experiments with real graphs}, it was 5-10 ms.  
($ii$)~Online Attribute Analysis's time is noticeable, between 15\% and 37\% of the total time, and increases with the number of measures: \sys{} must analyze them before deciding that they are not suitable dimensions. 
($iii$)~Aggregate Evaluation dominates the processing time; it increases with the number of distinct groups %(multidimensional cells containing facts) 
and the number of measures as each measure leads to a different aggregate. 
($iv$)~The time to select the best aggregates (evaluate their interestingness and pick the top-$k$) is also noticeable and grows as expected with the number of aggregates. ($v$)~Sparsity has a moderate impact.
From these results, we conclude that \emph{for a fixed CFS, Aggregate Evaluation dominates \sys{}'s execution, increasing with the number of distinct groups and the number of measures; Online Attribute Analysis has the second-highest cost, growing with the number of attributes} \textbf{(R8)}.

\begin{figure*}[t!]
	\centering
	\begin{subfigure}[t]{0.33\textwidth}
		\centering
		\includegraphics[scale=0.24]{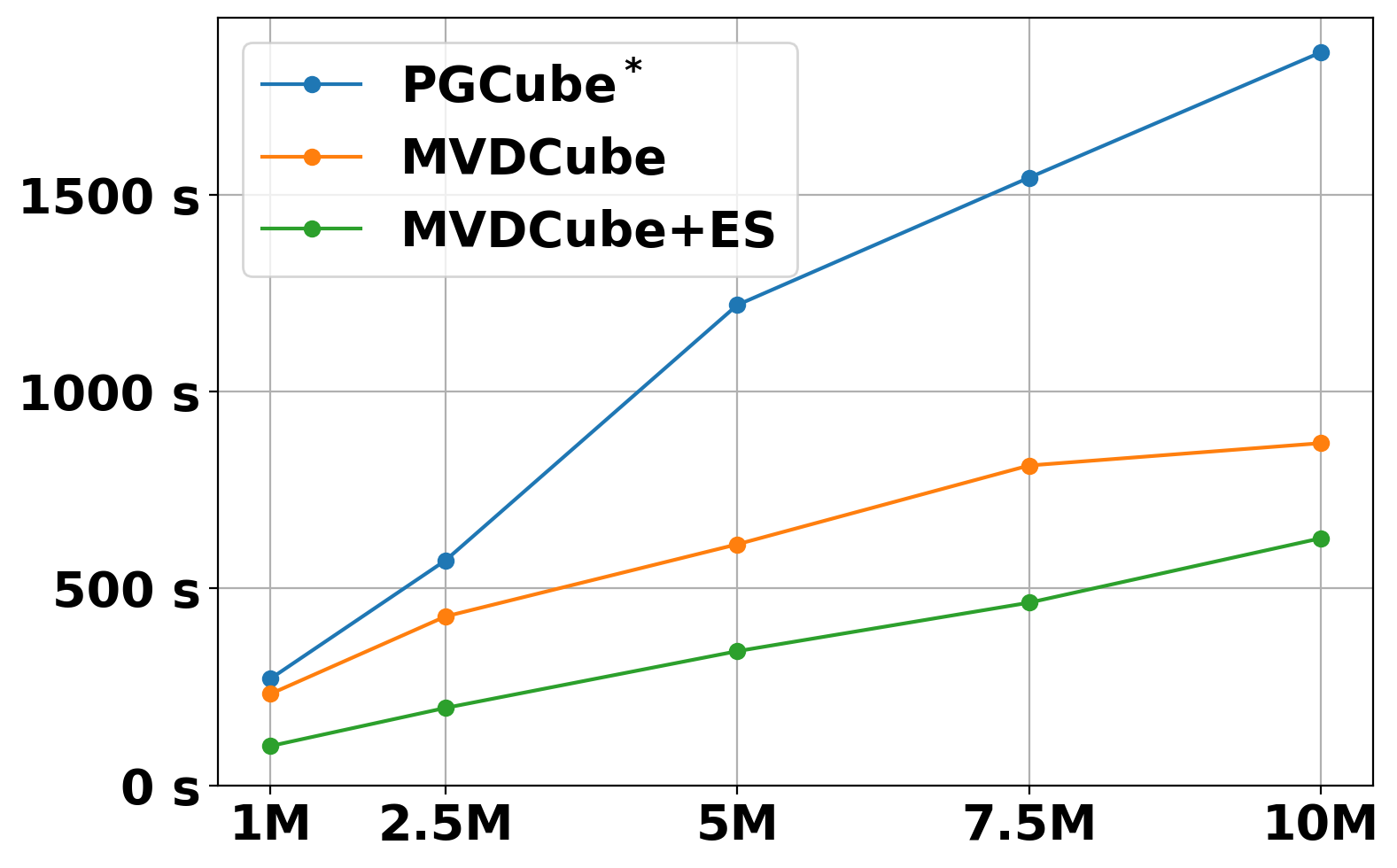}
		\veat{-2mm}
		\caption{\small Varying the number of facts $|CFS|$.\label{fig:scalability-varying-facts}}
	\end{subfigure}%
	~ 
	\begin{subfigure}[t]{0.33\textwidth}
		\centering
		\includegraphics[scale=0.24]{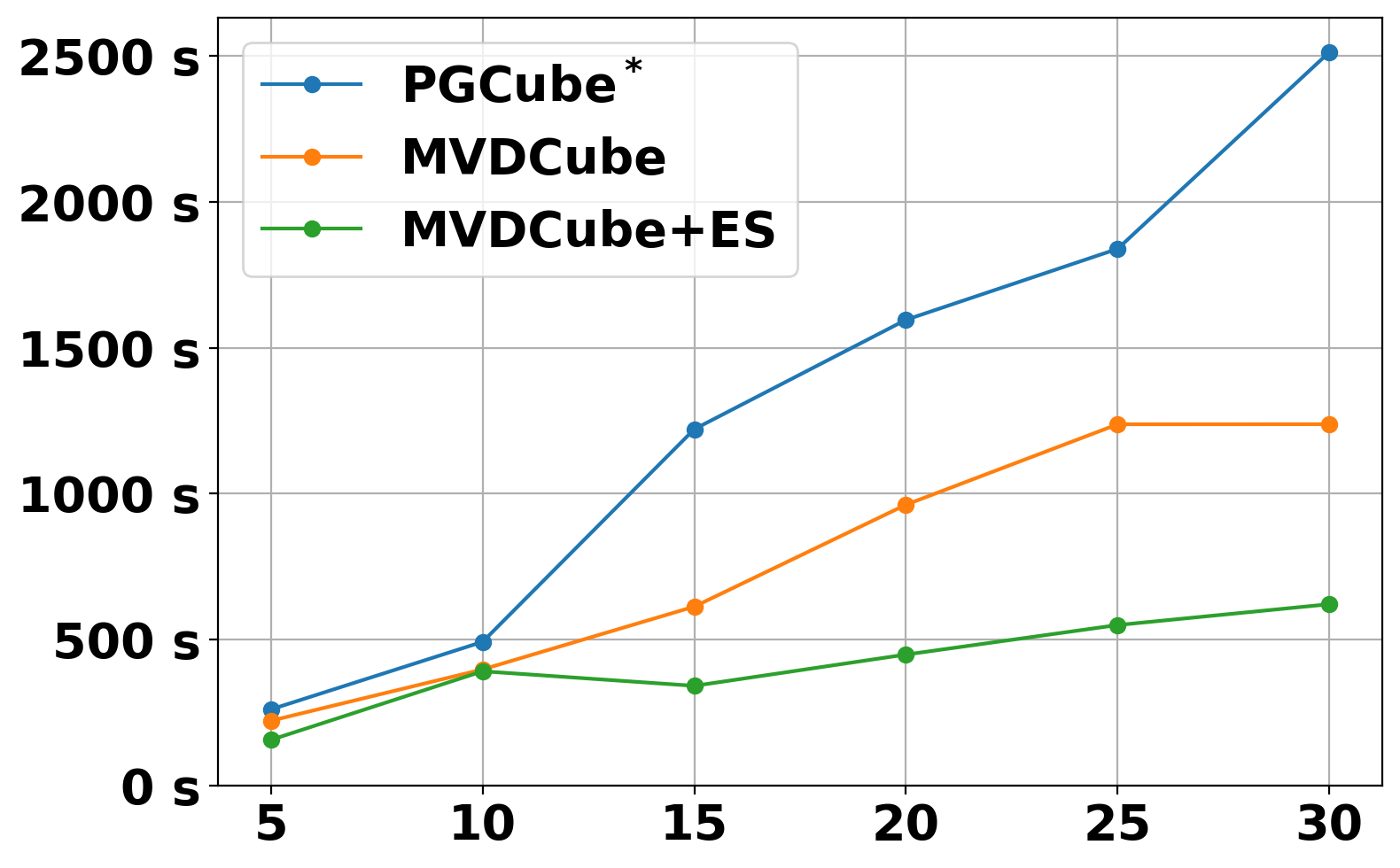}
		\veat{-2mm}
		\caption{\small Varying the number of measures $M$.\label{fig:scalability-varying-measures}}
	\end{subfigure}
	~
	\begin{subfigure}[t]{0.33\textwidth}
		\includegraphics[scale=0.24]{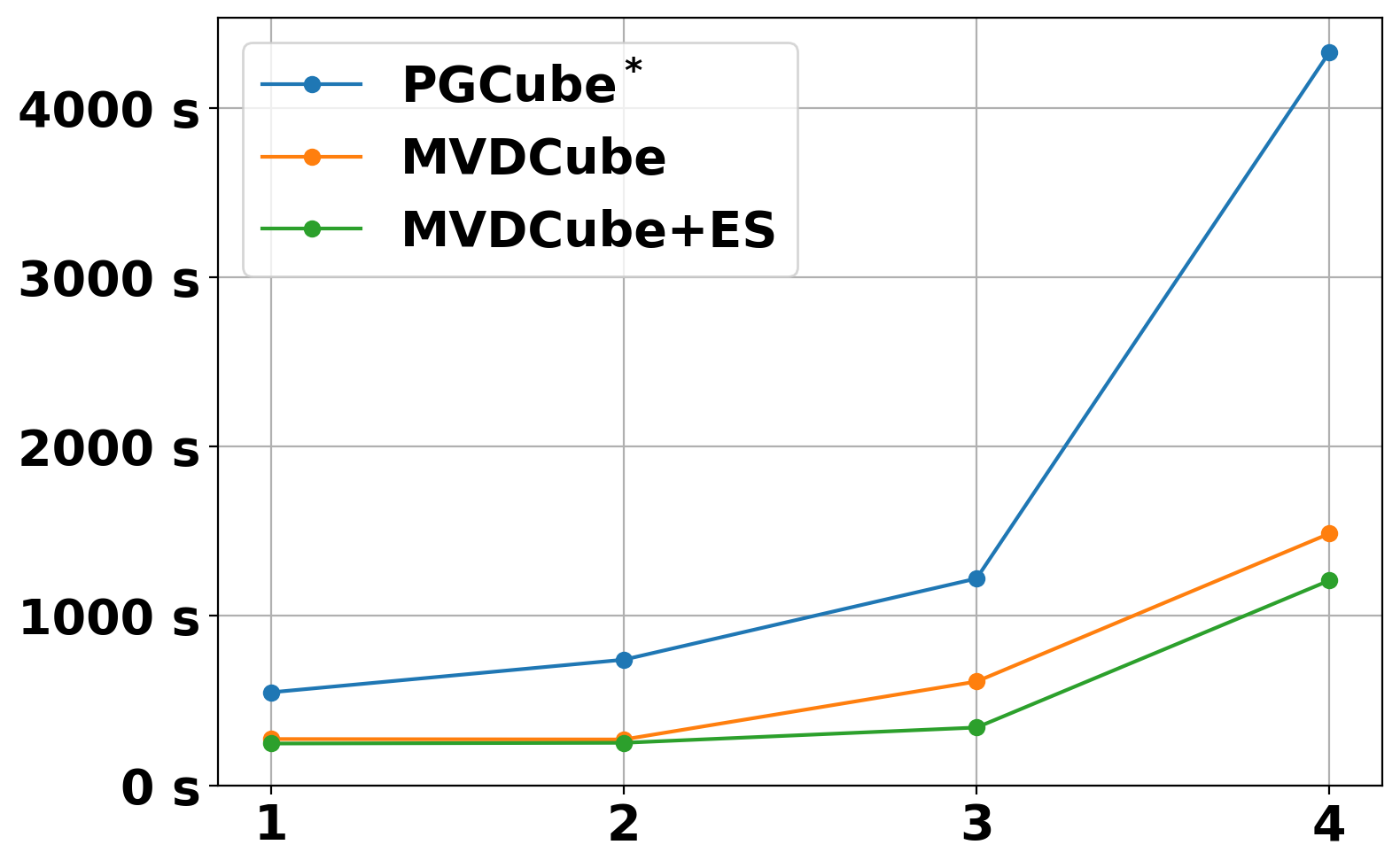}
		\veat{-2mm}
		\caption{\small Varying the number of dimensions $N$.\label{fig:scalability-varying-dimensions}}
	\end{subfigure}%
	\veat{-4mm}
	\caption{\small Scalability of \sys{} in the number of facts, measures, and dimensions.}
	\Description{Scalability of \sys{} in the number of facts, measures, and dimensions.}
	\veat{-3.5mm}
	\techreport{\veat{-1mm}}{}
\end{figure*}

% Experiment 6: vary the size of the candidate fact set: P1, P2, P3
\textbf{Experiment 6.} We now study the impact of $|CFS|$, $N$, and $M$ on the performance of \sys{}. As a base configuration, we fixed the synthetic graph with $|CFS|=5\text{M}$, 3 dimensions, and 15 measures (generated as above). For each dimension, we set the uniform value distribution (as above) and sparsity $0.1$, as Experiment 5 proved this configuration to be the most difficult.
Figures~\ref{fig:scalability-varying-facts}, \ref{fig:scalability-varying-measures}, \ref{fig:scalability-varying-dimensions} show the total execution time of \sys{}'s online pipeline when we vary $|CFS|\in\{1\text{M}, 2.5\text{M}, 5\text{M}, 7.5\text{M}, 10\text{M}\}$, $M\in\{5, 10, 15, 20, 25, 30\}$, and $N\in\{1,2,3,4\}$, respectively; the Aggregate Evaluation step was executed through \pgstar{}, \ouralgo{}, and \ouralgo{} with early-stop as evaluation modules. We chose \pgstar{} as on these graphs it is correct, and it is faster than \pgdist{}.
The figures show that \emph{\ouralgo{} scales linearly when $|CFS|$ and $M$ grow, and its run time increases more with $N$}; the latter is expected given the high number of lattices that are enabled by more dimensions. Further, \sys{} using \ouralgo{} \emph{is consistently faster} than using \pgstar{} by up to $2.9\times$;
%\PG{Checked the times for \pgstar{}: 4328.526s and \ouralgo{}: 1485.931, ratio 4328.526/1485.931 = 2.9130060547898924}
it also scales better as $|CFS|$, $N$ and $M$ grow, and  \emph{\ouralgo{} with early-stop is consistently the fastest} \textbf{(R9)}. Note that in Figure~\ref{fig:scalability-varying-measures}, \ouralgo{} with early-stop took slightly longer for $M\!\!=\!\!10$ than for $M\!\!=\!\!15$\cut{. The reason is that}: in these cases, the random samples drawn by early-stop (Section~\ref{sec:early-stop}) were less helpful for $M\!\!=\!\!10$ than for $M\!\!=\!\!15$.

\subsection{Experimental conclusions}
Our experimental results established, first, the need for a novel framework for finding interesting aggregates in RDF graphs: in heterogeneous graphs lacking well-defined facts, dimensions, and measures, Property Derivation increases significantly the space of interesting aggregates \textbf{(R1)}.
Due to multi-valued dimensions, relational aggregate evaluation algorithms often introduce errors \textbf{(R4)}, which can be very significant \textbf{(R5)}. On real-world graphs, our algorithm, \ouralgo{}, not only produces correct results but is also faster (by 20\% to 80\%) than the best comparable (PostgreSQL) baseline \textbf{(R2), (R3)}. 
Our novel early-stop technique reduces \ouralgo{}'s run time by 10\% to 43\% in many cases \textbf{(R6)}, while remaining accurate \textbf{(R7)}. In the entire online pipeline of \sys{}, the most time-consuming steps are Aggregate Evaluation, followed by Online Attribute Analysis \textbf{(R8)}. \ouralgo{} consistently outperforms \pgcube{} while scaling in the number of facts, measures, and dimensions; early-stop further improves the performance \textbf{(R9)}.

	\techreport{\veat{-3.5mm}}{}
\section{Related work}
\techreport{\veat{-1mm}}{}

%SHORTER VERSION
\textbf{Graph exploration.}
By providing visually meaningful, \cut{user-friendly, }interactive interfaces, RDF graph visualization~\cite{bookLDviz} allows casual users to access the data in RDF graphs. Based on the graph structure, content, and/or semantics, RDF summarization~\cite{cebiric:hal-01925496} computes a synopsis (summary) of the data, % a condensed data structure
 encapsulating the essential information of the graph from a given perspective. Example-based graph exploration, such as in~\cite{DBLP:conf/icde/LissandriniMPV18}, helps users discover data based on examples they specify. Our work is complementary to these approaches.

\textbf{Insight extraction from multidimensional data} is a common technique for data exploration. Research conducted in~\cite{botang}~and~\cite{quickinsights} provides automatic extraction of the top-$k$ insights from multidimensional \emph{relational} data. An insight is an observation derived from aggregation in multiple steps; it is considered interesting when it is remarkably different from others, or it exhibits a rising or falling trend. Multi-structural databases~\cite{DBLP:conf/pods/FaginGKNST05} distribute data across a set of dimensions, compare two sets of data along given dimensions, and separate the data into cohesive groups. %with respect to the known dimensions. 
A smart drill-down operator~\cite{DBLP:journals/tkde/JoglekarGP19} is proposed for interactively exploring a relational table to discover %the top-$k$ interesting 
groups of tuples that are frequent, specific,  %rather then general 
and diverse.
Works in this area assume a fixed relational schema; more recently, they consider graphs as in~\cite{BLECO201949}, but, unlike \sys{}, they require them to have a very regular \cut{and simple }structure.

\textbf{Visualization recommendation.} %Many works have been proposed on this topic. 
SeeDB~\cite{seedb} identifies, in relational data, the one-dimensional aggregates that exhibit the largest deviation between a target dataset and a reference dataset.
A study in~\cite{DBLP:journals/tkde/EhsanSC18} lays out a recommendation scheme for top-$k$ aggregate visualizations from relational data using a multi-objective utility function to prune as many low-utility views as possible.
%They focus on the impact of numeric dimension attributes in particular. 
Recent work~\cite{DBLP:conf/sigmod/Zhang20} %proposes a more general approach where the utility function is not defined a priori but 
shows how to automatically discover the utility function to match the user intentions.
DeepEye~\cite{deepeye} finds and ranks visualizations by combining a binary classifier, supervised learning, and expert rules.
%Similarly to our derivations, it applies operations that bin, group and/or aggregate the data to provide interesting visualizations. 
QAGView~\cite{qagview,DBLP:journals/pvldb/WenZRY18} provides summaries of high-valued aggregate query answers that ensure properties including coverage, diversity, and relevance, customized based on user preferences. LensXPlain~\cite{DBLP:journals/pvldb/MiaoLR19} helps users understand answers to aggregate queries by providing the top-$k$ explanations. 
%Explanations are predicates capturing a subset of tuples. Their interestingness depends on the impact ($i$)~they have on the query answer and ($ii$)~their removal has on the query answer.

In contrast to these works, \sys{} applies on a schemaless RDF graph, and hence must automatically derive those dimensions and measures that are good candidates to produce some insights.
%There is no universally accepted definition of interestingness; frequently, something unexpected (which differs from a certain reference) is considered interesting. The reference might be known a priori, come from historical data, or be the average behavior. So far we have experimented with variance, skewness and kurtosis; we are also working to use the entropy of data distribution along graph dimensions. Many different RDF visualizations techniques can be used, e.g.,~\cite{DBLP:conf/semweb/BenedettiBP15}.

\textbf{Cube computation} is at the heart of multidimensional data analysis and has been intensely studied~\cite{DBLP:journals/csur/MorfoniosKIK07}.
%It is also very expensive and implementations that compute each node separately are impractical given that the number of nodes is exponential with respect to the number of dimensions. 
%Much research has been conducted to provide efficient methods to implement the cube~\cite{DBLP:journals/csur/MorfoniosKIK07}. 
To limit the number of scans of the data and to share computation as much as possible, many algorithms %works proceed top-down, 
compute the aggregates in the lattice from one of their parents~\cite{DBLP:conf/vldb/AgarwalADGNRS96,molap,DBLP:conf/sigmod/ChenN05}. ArrayCube~\cite{molap} is a widely accepted algorithm in this category proposing a one-pass solution that simultaneously aggregates along multiple dimensions.

	\techreport{\veat{-1mm}}{}
\section{Conclusions and future work}
Discovering interesting insights from RDF graphs requires automatic,
expressive, and efficient methods. We presented \sys{}, an extensible
framework that enumerates a large and rich space of insights in 
the form of RDF aggregate queries and produces top-$k$ results that maximize a given
interestingness function. To efficiently explore the large
space of candidates aggregates, \sys{} introduces: ($i$)~\ouralgo{}, an efficient algorithm for evaluating many aggregates in a single pass
over the data, 20\% to 80\% faster than the best comparable method
implemented in PostgreSQL, and ($ii$)~a novel probabilistic technique that prunes uninteresting aggregates early. 
%allow reducing the aggregate evaluation effort. 
\sys{} scales well with the data size and the number of measures.

In future work, we plan to study more insight extraction methods to support numeric trends~\cite{botang}, time series, and geo-referenced data. Another research direction is  ``human-in-the-loop'' data exploration that allows the user to work synergistically with the system to broaden the set of insights discovered from large graphs.

	\techreport{\veat{-1.5mm}}{}
	\begin{acks}
		\techreport{\veat{-1mm}}{}
		Yanlei Diao and Pawe\l{} Guzewicz are supprted by the \grantsponsor{ERC}{European Research Council, H2020 research program}{} under GrantNo.:~\grantnum{ERC}{725561}{}, and by the \grantsponsor{}{Agence Nationale de la Recherche}{} under GrantNo.:~\grantnum{}{ANR-16-CE23-0010-01}{}.
		Mirjana Mazuran is supported by the \grantsponsor{ERC}{European Research Council, H2020 research program}{} under GrantNo.:~\grantnum{ERC}{800192}{}.
	\end{acks}
	
	\clearpage
	\bibliographystyle{ACM-Reference-Format}
	\bibliography{bibliography}

	\clearpage
	\techreport{\begin{appendices}
%\appendix

\section{Skewness and kurtosis as interestingness functions in early-stop\label{app:skewness-kurtosis}}
In case of variance $\frac{\partial\widehat{H}_r(\bm{y})}{\partial y_s}=\frac{2}{G-1}\left(\bm{y}_s-\frac{1}{G}\sum\limits_{i=1}^G \bm{y}_i\right)$ for $1\leq s\leq G$ (recall Section~\ref{sec:early-stop}).

In case of skewness, $\widehat{I}_r(\bm{y})=\left[\frac{1}{G}\sum\limits_{i=1}^G\left(y_i-\frac{1}{G}\sum\limits_{j=1}^G y_j\right)^3\right]\cdot\left[\widehat{H}_r(\bm{y})\right]^\frac{2}{3}$.
First, we derive $\frac{\partial\widehat{I}_r(\bm{y})}{\partial y_s}$:
\begin{align*}
	\frac{\partial\widehat{I}_r(\bm{y})}{\partial y_s}&=\frac{\partial}{\partial y_s}\left\{\left[\frac{1}{G}\sum\limits_{i=1}^G\left(y_i-\frac{1}{G}\sum\limits_{j=1}^G y_j\right)^3\right]\cdot\left[\widehat{H}_r(\bm{y})\right]^\frac{2}{3}\right\}\\
	&=\left\{\frac{\partial}{\partial y_s}\left[\frac{1}{G}\sum\limits_{i=1}^G\left(y_i-\frac{1}{G}\sum\limits_{j=1}^G y_j\right)^3\right]\right\}\cdot\left[\widehat{H}_r(\bm{y})\right]^\frac{2}{3}\\
	&+\left[\frac{1}{G}\sum\limits_{i=1}^G\left(y_i-\frac{1}{G}\sum\limits_{j=1}^G y_j\right)^3\right]\cdot\left\{\frac{\partial}{\partial y_s}\left[\widehat{H}_r(\bm{y})\right]^\frac{2}{3}\right\}\\
\end{align*}
Second, we derive the sub-expressions
\begin{align*}
	&\frac{\partial}{\partial y_s}\left[\frac{1}{G}\sum\limits_{i=1}^G\left(y_i-\frac{1}{G}\sum\limits_{j=1}^G y_j\right)^3\right]\\
	%&=\frac{1}{G}\left[\frac{\partial}{\partial y_s}\sum\limits_{i=1, i\neq k}^G\left(y_i-\frac{1}{G}\sum\limits_{j=1}^G y_j\right)^3+\frac{\partial}{\partial y_s}\left(y_s-\frac{1}{G}\sum\limits_{j=1}^G y_j\right)^3\right]\\
	%&=\frac{3}{G}\left[\sum\limits_{i=1, i\neq k}^G\left(y_i-\frac{1}{G}\sum\limits_{j=1}^G y_j\right)^2\cdot\left(-\frac{1}{G}\right)+\left(y_s-\frac{1}{G}\sum\limits_{j=1}^G y_j\right)^2\cdot\left(1-\frac{1}{G}\right)\right]\\
	%&=\frac{3}{G}\left[-\frac{1}{G}\sum\limits_{i=1}^G\left(y_i-\frac{1}{G}\sum\limits_{j=1}^G y_j\right)^2+\left(y_s-\frac{1}{G}\sum\limits_{j=1}^G y_j\right)^2\right]\\
	%&=\frac{3}{G}\left[-\frac{1}{G}\sum\limits_{i=1}^G\left(y_i^2-\frac{2y_i}{G}\sum\limits_{j=1}^G y_j+\frac{1}{G^2}\left(\sum\limits_{j=1}^G y_j\right)^2\right)+y_s^2-\frac{2y_s}{G}\sum\limits_{j=1}^G y_j+\frac{1}{G^2}\left(\sum\limits_{j=1}^G y_j\right)^2\right]\\
	%&=\frac{3}{G}\left[y_s^2-\frac{1}{G}\left(\sum\limits_{i=1}^G\left(y_i^2-\frac{2y_i}{G}\sum\limits_{j=1}^G y_j\right)+\frac{G}{G^2}\left(\sum\limits_{j=1}^G y_j\right)^2+2y_s\sum\limits_{j=1}^G y_j-\frac{1}{G}\left(\sum\limits_{j=1}^G y_j\right)^2\right)\right]\\
	&=\frac{3}{G}\left[y_s^2-\frac{1}{G}\left(\sum\limits_{i=1}^G\left(y_i^2-\frac{2y_i}{G}\sum\limits_{j=1}^G y_j\right)+2y_s\sum\limits_{j=1}^G y_j\right)\right]\\
\end{align*}
and
$$\frac{\partial}{\partial y_s}\left[\widehat{H}_r(\bm{y})\right]^\frac{2}{3}=\frac{2}{3}\left[\widehat{H}_r(\bm{y})\right]^{-\frac{1}{3}}\cdot\frac{\partial\widehat{H}_r(\bm{y})}{\partial y_s}$$
Then, coming back to the original equation, we have that
\begin{align*}
	\frac{\partial\widehat{I}_r(\bm{y})}{\partial y_s}&=\frac{3}{G}\left[y_s^2-\frac{1}{G}\left(\sum\limits_{i=1}^G\left(y_i^2-\frac{2y_i}{G}\sum\limits_{j=1}^G y_j\right)+2y_s\sum\limits_{j=1}^G y_j\right)\right]\cdot\left[\widehat{H}_r(\bm{y})\right]^\frac{2}{3}\\
	&+\frac{2}{3}\left[\frac{1}{G}\sum\limits_{i=1}^G\left(y_i-\frac{1}{G}\sum\limits_{j=1}^G y_j\right)^3\right]\cdot\left[\widehat{H}_r(\bm{y})\right]^{-\frac{1}{3}}\cdot\frac{\partial\widehat{H}_r(\bm{y})}{\partial y_s}
\end{align*}
Therefore, $\frac{\partial\widehat{I}_r(\bm{y})}{\partial y_s}$, as a combination of:
\begin{enumerate}
	\item $\widehat{H}_r(\bm{y})$, which is itself a combination of elementary (thus continuous) functions
	\item $\frac{\partial\widehat{H}_r(\bm{y})}{\partial y_s}$, which we showed previously to be continuous
	\item other elementary (thus continuous) functions
\end{enumerate}
is also continuous.

In case of kurtosis, $\widehat{J}_r(\bm{y})=\left[\frac{1}{G}\sum\limits_{i=1}^G\left(y_i-\frac{1}{G}\sum\limits_{j=1}^G y_j\right)^4\right]\cdot\left[\frac{G-1}{G}\widehat{H}_r(\bm{y})\right]^{-2}-3$.
First, we derive $\frac{\partial\widehat{J}_r(\bm{y})}{\partial y_s}$:
\begin{align*}
	\frac{\partial\widehat{J}_r(\bm{y})}{\partial y_s}&=\frac{\partial}{\partial y_s}\left\{\left[\frac{1}{G}\sum\limits_{i=1}^G\left(y_i-\frac{1}{G}\sum\limits_{j=1}^G y_j\right)^4\right]\cdot\left[\frac{G-1}{G}\widehat{H}_r(\bm{y})\right]^{-2}-3\right\}\\
	&=\left\{\frac{\partial}{\partial y_s}\left[\frac{1}{G}\sum\limits_{i=1}^G\left(y_i-\frac{1}{G}\sum\limits_{j=1}^G y_j\right)^4\right]\right\}\cdot\left[\frac{G-1}{G}\widehat{H}_r(\bm{y})\right]^{-2}\\
	&+\left[\frac{1}{G}\sum\limits_{i=1}^G\left(y_i-\frac{1}{G}\sum\limits_{j=1}^G y_j\right)^4\right]\cdot\left\{\frac{\partial}{\partial y_s}\left[\frac{G-1}{G}\widehat{H}_r(\bm{y})\right]^{-2}\right\}\\
\end{align*}
Second, we derive the sub-expressions
\begin{align*}
	&\frac{\partial}{\partial y_s}\left[\frac{1}{G}\sum\limits_{i=1}^G\left(y_i-\frac{1}{G}\sum\limits_{j=1}^G y_j\right)^4\right]\\
	%&=\frac{1}{G}\left[\frac{\partial}{\partial y_s}\sum\limits_{i=1, i\neq k}^G\left(y_i-\frac{1}{G}\sum\limits_{j=1}^G y_j\right)^4+\frac{\partial}{\partial y_s}\left(y_s-\frac{1}{G}\sum\limits_{j=1}^G y_j\right)^4\right]\\
	%&=\frac{4}{G}\left[\sum\limits_{i=1, i\neq k}^G\left(y_i-\frac{1}{G}\sum\limits_{j=1}^G y_j\right)^3\cdot\left(-\frac{1}{G}\right)+\left(y_s-\frac{1}{G}\sum\limits_{j=1}^G y_j\right)^3\cdot\left(1-\frac{1}{G}\right)\right]\\
	%&=\frac{4}{G}\left[-\frac{1}{G}\sum\limits_{i=1}^G\left(y_i-\frac{1}{G}\sum\limits_{j=1}^G y_j\right)^3+\left(y_s-\frac{1}{G}\sum\limits_{j=1}^G y_j\right)^3\right]\\
	%&=\frac{4}{G}\left[-\frac{1}{G}\sum\limits_{i=1}^G\left(y_i^3-\frac{3y_i^2}{G}\sum\limits_{j=1}^G y_j+\frac{3y_i}{G^2}\left(\sum\limits_{j=1}^G y_j\right)^2-\frac{1}{G^3}\left(\sum\limits_{j=1}^G y_j\right)^3\right)\right.\\
	%&\left.+y_s^3-\frac{3y_s^2}{G}\sum\limits_{j=1}^G y_j+\frac{3y_s}{G^2}\left(\sum\limits_{j=1}^G y_j\right)^2-\frac{1}{G^3}\left(\sum\limits_{j=1}^G y_j\right)^3\right]\\
	%&=\frac{4}{G}\left[y_s^3-\frac{1}{G}\left(\sum\limits_{i=1}^G\left(y_i^3-\frac{3y_i^2}{G}\sum\limits_{j=1}^G y_j+\frac{3y_i}{G^2}\left(\sum\limits_{j=1}^G y_j\right)^2\right)-\frac{G}{G^3}\left(\sum\limits_{j=1}^G y_j\right)^3\right.\right.\\
	%&\left.\left.+3y_s^2\sum\limits_{j=1}^G y_j-\frac{3y_s}{G}\left(\sum\limits_{j=1}^G y_j\right)^2+\frac{1}{G^2}\left(\sum\limits_{j=1}^G y_j\right)^3\right)\right]\\
	&=\frac{4}{G}\left[y_s^3-\frac{1}{G}\left(\sum\limits_{i=1}^G\left(y_i^3-\frac{3y_i^2}{G}\sum\limits_{j=1}^G y_j+\frac{3y_i}{G^2}\left(\sum\limits_{j=1}^G y_j\right)^2\right)\right.\right.\\
	&\left.\left.+3y_s^2\sum\limits_{j=1}^G y_j+\frac{3y_s}{G}\left(\sum\limits_{j=1}^G y_j\right)^2\right)\right]\\
\end{align*}
and
\begin{align*}
\frac{\partial}{\partial y_s}\left[\frac{G-1}{G}\widehat{H}_r(\bm{y})\right]^{-2}&=\frac{2(G-1)}{G[\widehat{H}_r(\bm{y})]^3}\cdot\frac{G-1}{G}\frac{\partial\widehat{H}_r(\bm{y})}{\partial y_s}\\
&=\frac{2(G-1)^2}{G^2[\widehat{H}_r(\bm{y})]^3}\cdot\frac{\partial\widehat{H}_r(\bm{y})}{\partial y_s}
\end{align*}
Then, coming back to the original equation, we have that
\begin{align*}
	\frac{\partial\widehat{J}_r(\bm{y})}{\partial y_s}&=\frac{4}{G}\left[y_s^3-\frac{1}{G}\left(\sum\limits_{i=1}^G\left(y_i^3-\frac{3y_i^2}{G}\sum\limits_{j=1}^G y_j+\frac{3y_i}{G^2}\left(\sum\limits_{j=1}^G y_j\right)^2\right)\right.\right.\\
	&\left.\left.+3y_s^2\sum\limits_{j=1}^G y_j+\frac{3y_s}{G}\left(\sum\limits_{j=1}^G y_j\right)^2\right)\right]\cdot\left[\frac{G-1}{G}\widehat{H}_r(\bm{y})\right]^{-2}\\
	&+\frac{2(G-1)^2}{G^2[\widehat{H}_r(\bm{y})]^3}\left[\frac{1}{G}\sum\limits_{i=1}^G\left(y_i-\frac{1}{G}\sum\limits_{j=1}^G y_j\right)^4\right]\cdot\frac{\partial\widehat{H}_r(\bm{y})}{\partial y_s}
\end{align*}
Therefore, $\frac{\partial\widehat{J}_r(\bm{y})}{\partial y_s}$, as a combination of:
\begin{enumerate}
	\item $\widehat{H}_r(\bm{y})$, which is itself a combination of elementary (thus continuous) functions
	\item $\frac{\partial\widehat{H}_r(\bm{y})}{\partial y_s}$, which we showed previously to be continuous
	\item other elementary (thus continuous) functions
\end{enumerate}
is also continuous.

\section{Sum as an aggregate function in early-stop\label{app:sum}}
\rev{To obtain the $sum$ estimate, we compute the product of the size of the $i$-th aggregate group $c_i$ and the sample mean. We estimate $c_i$ while sampling during Data Translation: the count in the root node of the lattice is always correct, whereas in the other lattice nodes, depending on the presence of multi-valued dimensions, it may be overestimated. Recall from Section~\ref{sec:estimating-interestingness} the sample mean estimator $\bm{\bar{Y}}_i=\frac{1}{r}\sum\limits_{j=1}^{r} X_j$, and that $\bm{\bar{Y}}_i\sim\mathcal{N}(\mu_i,\frac{\sigma_i^2}{r})$ as ${r\to\infty}$. We now construct a new estimator $\bm{S}_i=\frac{c_i}{r}\sum\limits_{j=1}^{r} X_j=c_i\bar{\bm{Y}_i}$. As a consequence, we have that $\bm{S}_i\sim\mathcal{N}(c\mu_i,\frac{c_i^2\sigma_i^2}{r})$ as ${r\to\infty}$. This leads to the correct $sum$ estimate thanks to the estimator mean equal to $c_i\mu_i$. While deriving the CI bounds in the proof, we account for the different variance of the estimator by applying $\Var(\bm{S}_s)=\frac{c_s^2\sigma_s^2}{r}$ to obtain $\tau^2$.}

\rev{Finally, the CI bounds are scaled by the constant factor of $c_s^2$ for each aggregate group w.r.t. the case of the average estimate: the impact of the scaling is hidden within $\widehat{\tau^2}$, the estimator of $\tau^2$. We thus obtain the formula for our $sum$-estimate confidence interval:
\begin{center}
	$P\left(\left|\widehat{H}_r(\bm{S})-\widehat{H}_r(\bm{c}\bm{\mu})\right|\leq\sqrt{\frac{z_{1-\alpha}^2 \widehat{\tau^2}}{r}}\right)\approx (1-\alpha)$
\end{center}
\noindent where $\bm{S}=\left(\bm{S}_1,\bm{S}_2,\ldots,\bm{S}_G\right)^\intercal$ and $\bm{c}=\left(c_1,c_2,\ldots,c_G\right)^\intercal$ (the correct aggregate group sizes).}

\section{Min and max as aggregate functions in early-stop\label{app:min-max}}
Point estimates for min, and max are sample min, respectively, max: the function applied over the sample, i.e., $\widehat{Z}_r(\bm{x})=\min\limits_r(\bm{x})$ or $\widehat{Z}_r(\bm{x})=\max\limits_r(\bm{x})$. We then bound $\widehat{H}_r(\bm{y})$, the variance of $\bm{y}=\widehat{Z}_r(\bm{x})$, with Popoviciu's inequality for the upper bound: $\widehat{H}_r(\bm{y})\leq\frac{1}{4}(\widehat{Z}_r(\bm{x})-b)^2$, where $b$ is the lower bound on min (respectively the upper bound on max).

Analogically, we apply Sz\H{o}kefalvi-Nagy's inequality for the lower bound: $\widehat{H}_r(\bm{y})\leq\frac{(\widehat{Z}_r(\bm{x})-b)^2}{2r}$. We obtain the global statistics for $b$ for each attribute during Online Attribute Analysis step (Section~\ref{sec:overview}).

% interesting aggregates
%\input{appendix-intresting-MDAs}

\end{appendices}}{}

\end{document}